\documentclass[preprint, superscriptaddress, amsmath, amssvymb, prfluids, longbibliography]{revtex4-2}

\usepackage{graphicx}
\usepackage{upgreek}
\usepackage{amsfonts}
\usepackage{mathrsfs}
\usepackage{amsthm}
\usepackage[figuresright]{rotating}
\usepackage{appendix}
\usepackage[T1]{fontenc}
\usepackage{newtxtext}
\usepackage{newtxmath}
\usepackage{textcomp}
\usepackage{xcolor}
\usepackage{onimage}
\usepackage{caption}
\usepackage{subcaption}

\usepackage{tikz}

\usetikzlibrary{arrows,positioning} 
\tikzset{
    >=stealth',
    punkt/.style={
           rectangle,
           rounded corners,
           draw=black, very thick,
           text width=6.5em,
           minimum height=2em,
           text centered},
    pil/.style={
           ->,
           thick,
           shorten <=2pt,
           shorten >=2pt,}
}

\newcommand{\BibitemShut}[1]{}
\usepackage[colorlinks,allcolors=blue]{hyperref}
\usepackage{color}
\usepackage{mathtools}
\usepackage{tabls}
\usepackage{bm}
\usepackage{bbm}

\usepackage{tikz}
\usetikzlibrary{matrix}
\usetikzlibrary{calc}
\usetikzlibrary{patterns,fadings}
\usetikzlibrary{positioning}
\usetikzlibrary{decorations.markings}
\usetikzlibrary{fadings}

\usetikzlibrary{arrows}

\newcommand\uu{{\bf{u}}}

\newcommand\x{{\bf{x}}}
\newcommand\y{{\bf{y}}}

\newcommand\DDt{\ensuremath{\frac{d}{dt}}}
\newcommand\ddt{\ensuremath{\frac{\partial}{\partial t}}}

\definecolor{indi}{rgb}{0,0.255,0.4157}
\definecolor{deepblue}{rgb}{0,0,0.5}
\definecolor{deepred}{rgb}{0.6,0,0}
\definecolor{deepgreen}{rgb}{0,0.5,0}
\definecolor{lightblue}{rgb}{0.95,0.95,1}
\definecolor{lightgrey}{rgb}{0.6,0.6,0.6}
\usepackage{listings}
\usetikzlibrary{decorations.pathreplacing}

\begin{document}

\preprint{APS/123-QED}

\title{Mixing by stirring:\\optimizing shapes and strategies}

\author{Maximilian F. Eggl{{\href{https://orcid.org/0000-0001-5815-1045}{\includegraphics{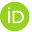}}}}}
 \email{maximilian.eggl@uni-mainz.de}
\affiliation{%
Institute of Physiological Chemistry, Johannes Gutenberg University of Mainz, Mainz, Germany
}%
\author{Peter J. Schmid{{\href{https://orcid.org/0000-0002-2257-8490}{\includegraphics{orcid_logo-eps-converted-to}}}}}%

\affiliation{%
Department of Mechanical Engineering, KAUST, Thuwal, Makkah, Saudi Arabia
}%

\date{\today}

\begin{abstract}
The mixing of binary fluids by stirrers is a commonplace procedure in many industrial and natural settings, and mixing efficiency directly translates into more homogeneous final products, more enriched compounds, and often substantial economic savings in energy and input ingredients. Enhancements in mixing efficiency can be accomplished by unorthodox stirring protocols as well as modified stirrer shapes that utilize unsteady hydrodynamics and vortex-shedding features to instigate the formation of fluid filaments which ultimately succumb to diffusion and produce a homogeneous mixture. We propose a PDE-constrained optimization approach to address the problem of mixing enhancement for binary fluids. Within a gradient-based framework, we target the stirring strategy as well as the cross-sectional shape of the stirrers to achieve improved mixedness over a given time horizon and within a prescribed energy budget. The optimization produces a significant enhancement in homogeneity in the initially separated fluids, suggesting promising modifications to traditional stirring protocols.
\end{abstract}

\maketitle


\section{Introduction}

The mixing of binary fluids is a fundamental process that underpins a wide array of applications in fluid mechanics. From tracking ocean pollution~\citep{kukulka2012effect} where wind-driven mixing affects the distribution of buoyant plastic debris, the study of the ventilation of buildings ~\citep{Hunt1999,Linden1999} to improved mixing of fuel in combustion engines~\citep{peng2020mixing}, the importance of mixing -- and the incentive to do it well -- can hardly be overstated.  Particularly in industrial settings, mixing has become increasingly important as most chemical engineering processes are constantly optimized for energy efficiency, yield, output quality or time-to-product. Recent examples of this effort can be seen in \citep{thompson2017use} using computational tools to study industrial pre-mixers, in \citep{mihailova2018optimisation} investigating helical mixers and in \citep{mazumdar2017modeling} modeling and optimizing gas stirred ladle systems. Far from exhaustive, these examples illustrate the substantial interest and research directed towards a better understanding of mixing and its optimization. 

In particular the physical mechanisms responsible for good or acceptable mixing have been the focus of a great many studies, in an effort to enhance mixing efficiency and homogeneity of the final mixture. In an early investigation \cite{Spencer1951}  identified stretching, cutting and folding as the primary building blocks of mixing, while \cite{Ottino1989} pioneered a theoretical framework for a simplified analysis of viscous mixing by ignoring interfacial tension. In a more recent effort, the fundamental process and the definition of mixing has attracted the attention of mathematicians \citep{Mathew2005,mathew2007,Thiffeault2012} who derived necessary conditions for optimality in the viscous regime and proposed appropriate measures of mixedness.

While earlier studies have concentrated on the high-viscosity regime, a significant effort has also been directed towards understanding mixing in an inertially dominated, chaotic and turbulent setting. Explorations into this regime include work using direct numerical simulations to study passive scalar mixing via shock-wave interactions over a wide array of governing parameters such as Mach numbers, Reynolds number and Schmidt numbers \citep{gao2020Parametric}. Based on the early landmark work of \cite{Aref1984} who isolated the process of filament formation through the stirrers' shear action which subsequently fed into an effective diffusion process, numerous studies suggested schemes and mechanistic procedures to enhance mixing. A modern account of mixing by vortical motion using a network formalism of interacting vortices has been given by \cite{meena2021identifying}. Furthermore, laminar and turbulent fluid motion has been used and analyzed in 
\cite{chien1986laminar}, where oscillatory wall motion has been used to enhance mixing, and, more recently, \cite{kadoch2020efficiency} found that turbulent flow in the near-wall region accomplishes a markedly higher degree of mixing in a passive scalar field than than either chaotic or laminar flow.

A better understanding of the underlying mechanisms for effective mixing have further encouraged investigations into optimizing, or at least enhancing, the overall mixing process within prescribed bounds and constraints. Treatises on this topic include the use of boundary feedback control based on open-loop system characteristics \citep{Balogh2005,Gubanov2010} to optimize the velocity profile imposed by a double-periodic sine-mixer. The newly proposed mixing measures (for example, the $H^{-1}$-norm, equivalent to the square-root of the variance of a low-pass filtered passive tracer field) have been used in \cite{Lin2011} to achieve improved mixing. These studies corroborate earlier work by \cite{Liu2008} who proved the existence of an optimal flow and derived associated optimality conditions for achieving it. In a further study of mixing enhancement, and building on fundamental preparatory work \citep{Foures2012,Foures2013}, \cite{Foures2014} demonstrated the use of wall-mounted blowing-and-suction strategies in a channel, coupled to an adjoint-based optimization formalism, as an effective tool to enhance mixing -- if an appropriate mixing measure~\citep[see][]{Mathew2005} is considered as an cost objective. This study has been extended to three-dimensional and stratified flows by \cite{Marcotte2018} and~\cite{Vermach2018} with similar findings.  In a recent article~\citep{wu2018jet}, mixing by jets has been demonstrated in an experimental setup, using machine learning to arrive at enhanced mixing.  Optimization of mixing via jet actuators has also been pursued in \cite{hilgers2001optimization}, finding that a combined axial and helical actuation is far more effective of reaching a homogeneous mixture than manipulating solely by helical actuation. Finally, and similar in approach to this study, \citep{alexias2020shape} employed a gradient-based approach to enhance passive mixing by optimizing static baffles.

In this article, we will pursue stirrer-based mixing enhancement by combining a gradient-based and PDE-constrained optimization formulation with a modern description of mixing measures. This approach continues earlier work \citep{EgglSchmid2018,eggl2020mixing} by extending the optimization space to (i) topology changes for the stirrers' cross-sections and to (ii) the temporal velocity protocol along given paths. In addition, we explore a parameter regime that has received less attention from the mixing community: the inertial, but laminar regime, described by a Reynolds number above the Stokes-regime, but below a critical value that would induce turbulent fluid motion. In this chosen regime, we expect and observe a rich flow behavior, characterized by strong shear and robust vortical motion, ideal for seeking a sophisticated and perhaps non-intuitive mixing strategy, while at the same time avoiding exceeding sensitivity due to turbulent fluctuations (which would run counter to our gradient-based optimization approach). We anticipate an optimized stirring protocol that, simultaneously and/or progressively, takes advantage of advective, unsteady and diffusive processes.
Despite the restriction, this parameter regime covers many applications of industrial and technological interest. 

We will concentrate on an active stirrer system and utilize the computational framework introduced in previous work to optimize the velocity of the stirrers along a prescribed path as well as their cross-sectional area to obtain mixing protocols that enhance mixing in a binary, incompressible and two-dimensional fluid system. We enforce constraints driven by mechanical and material limitations of the stirrer system, which translate into bounds on the stirrers' velocities and accelerations; moreover, we restrict the stirrers' shapes to avoid multiply-connected cross-sections or delicate shapes. The focus of this work is in the enhancement of mixing using a fixed time horizon and energy budget and the description of the identified stirring protocol in terms of a dynamic interplay of the vortical structures with themselves and with the embedded stirrers. 

The rest of this article is organized as follows; in \S \ref{sec:theory} we present the governing equations, motivate our chosen measure of mixing, and then describe mixing enhancement as an optimization problem. Following on from this, in \S \ref{sec:comp} we briefly discuss the numerical implementation that arises from our governing equations; this section is included for completeness, and special emphasis is drawn on computational concepts arising in the present study. Subsequently in \S \ref{sec:results} we introduce three distinct test cases, present the optimization results and discuss the key features responsible for an enhancement in mixing. The three optimization experiments have been designed as follows: case \emph{(i)} aims to optimize the cross-sectional shape of the stirrers while keeping the velocities of the stirrers on their respective paths fixed to a prescribed protocol. The resulting enhanced shapes are then fixed for case \emph{(ii)}, which then targets the velocities along the paths in an effort to further enhance mixing. Lastly, case \emph{(iii)} presents an optimization where both the cross-sectional shapes {\it{and}} path velocities are optimized simultaneously, rather than successively. In \S \ref{sec:Conc} we draw conclusions and provide an outlook for future optimization studies.

\section{Theoretical background}
\label{sec:theory}
\subsection{Governing equations}

The focus of this study is the mixing process of a binary, miscible and Newtonian fluid by multiple stirrers on prescribed paths, and its optimization by manipulating the cross-sectional shape of the stirrers and/or their stirring strategy within specified constraints. A two-dimensional configuration is considered. The process can be simulated by solving the fluid equations of motion, augmented by a transport equation for a passive scalar $\theta$ that identifies the binary fluid. We have

\begin{align}
 \ddt \uu + \uu \cdot \nabla \uu + \frac{1}{C_\eta}\left(\chi \uu -
 \chi_k \uu_{s,k} \right) + \nabla p - \frac{1}{Re} \nabla^2 \uu &=
 0, \label{eq:GovEqu1} \\[4pt]
 \nabla \cdot \uu &= 0, \label{eq:GovEqu2} \\[4pt]
 \ddt \theta + \left( (1-\chi) \uu + \chi_k \uu_{s,k} \right) \cdot
 \nabla \theta - \nabla \cdot \left( \left[ \frac{1}{Pe} (1-\chi) +
   \frac{\chi}{C_{\eta}} \right] \nabla \theta \right) &= 0.
  \label{eq:GovEqu3}
\end{align}

\noindent with $\uu$ as the velocity vector, $p$ as the pressure field and $\theta$ as a passive scalar (ranging from zero in one fluid to one in the other). The governing equations have been cast in non-dimensional form using a characteristic length $L_0$ and a velocity scale $u_0.$ In our case, we choose the stirrer diameter and the initial control velocity of the stirrer as our characteristic quantities. This choice introduces the Reynolds number $Re$ and the P\'eclet number $Pe,$ which express the kinematic viscosity and the diffusion coefficient of the mixing fluid in non-dimensional form. Furthermore, the system~(\ref{eq:GovEqu1})-(\ref{eq:GovEqu3}) contains terms that model the embedded stirrers via a Brinkman penalization approach \citep[see][]{Angot1999}. The multiple solid stirrers, indexed by the subscript $k$, are described in shape by masks $\chi_k,$ equal to one for points occupied by the $k$-th stirrer and zero outside of it. The mask $\chi$ (without subscripts) accounts for the overall geometry, such as the domain boundaries. In our configuration, the stirrers travel on circular paths throughout our domain. It thus seems prudent to redefine the velocity of the $k$-th stirrer $\bm{u}_{s,k}$ using a polar coordinate system. To this end, we introduce the vector-valued function
$\bm{l}$, which transforms between the polar and Cartesian coordinate systems
according to $\bm{l}(\phi) = (l_1(\phi),l_2(\phi)) = (-\sin(\phi),\cos(\phi))$ with
$\phi$ as the angle traversed along the path of the circle. The
parameterization of the velocity of the $k$-th stirrer thus becomes
\begin{eqnarray}
  \bm{u}_{s,k}
  &=&\omega_{k}(t)r_k\bm{l}(\varphi_k(t)), \label{FLUSI:US}
\end{eqnarray}
where $r_k$ denotes the radius of the circular path for stirrer $k,$ and $\varphi$ denotes the sum of the angles traveled along the path according to 
\begin{align}
    \varphi_k(t) = \int_0^t\omega_{k}({\tau}) \ \text{d}{\tau}.
\end{align}
The constant $C_\eta$ in the governing equations ensures the rapid relaxation of the fluid variables towards the respective values imposed by the stirrers or the geometry. The above formalism allows for an efficient treatment of objects moving through a background grid on which the motion of the surrounding fluid is described and solved. Details of the above approach, together with its numerical implementation, can be found in \cite{EgglSchmid2018}. The particular set-up, shown in~(\ref{eq:GovEqu1})-(\ref{eq:GovEqu3}), imposes no-slip velocity boundary conditions on the stirrers and no-flux Neumann conditions for the passive scalar on the solid bodies.

The above governing equations provide a constraint on the optimization of mixedness of the binary fluid via the stirrers. Besides this constraint, a cost objective has to formulated that quantifies the degree of homogeneity of the binary fluids. We follow earlier studies and use the mix-norm as a measure of mixedness. Representing a Sobolev norm of negative fractional index, it is defined as 

\begin{align}
  \Vert \theta \Vert_{\rm{mix}} &\equiv \frac{1}{\vert V
    \vert}\int_{V} \Vert \nabla^{-2/3} \theta(\x,t) \Vert \text{
    d}V,
    \label{eq:SobolevNorm}
\end{align}
with $V$ denoting our computational domain, and $\vert
V \vert$ representing its size (volume or area). This type of norm gives weight to larger structures and de-emphasizes small-scale structures \citep{Mathew2005,Thiffeault2012,Foures2014}. Used within an optimization framework, it targets and promotes the breakdown of large scales into smaller structures, as well as the generation of thin filaments. Smaller values of the mix-norm tend to indicate a more homogeneous mixture. 

To complete the optimization problem we have to decide on a time horizon $T$ over which the stirrers will be active. In addition, we have to specify the total amount of energy $E_0$ exerted by the stirrers, as well as the limits on the stirrers' velocities, $\bm{u}_{s,k}$, and accelerations, $\bm{a}_{s,k}$. For an argument for this set of constraints, the reader is referred to \cite{Eggl2020}. The full optimization problem can then be stated as 

\begin{align}
  & \min \left\{ \int_0^T \Vert \theta \Vert_{\rm{mix}} \ dt \right\}
  \label{eq:Cost}
  \\[4pt]
  & \hbox{subject to equations}~(\ref{eq:GovEqu1})-(\ref{eq:GovEqu3}) \\[4pt]
  & \hbox{ and} \quad \int_0^T \sum_k \Vert \bm{u}_{s,k} \Vert^2 \ dt \leq E_0 \\[4pt]
  & \hbox{ and} \quad \bm{u}_{s,\rm{lower}} \leq \bm{u}_{s,k} \leq \bm{u}_{s,\rm{upper}} \qquad k=1,2, \\[4pt]
  & \hbox{ and} \quad \bm{a}_{s,\rm{lower}} \leq \bm{a}_{s,k} \leq \bm{a}_{s,\rm{upper}} \qquad k=1,2.
  \label{eq:constropt}
\end{align}
This system has to be discretized, embedded within a gradient-based (direct-adjoint) framework and coupled to a standard unconstrained optimization routine. 

\section{Computational framework}
\label{sec:comp}

Since the computational framework, including the discretization of the governing equations, has been presented in full in \cite{EgglSchmid2018}, we will only briefly discuss the numerical details here for completeness. The starting point for the computational implementation was the open-source software FLuSI \citep{Kolomenskiy2009,Engels2015}, a parallelized Fourier-based pseudo-spectral code. The three-dimensional incompressible Navier-Stokes equations are solved on a triple periodic domain using an equidistant mesh. Time-stepping is performed adaptively by multi-stage schemes and is based on numerical stability constraints. The interaction between fluid and solid cells is modeled using a Brinkman-style penalization method, which was discussed above in relation to the governing equations.

Based on the Fourier representation of the flow field variables, we can formulate spatial derivatives as a multiplication by a Fourier discretization matrix. We thus replace the derivative terms in the governing equation as follows 
\begin{align}
    \frac{\partial}{\partial x_i} \rightarrow \bf{A}_i
\end{align}
where the index $i$ denotes the respective coordinate direction or the associated spatial wavenumber. Under this transformation we can state the semi-discretized equations, where we have assumed the Einstein summation convention over repeated indices, as 

\begin{align}
  \DDt \uu_i + \uu_j \circ \left[ {\bf{A}}_j \uu_i \right] +
  \frac{1}{C_\eta}\left(\chi \circ \uu - \chi_k \circ
  {\uu_{s,k}}\right)_i + {\bf{A}}_i p - \frac{1}{Re} {\bf{A}}_j {\bf{A}}_j \uu_i &= 0, \\[4pt]
  {\bf{A}}_j \uu_j &= 0, \\[4pt]
  \DDt \theta + \left( ({\bf{1}} - \chi) \circ \uu + \chi_k \circ {\uu_{s,k}} \right)_j \circ \left[ {\bf{A}}_j \theta \right] - {\bf{A}}_j \left[ \frac{1}{Pe} \left( {\bf{1}} - \chi \right) + \kappa \chi \right] \circ {\bf{A}}_j \theta &= 0. \phantom{123456}
\end{align}
In the above expressions we denote by $\circ$ the Hadamard (elementwise) product. The solution of this set of equations follows a typical operator-splitting approach, where a pressure-delayed solution is computed, which subsequently is corrected by a pressure Poisson equation of the form  
\begin{equation}
  {\bf{A}}_j {\bf{A}}_j p + {\bf{A}}_i \left( \bm{u}_j
  \circ \left[ {\bf{A}}_j \bm{u}_i \right] \right) +
        {\bf{A}}_j \left[ \frac{\chi}{C_{\eta}} \circ
          \bm{u}- \frac{\chi_k}{C_{\eta}} \circ
          \bm{u}_{s,k} \right]_j = 0.\label{FLUSI:Eq3}
\end{equation}
The iterative progression of this system of equations forms the forward problem for our mixing setup, resulting in temporal snapshots of the flow field and the passive scalar.  

\subsection{Optimization framework using adjoint techniques}

The above semi-discretized system has to be embedded in an overall optimization scheme. This is accomplished by formulating an unconstrained optimization problem consisting of our cost objective ${\cal{J}}$ (based on the mix-norm, and defined by \eqref{eq:Cost}) and Lagrange-multiplier terms enforcing all our constraints. The so-called augmented Lagrangian is given as 

 \begin{align}
 \mathcal{L} = \mathcal{J} - &\int_0^T \bm{u}_i^{\dag,H}M\left\{\partial_t\bm{u} + \bm{u}_j\circ[\bm{A}_j\bm{u}] + \frac{\chi}{C_\eta}\circ\bm{u}-\frac{\chi_k}{C_\eta}\circ\bm{u}_{s,k}+
 \mathbf{A}p - Re^{-1}[\bm{A}_j\bm{A}_j\bm{u}]\right\}_i \nonumber\\
 &+ p^{\dag,H}M\left\{[\bm{A}_j\bm{A}_j]p + \bm{A}_i(\bm{u}_j\circ[\bm{A}_j \bm{u}_i]) + \bm{A}_j\left[\frac{\chi}{C_\eta}\circ\bm{u}-\frac{\chi_k}{C_\eta}\circ\bm{u_{s,k}}\right]_j\right\} \nonumber \\
 & +\theta^{\dag,H}M\left\{\partial_t\theta + (\bm{1}-\chi)\circ\bm{u}_j\circ[\bm{A}_j\theta]-\chi\circ(\bm{u}_{s,k})_j\circ[\bm{A}_j\theta]-\bm{A}_j([Pe^{-1}(\bm{1}-\chi)+\kappa\chi]\circ \bm{A}_j\theta)\right\} \nonumber \\
 &+\chi_k^{\dag,H}M[\chi_k-g_k(\bm{x},t)] +\omega_{k}^{\dag,H}M[\omega_{k} - z_k(t)] \ \text{d} t 
 \label{Eq:AugLagrange}
 \end{align}
where we recognize the various terms of the governing equations, each premultiplied by a Lagrange multiplier or an adjoint variable. We denote by $z_k,g_k$ the given previous values of $\omega_{k}$  and $\chi_{k}$, respectively. All adjoint variables act as independent variables in the augmented Lagrangian ${\cal{L}},$ and can be linked to sensitivities of the cost functional ${\cal{J}}$ with respect to respective flow variables or parameters. The scalar product chosen in the above expression is defined by the positive-definite weight matrix $M$ which accounts for mesh geometry or user-specified weights.
The next step consists of taking the first variation of ${\cal{L}}$ with respect to all independent variables and rendering these variations zero (for an optimum). The first variation with respect to the adjoint variables recovers the governing equations stated above. The first variation with respect to the direct variables (i.e., flow velocities, pressure and passive scalar) produces -- after some cumbersome algebraic manipulations -- governing equations for the adjoint variables. They read  

\begin{align}
  \DDt \uu^{\dag}_i - \Pi^{\dag}_j \circ [{\bf{A}}_i \uu_j ] -
       {\bf{A}}_j^H [\uu_j \circ \Pi^{\dag}_i] - \frac{\chi}{C_{\eta}} \circ \Pi^{\dag}_i + \frac{1}{Re} {\bf{A}}_j^H {\bf{A}}_j^H \uu_i^{\dag} -({\bf{1}}-\chi) \circ \theta^{\dag} \circ[ {\bf{A}}_i \theta]
   &= 0, \\[4pt]
  {\bf{A}}_j^H \Pi_j^\dag &= 0 \\[4pt]
  \DDt \theta^{\dag} - {\bf{A}}_j^H [({\bf{1}}-\chi) \circ \uu_j
    \circ \theta^{\dag}] + {\bf{A}}_j^H ([\frac{1}{Pe}({\bf{1}}-\chi) +
    \kappa\chi] \circ {\bf{A}}_j^H \theta^{\dag})
  -{\bf{A}}_j^H [ \chi_k \circ (\uu_{s,k})_j\ \circ \theta^{\dag}]
  &= 0
  \label{Adjoint:EqEnd}
\end{align}
where we have introduced the abbreviation $\Pi^{\dag}_i = \uu^{\dag}_i + {\bf{A}}^{H}_i p^{\dag}$. We also obtain the terminal conditions (which result from integration by parts in the time-domain) ,
\begin{align}
  \bm{u}^{\dag}(\bm{x},T) = 0, \qquad \qquad
  \theta^{\dag}(\bm{x},T) =  \frac{2}{|V|}\left({\bf{A}}_i^{-2/3}\right)^H\left({\bf{A}}_i^{-2/3}\theta(\bm{x},T)\right).
\end{align}
This set of equations has to be solved backward in time using the same computational framework as introduced for the forward problem. 

The final ingredient for the optimization problem is the first variation of the augmented Lagrangian ${\cal{L}}$ with respect to the control variable. Two cases shall be considered in this article: (i) the velocity of the stirrers along their circular path, and (ii) the cross-sectional shape of the stirrers. These two cases are treated by altering the variables $\omega_k$ and $\chi_k$, respectively. Either case produces a gradient of the cost functional that ultimately will be used to guide an unconstrained, gradient-based optimization algorithm towards an enhanced solution with an improved value of the cost functional (mixedness measure). 

\subsection{Gradient for velocity optimisation}

The first case, optimizing the velocity of the stirrers along their circular path, results from the first variation of the augmented Lagrangian with respect to $\omega_{k}$ and yields an expression for the gradient that involves adjoint and direct variables. After some algebra we obtain 

\begin{align}
  \omega_{k}^\dag = r_k\left[l_j(\varphi(t))
    +\frac{\omega_{k}}{\dot{\omega}_{k}} \frac{\partial l_j}{\partial
      \varphi}\right]\chi_k^H\left(\frac{\Pi^{\dag}_j}{C_\eta}-
  (\theta^{\dag}\circ[\bm{A}_j\theta])\right)
  \label{eq:velgrad}
\end{align}
for the adjoint rotational speed, which is directly linked to the sought-after cost-functional gradient. Equation~\ref{eq:velgrad} then provides the gradient information to our iterative optimization scheme, and, together with a line-search routine, we march the current stirring protocol towards a more effective one. The optimization terminates when no further progress can be made, or the magnitude of the cost functional gradient falls below a user-prescribed threshold.

\subsection{Gradient for shape optimization}

The optimization of the cross-sectional stirrer shape is somewhat more complex and requires more computational steps. A detailed discussion of shape parameterization, regularization techniques and their embedding into an optimization scheme, was presented in \cite{Eggl2020}, and we refer the reader to this reference for further details. Here, we will briefly touch upon the main concepts. 

We pursue a shape parameterization that is computationally efficient, yet sufficiently flexible to cover a rich design space. In view of the Fourier-based computational framework for the governing equations, we also express the cross-sectional shapes in terms of truncated Fourier series. We define each shape by a two-dimensional parametric curve $f$ given by  
\begin{align}
f_x(\alpha) &= \frac{a_0}{2}+\sum_{k=1}^n a_k\cos\left(k \alpha \right) - b_k \sin\left(k \alpha \right), \\
f_y(\alpha) &= \frac{c_0}{2}+\sum_{k=1}^n c_k\cos\left(k \alpha \right) - d_k \sin\left(k \alpha \right),
\label{Eq:FourierShape}
\end{align}
where $\alpha$ is the independent variable with $\alpha \in [0,\ 2\pi ].$ The cross-section is thus determined (and low-dimensionally parameterized) by choosing sets of four coefficients according to, i.e., $f(\{a_i\},\{b_i\},\{c_i\},\{d_i\}) = [f_x,f_y].$
With the curve thus defined, we need to determine the mask $\chi$, or, more specifically, design an efficient way of  determining whether a point on the computational grid is contained within the closed parametric curve $f.$ To this end, we introduce the winding number $w_k(\bm{x})$ of the $k$-th stirrer which, for a discrete setting, is defined as 
\begin{align}
w_k(\bm{x}) &= \sum_{i=1}^{n} \phi_{i,i-1}. 
\end{align}
In the above expression $\bm{x}$ defines a point of interest, and $\phi_{i,j}$ denotes the angle of the arc that runs along the parametric curve of stirrer $k,$ from $\bm{x}_i$ to $\bm{x}_j$ on the curve. As we sweep the full parametric path, and obtain the full sum of angles, the total angle traversed will be non-zero multiples of $2\pi$ if the point $\bm{x}$ is inside the curve, or zero otherwise. When coupled to our stirrers' mask $\chi_k,$ the grid projection can be stated as  

\begin{equation}
\chi_k(\bm{x},t) =
\begin{cases}
 1, \hspace{12pt} &w_k(\bm{x}) = 2 \pi, \\
 0, \hspace{12pt} &w_k(\bm{x}) = 0,\\
\end{cases}
\end{equation}
where we have neglected the case of winding numbers higher than one. 

As this formulation is discontinuous, simulation relying solely on this definition of the mask $\chi$ would result in spurious oscillations and numerical instabilities. Hence, one final modification consists of a smoothing layer and, by extension, an applicable distance function. For simplified geometries, such as in \cite{Engels2015} or \cite{EgglSchmid2018}), for example, it is sufficient to simply measure the distance from the centroid of the cross-sectional shape. For more complicated shapes, however, we must explicitly compute the minimum distance from our chosen point to our curve, following
\begin{align}
d = \min_{\alpha} \sqrt{(x-f_x)^2+(y-f_y)^2}. \label{eq:minDist}
\end{align}
The expression for the full mask $\chi_k$ is thus given as
\begin{equation}
\chi_k(x,t) =
\begin{cases}
 1, \hspace{12pt} &w_k=2\pi \ \text{and} \ d > h, \\
  \frac{1}{2}\left(1+\cos\left(\frac{\pi (h-d)}{h}\right)\right), & w_k=2\pi \ \text{and} \ d < h, \\
 0, \hspace{12pt} &w_k = 0.\\
\end{cases}
\label{eq:fullchi}
\end{equation}
with $h$ as the grid size. With this geometric representation in place, we now have all necessary ingredients to formulate the shape gradient for our adjoint-based optimization framework. 

For clarity, we present the formulation for only one of the four Fourier coefficient $a_i$, but stress that the effectiveness of the shape optimization rests in the ability to act on all control parameters simultaneously. Since the augmented Lagrangian is not explicitly dependent on the Fourier coefficients, we invoke a chained sequence of gradients to link the mask $\chi_k$ to the Fourier coefficient $a_i.$ We have

\begin{align}
\frac{\partial \mathcal{L}}{\partial a_i} &= \frac{\partial \mathcal{L}}{\partial \chi_k}\frac{\partial \chi_k}{\partial d}\frac{\partial d}{\partial a_i}.
\label{eq:LagrangeFourier}
\end{align}

The first term can be obtained from equation~\eqref{Eq:AugLagrange}. The second term in this equation derives from~\eqref{eq:fullchi} and can be seen in analogous form in \cite{EgglSchmid2018}. For the last term, we note that equation~(\ref{eq:minDist}) does include a minimisation, however the derivative term in equation (\ref{eq:LagrangeFourier}) is independent of this minimum, which consequently does not affect the subsequent derivation. Given the form of the parameterization, the derivative is in essence isolating the parameter $\tilde{\alpha}$ within the minimum distance function, i.e.,
\begin{align}
\frac{\partial d}{\partial a_i} = \frac{(f_x-x)\cos(i\alpha)}{d}\biggr|_{\tilde{\alpha}}.
\end{align}

Combining these terms, we arrive at the full gradient. The composite nature of the above formulation means that many Fourier coefficients (and thus complex shapes) can be computed with relative ease and little computational effort: the first two terms can be computed off-line for all Fourier coefficients, while only the last term changes during the optimization and needs to be continuously updated. Consequently, the optimization procedure is numerically efficient and readily scalable, as the majority of the terms can be calculated a-priori.

Following the computation of the shape gradient, we may have to impose further constraints to avoid unrealistic or undesired solutions. These additional constraints on the cross-sectional shapes will be dealt with via projections: once an optimized shape has been determined, the resulting parametric curve is projected onto the set of permissible solutions. This enforces feasibility of the enhanced stirrers -- of course, at the cost of full optimality. Perhaps the most obvious constraint is that of area conservation. Both extremes, i.e., vanishingly thin and fragile stirrers or stirrers that occupy the bulk of the computational domain, must be avoided. This is easily accomplished by dividing the Fourier coefficients of the optimized shape, with an area of $A,$ by a factor of  $\sqrt{\frac{A}{A_0}}$, where $A_0$ refers to the area of the original starting shape. In this manner, we only allow a re-arrangement of material, rather than a reduction/augmentation. 

A further imposed constraints strongly penalizes self-intersecting parametric curves which would lead to topological pinch-off. In fact, a related constraint avoids thin and frail features in the cross-sectional shapes that may fracture in real-world settings. This latter constraint is enforced by an untwisting and expansion routine based on intramolecular (e.g., Buckingham) potentials distributed along the optimized parametric curve. 

\subsection{Summary of optimization procedure}

Before turning to the presentation of the result we summarize the overall optimization procedure. We start by initializing the mixing strategy with an initial guess for the velocity (commonly a constant velocity along the circular path) and shape of the embedded stirrers (commonly a circular cross-section). Using this guess, we then solve the governing equations \eqref{eq:GovEqu1} - \eqref{eq:GovEqu3} forward in time over the chosen time horizon $[0,T]$. Due to the nonlinear nature of our governing equations, which results in a dependence of the adjoint variables on the corresponding direct state, it is necessary to retain or recover state variables at every time step (see comments below). Once we have arrived at $t=T$, we initialize the adjoint variables with the appropriate terminal condition and commence their backward propagation from $t=T$ to $t=0$, injecting the direct state variables at the appropriate times. The trajectory of the adjoint variables is then used to evaluate the optimality condition and retrieve the gradient information of the cost functional with respect to either the  path velocity and cross-sectional shape for each stirrer. These gradients can then be utilized by a standard optimization routine, such as, e.g., steepest descent or conjugate gradient, to arrive at updated, and improved, stirrer shapes and/or  path velocity signals. Finally, these updated variables are then projected onto the feasible design space, such that the accepted variables do not violate any of our imposed constraints. This forward-backward-update-projection loop is repeated until either no further improvements can be achieved or further steps would unavoidably violate constraints.

Finally, it is worth commenting on the computational cost associated with the iterative direct-adjoint part of the optimization scheme. As alluded to above, the nonlinear nature of the governing equations produces a link between the adjoint governing equations and the direct state variables, which appear in the form of varying coefficients in the otherwise linear adjoint equations. Saving a large number of multi-dimensional arrays may exceed the available memory resources. To mitigate this bottleneck we employ checkpointing, which allows us to trade excessive memory requirements for slightly longer simulation times. This strategy involves saving a set of state variables at particular points, referred to as checkpoints, which are subsequently used as initial conditions to recover the necessary flow fields in-between the checkpoints. Only storage of flow fields for a typical interval between checkpoints and the checkpoints themselves is required, which is typically far smaller than storing the flow fields over the entire simulation horizon. An efficient implementation of an optimized placement of checkpoints and the flow field management during the backward sweep is given by the {\tt{revolve}} library  \citep{GriewankWalther2000}.

\section{Results \label{sec:results}}

The iterative optimization algorithm is now applied to enhance mixing of a binary fluid by stirrers on circular paths. We consider a two-dimensional configuration where the binary fluid fully occupies a circular vessel, is initially separated and confined to the lower and upper half, respectively. Two stirrers on circular paths of different radii then move through the fluid and cause the binary fluid to mix. For the simulations, we choose a Reynolds number of $Re=100$ and a P\'eclet number of $Pe=10000.$ This choice of parameters is a departure from earlier studies \citep{EgglSchmid2018,Eggl2020} where advective mechanisms dominated viscous diffusion in the optimal strategies. The current choice requires a far greater effort to achieve mixing enhancement and thus poses greater challenges to the optimization scheme. The optimization horizon has been fixed to $T=12$ non-dimensional time units. Over this time span, information on the evolving flow field and passive scalar is available for an optimization of the mix-norm at the end of the time horizon. Beyond $t=12,$ the stirrers will cease to move and become passive elements within the mixing vessel; rest momentum will continue to prevail, but increasingly be dissipated by viscous diffusion. Despite our focus on this single parameter configuration, we stress that the direct-adjoint optimization proves efficient over a wide range of alternative $Re$-$Pe$-values. There are, however, limitations on the choice of the optimization horizon $T,$ which we will discuss later. 

\begin{figure}[ht!]
     \centering
     \begin{tabular}{cc}
     \begin{subfigure}[b]{0.4\textwidth}
         \centering
         \includegraphics[trim=120 40 90 30 ,clip,width=\textwidth]{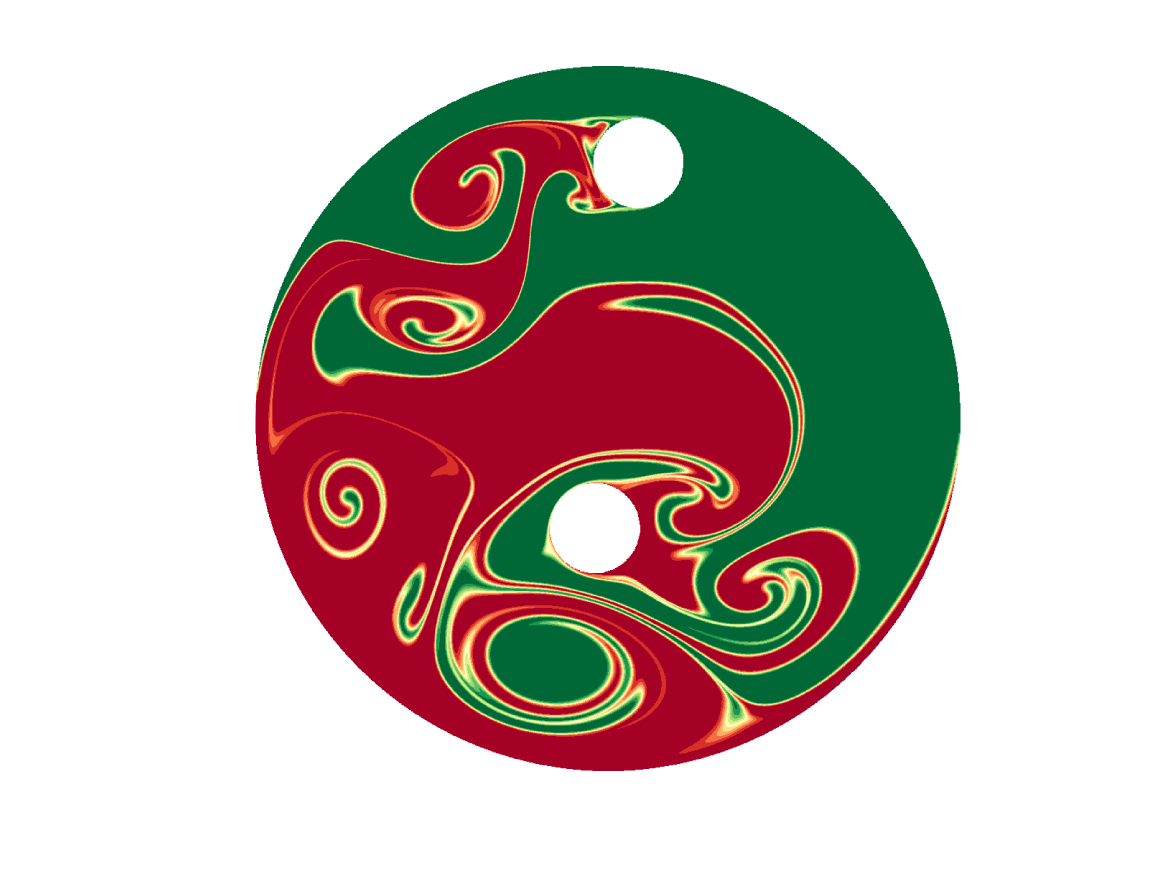}
         \caption{unoptimized}
         \label{fig:Unoptized_12}
     \end{subfigure} & 
     \begin{subfigure}[b]{0.4\textwidth}
         \centering
         \includegraphics[trim=120 40 90 30 ,clip,width=\textwidth]{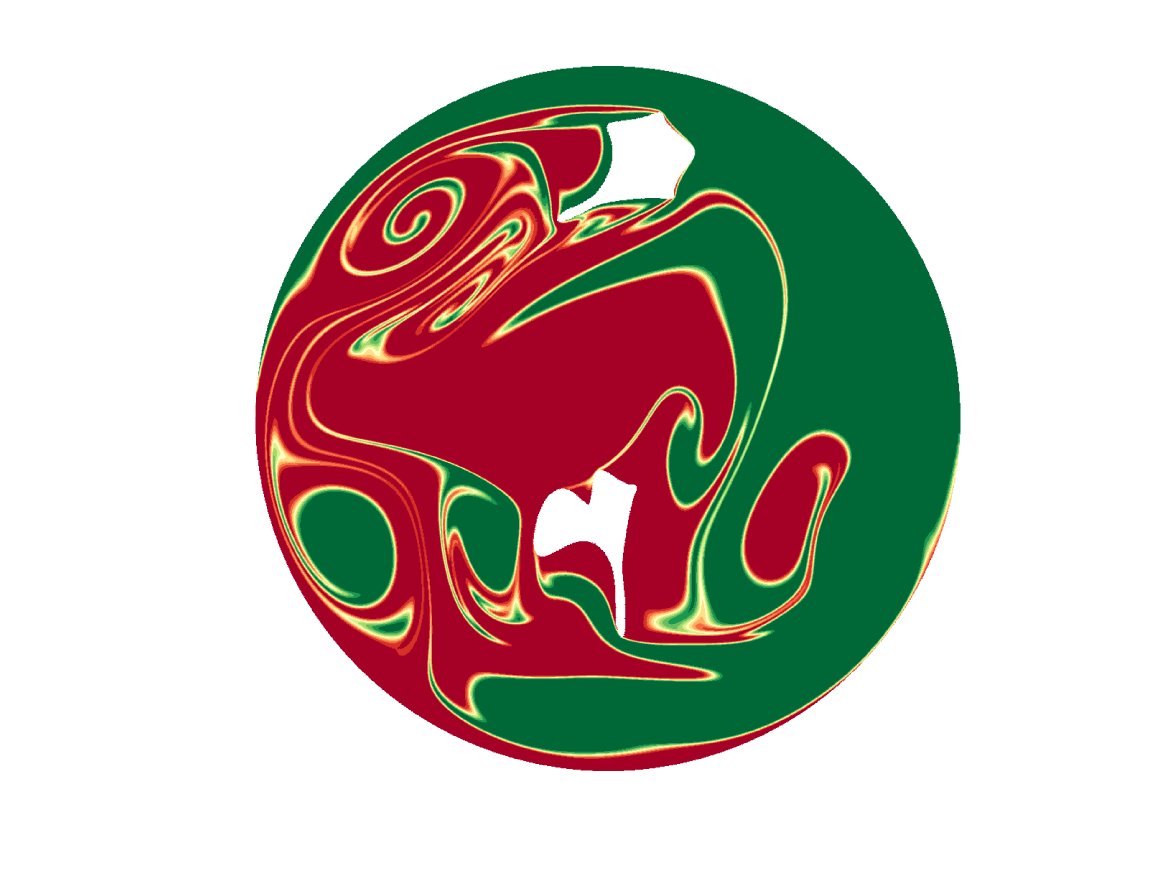}
         \caption{shape optimization}
         \label{fig:ShapeOpt_12}
     \end{subfigure}
     \\
     \begin{subfigure}[b]{0.4\textwidth}
         \centering
         \includegraphics[trim=120 40 90 30 ,clip,width=\textwidth]{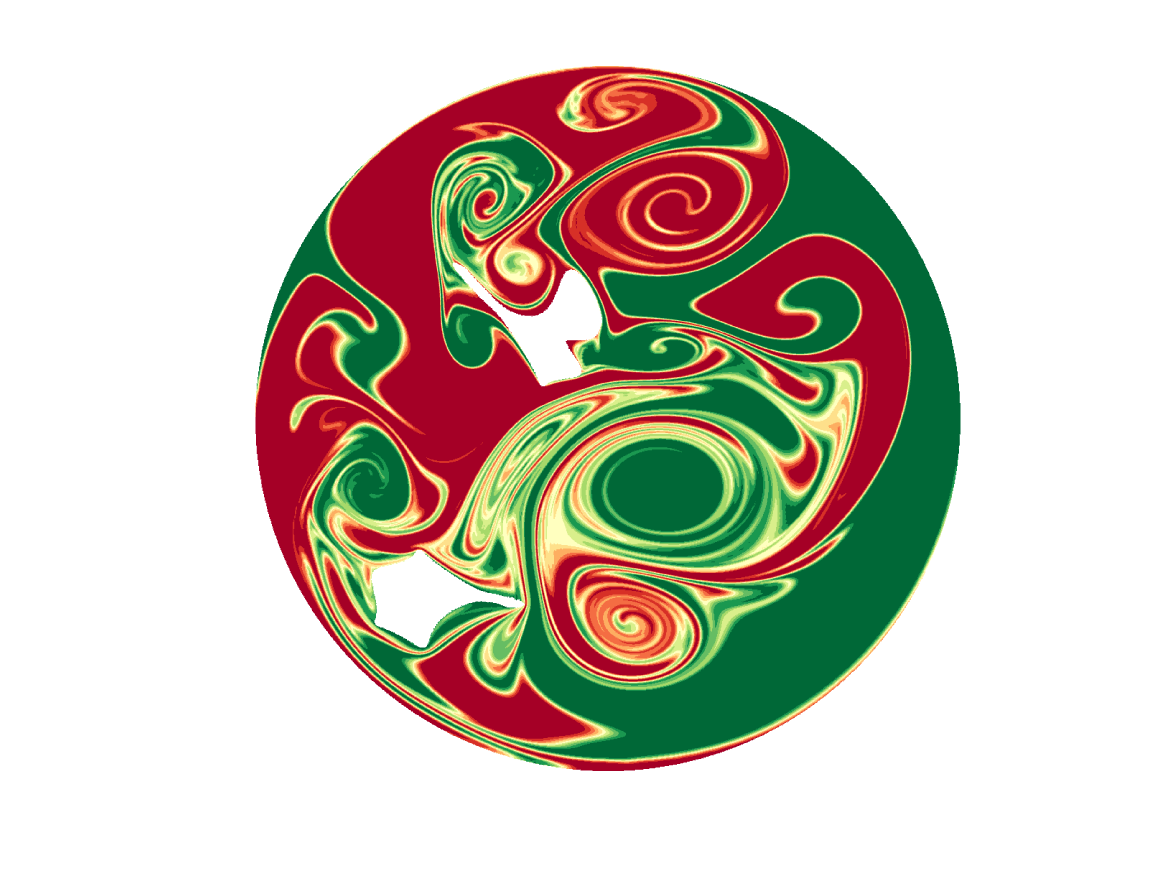}
         \caption{velocity optimization}
         \label{fig:VelOpt_12}
     \end{subfigure} & 
      \begin{subfigure}[b]{0.4\textwidth}
         \centering
         \includegraphics[trim=120 40 90 30 ,clip,width=\textwidth]{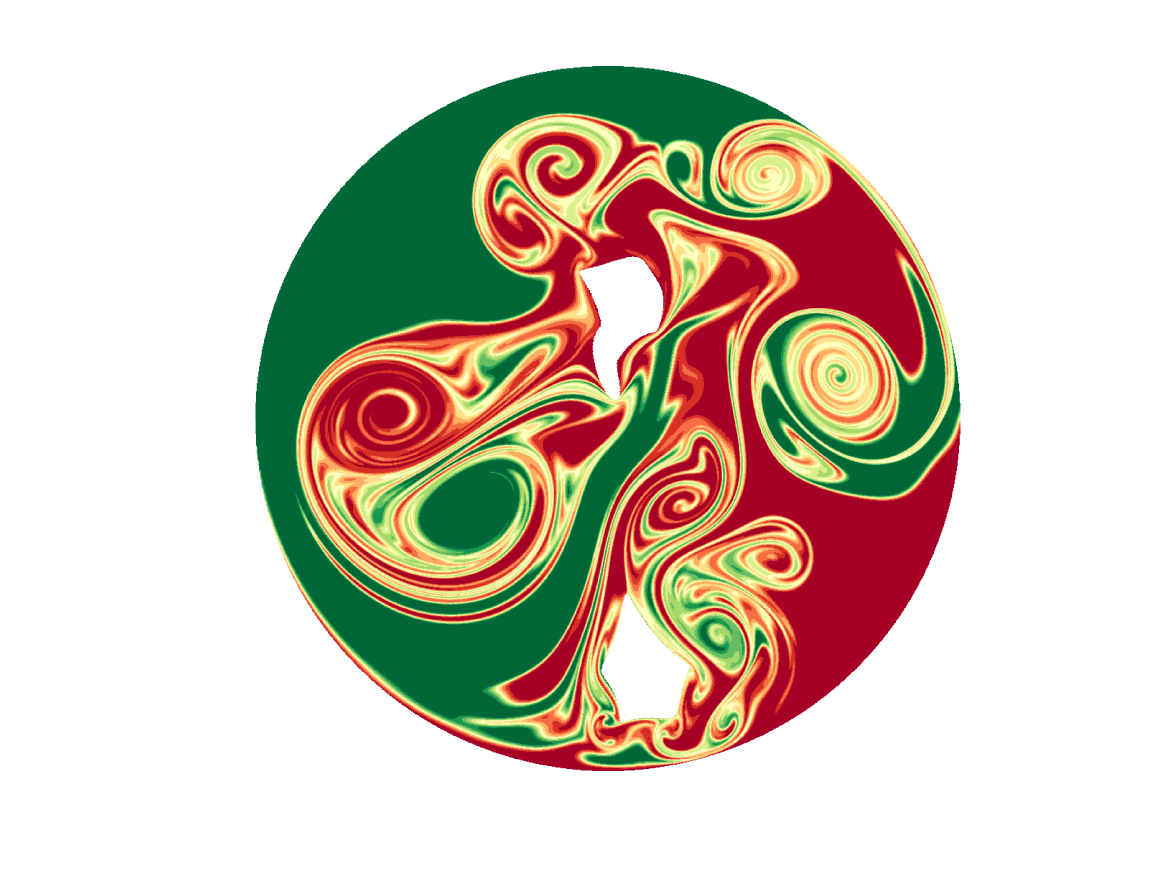}
         \caption{combined optimization}
         \label{fig:CombOpt_12}
     \end{subfigure}
     \end{tabular}
        \caption{Comparison of final snapshots at the end of the optimization horizon ($t=T=12$), visualized by the passive scalar $\theta:$ (a) unoptimized base case, (b) area-preserving shape optimization of stirrers' cross-sections, (c) path-velocity optimization of the shape-modified stirrers along circular paths, (d) composite optimization of cross-sectional shapes and path-velocities.}
        \label{fig:Comp}
\end{figure}

\begin{table}
  \centering
  \begin{tabular}{rlcrcl}
    & &\phantom{1} & iterations &\phantom{1}& 
    $\Vert \theta \Vert_{{\rm{mix}},t=12}$  \\
    \hline
    (i) & unoptimized base case & & n/a && 0.168  \\
    (ii) & shape optimization & & 11 &&  0.161  \\
    (iii) & velocity optimization & & 8 &&   0.105  \\
    (iv) & combined optimization  && 8 &&  0.072 \\
    \hline & & & & & \\
  \end{tabular}
  \caption{\label{tab:Cases} Summary of results for all numerical experiments. Number of iterations taken by the direct-adjoint optimization algorithm are displayed, where applicable, as well as the values of the final mix-norm of the passive scalar $\theta$ at $t=12.$}
\end{table}

To showcase the efficacy and applicability of the direct-adjoint framework, we present three distinct optimization cases: \emph{(i)} optimization of the cross-sectional shape of the moving stirrers, \emph{(ii)} optimization of the velocity along a circular path of the shape-modified stirrers (from step \emph{(i)}) and \emph{(iii)} compound and simultaneous optimization of both the shape and path velocity for both stirrers. The constraints include energy, velocity and energy bounds for any path velocity optimization~\citep[see, e.g.,][]{eggl2020mixing}, area-conservation and untwisting for the shape optimization~\citep[see, e.g.][]{Eggl2020} and both sets of constraints for the third case. In all cases presented here, we achieve enhanced mixing and a significant decrease in the mix-norm at the end of the optimization horizon, attesting to the effectiveness of an objective optimization approach to reach a more homogeneous mixture through unorthodox stirrer designs and improved stirring protocols. The final snapshots (at $t=T=12$) for each of the four scenarios -- unoptimized base case, shape optimization, path-velocity optimization of the optimized shape, and compound optimization -- are presented in figure~\ref{fig:Comp}. These snapshots show an increasing amount of thin filaments appearing as a result of stirring action. This qualitative analysis is further corroborated by the quantitative values listed in table~\ref{tab:Cases}, where we observe a monotonic decrease of the mix-norm of the passive scalar $\theta$ from the initial (unoptimized) base case to the combined shape/velocity optimization. In what follows, we will present the results of each optimization case in more detail and analyze the stirrer-induced vortex motion that yields enhanced mixing strategies. 

\subsection{The base case: cylindrical stirrers with uniform path velocities}

Our reference case, against which we measure any mixing improvement, consists of two stirrers of circular and identical cross-section that move along concentric circular paths (of given outer and inner radius) with a constant speed. Starting in the twelve-o-clock (for the outer radius) and six-o-clock (for the inner radius) position, the two stirrers complete a full revolution over the chosen time horizon. As the two stirrers plunge through the interface, separating the two components of the binary fluid, they deform the interface. Moreover, wake vortices, shedding continuously along their paths, further cause and promote the breakdown into small-scale features which ultimately dissipate (see {\tt{Control.mp4}} from the supplemental material). Snapshots of the base-case scenario can be seen in figure~\ref{fig:Shape} as the reduced-size subplots attached to the corresponding optimized configurations. They confirm a notable mixing of the binary fluid -- but simultaneously leave room for improvement. 

\subsection{Optimization of the stirrers' shape}

\begin{figure}[ht!]
  \centering
  \begin{tabular}{cc}
    \begin{tikzpicture}
      \clip (0,0) rectangle (6cm,6cm);
      \node[anchor=center] at (3,3)
      {\includegraphics[width=8cm]{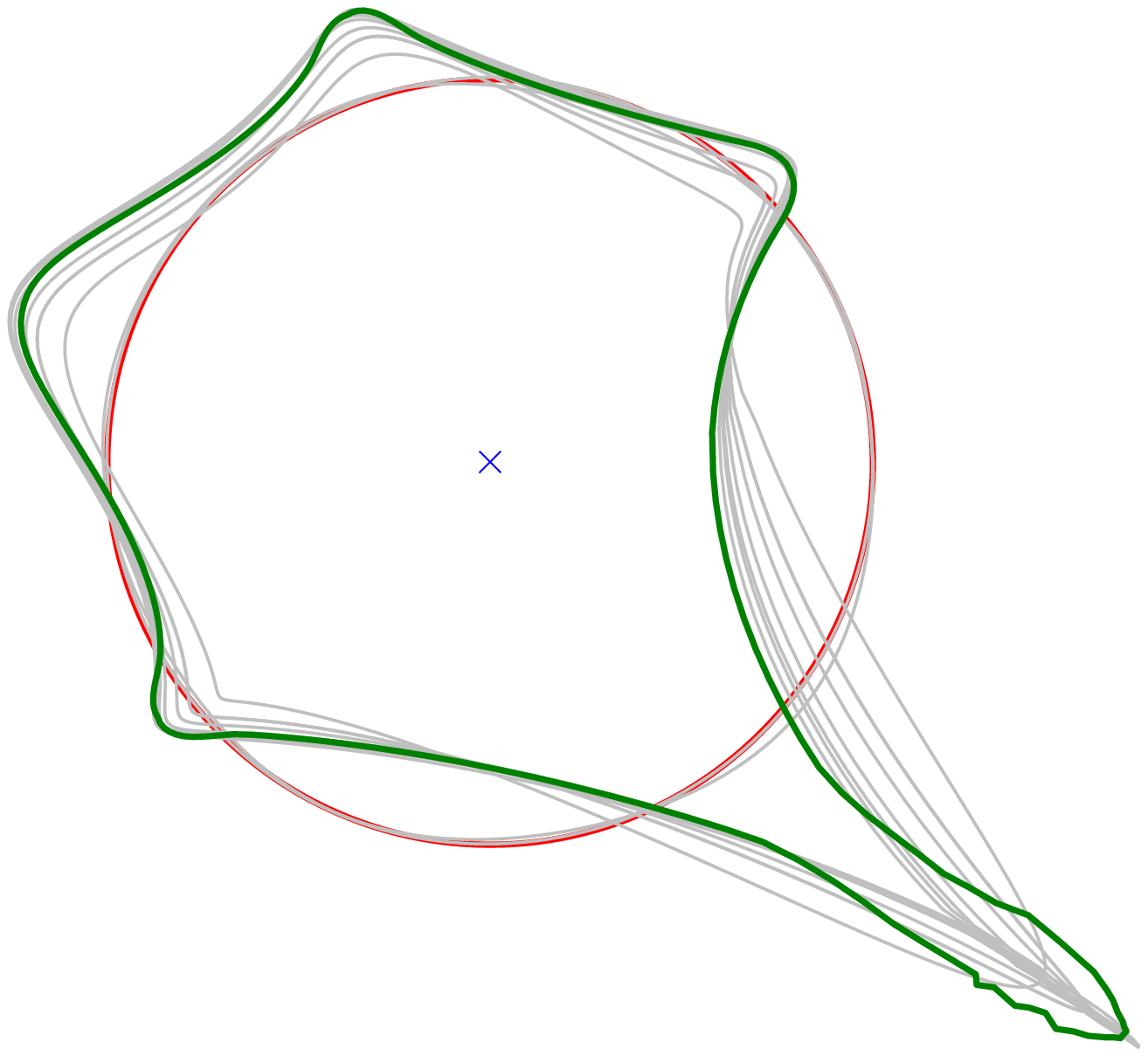}};
       \node (centre) at (2.9,3.3) {};
       \node (Mot) at (0.5,4.8) {};
        \draw[->,blue,thick]  (centre) to (Mot) {};
      \node (ar1a) at (4.2,2.6) {};
      \node (ar1b) at (5.8,0.6) {};
      \node (ar2a) at (1.1,3.9) {};
      \node (ar2b) at (0.4,4.2) {};
      \node (ar3a) at (2.2,5.0) {};
      \node (ar3b) at (2.1,5.7) {};
      \node (ar4a) at (1.4,2.5) {};
      \node (ar4b) at (1.1,2.0) {};
      \node (ar5a) at (3.8,4.5) {};
      \node (ar5b) at (4.2,5.0) {};
      \draw[->,black,thick]  (ar1a) to (ar1b) {};
      \draw[->,black,thick]  (ar2a) to (ar2b) {};
      \draw[->,black,thick]  (ar3a) to (ar3b) {};
      \draw[->,black,thick]  (ar4a) to (ar4b) {};
      \draw[->,black,thick]  (ar5a) to (ar5b) {};
      \node at (2,1) {(a)}; 
    \end{tikzpicture} & 
    \begin{tikzpicture}
      \clip (0,0) rectangle (6cm,6cm);
      \node[anchor=center] at (3,3) {\includegraphics[width=8cm]{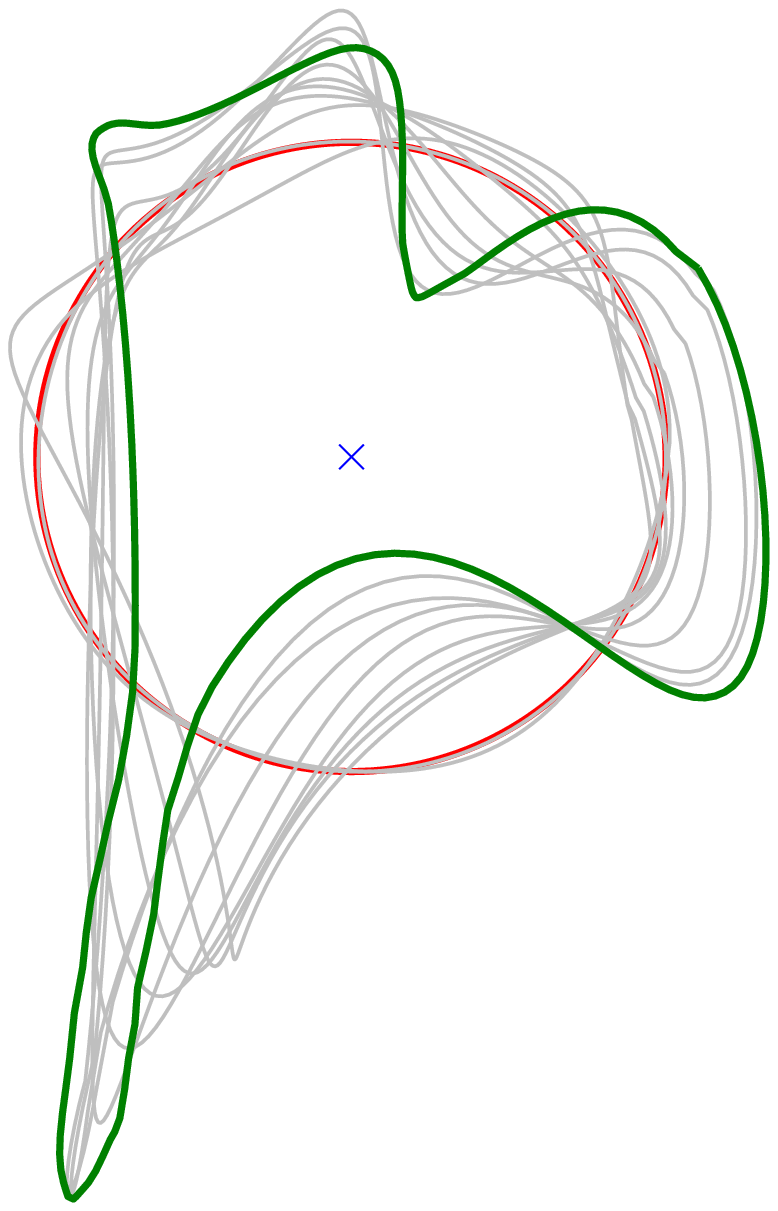}};
      \node (centre) at (2.85,3.8) {};
      \node (Mot) at (5.3,2.8) {};
      \node (ar1a) at (2.8,2.6) {};
      \node (ar1b) at (1.8,0.5) {};
      \node (ar2a) at (3.6,2.4) {};
      \node (ar2b) at (2.85,3.4) {};
       \node (ar3a) at (4.1,3.8) {};
      \node (ar3b) at (4.8,3.6) {};
      \node (ar4a) at (2.6,4.9) {};
      \node (ar4b) at (1.8,5.1) {};
      \node (ar5b) at (3.2,5.5) {};
      \draw[->,blue,thick]  (centre) to (Mot) {};
        \draw[->,black,thick]  (ar1a) to (ar1b) {};
    \draw[->,black,thick]  (ar2a) to (ar2b) {};
    \draw[->,black,thick]  (ar3a) to (ar3b) {};
    \draw[->,black,thick]  (ar4a) to (ar4b) {};
    \draw[->,black,thick]  (ar4a) to (ar5b) {};
    \node at (3,1) {(b)};
    \end{tikzpicture} 
    \end{tabular}
    \caption{Optimization of the cross-sectional shape of both stirrers, leaving the velocity along the circular paths unchanged. The red and green curves signify the initial and final configuration, respectively, with gray shapes denoting the intermediate steps of the optimization procedure. Blue arrows have been added to indicate the travel direction of the two stirrers. Black arrows pinpoint features of interest, as they develop during the optimization process. (a) The outer stirrer shows edges on the leading edge, up to the location perpendicular to the travel direction, and a pronounced splitter-plate-like appendage at the trailing egde. (b) The inner stirrer shows a more complex shape, with a marked appendage perpendicular to its travel direction, causing a significant increase in the cross-sectionally exposed area.}
    \label{fig:ShapeEvo}
\end{figure}

\begin{figure}[ht!]
  \centering
  \begin{tikzpicture}[scale=1]   
    \node at (0,0){\includegraphics[trim=120 40 90 30,clip,width=0.275\textwidth]{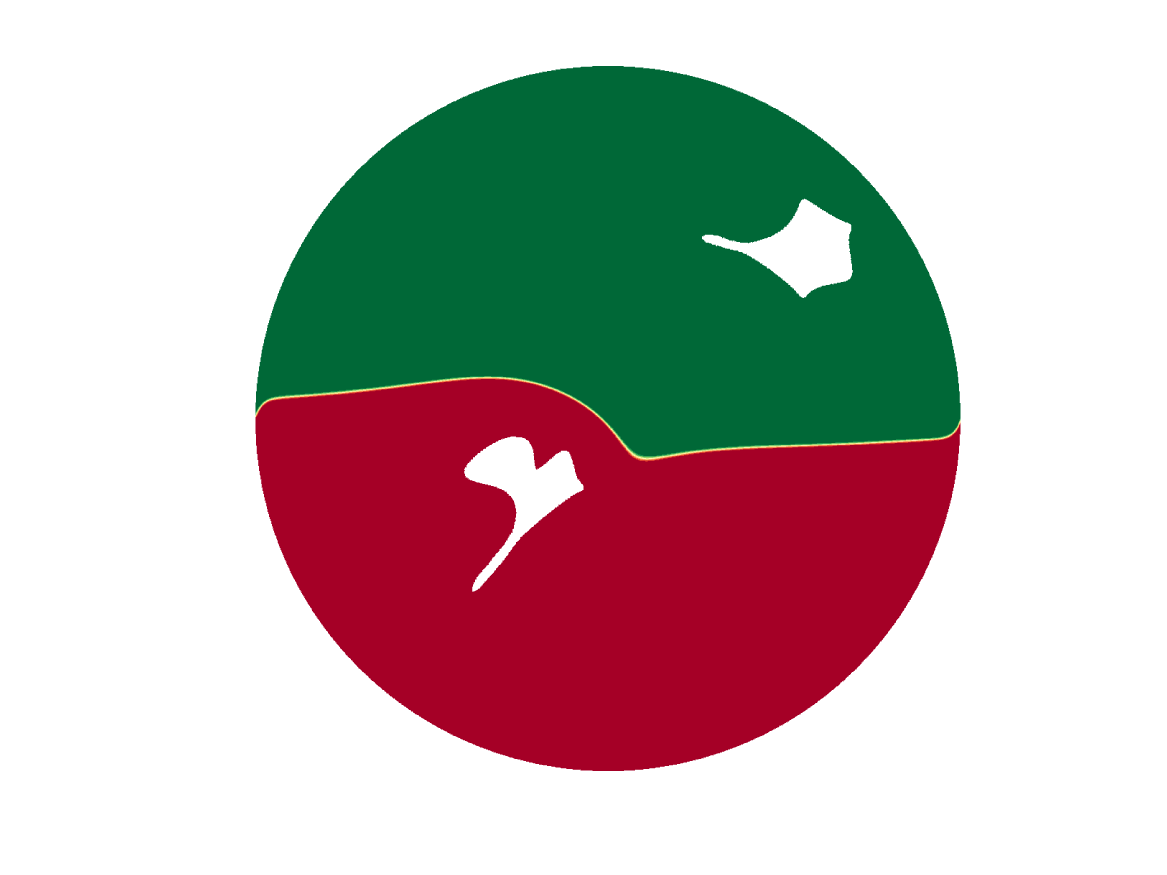}};
    \node at (5.5,0){\includegraphics[trim=120 40 90 30,clip,width=0.275\textwidth]{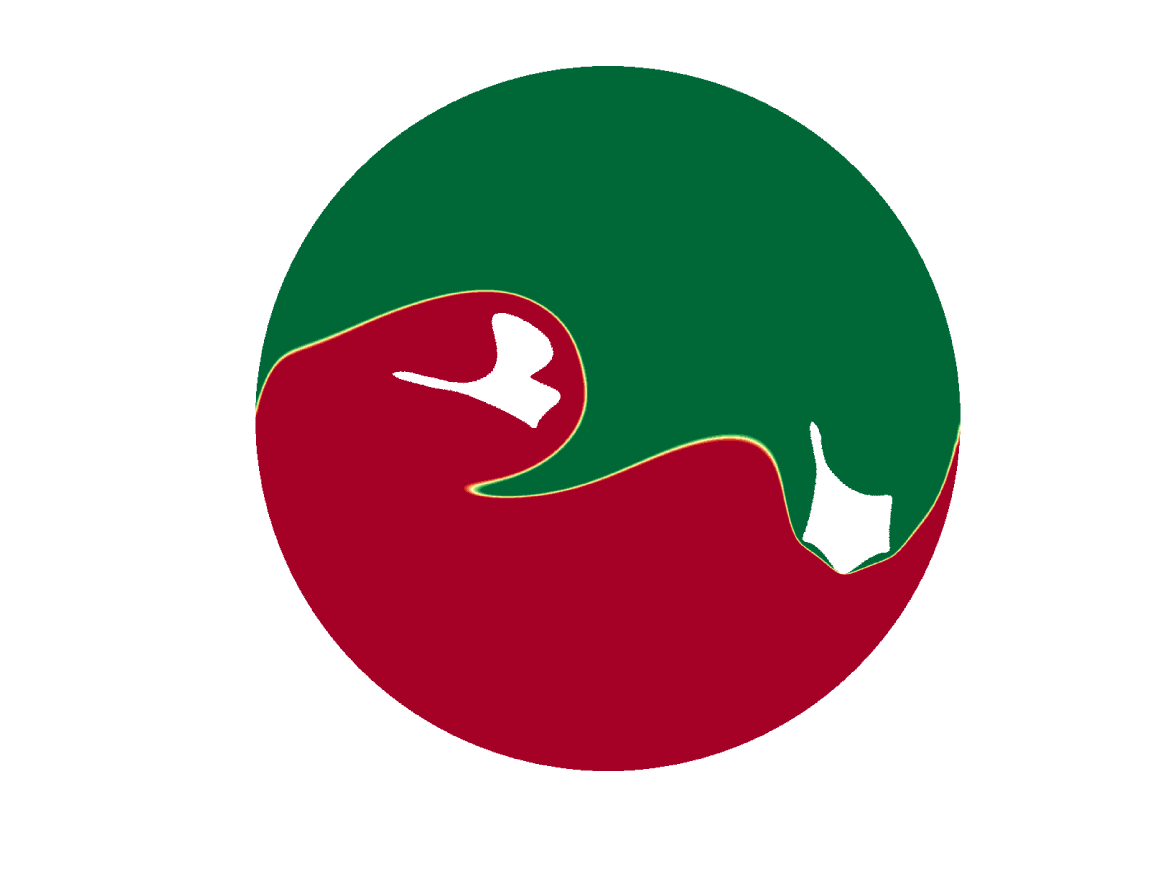}};
    \node at (11,0){\includegraphics[trim=120 40 90 30,clip,width=0.275\textwidth]{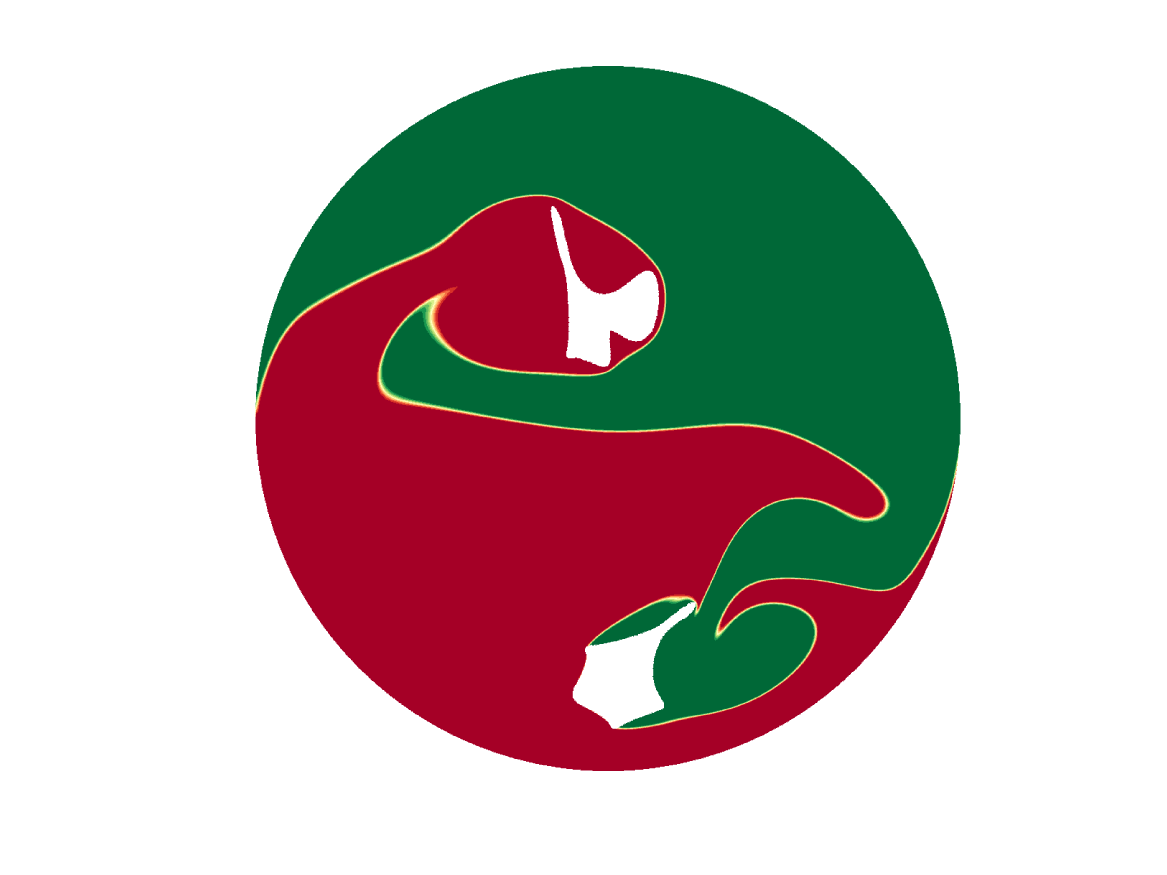}};
    \node at (2.7,-2.25){\includegraphics[trim=120 40 90 30,clip,width=0.12\textwidth]{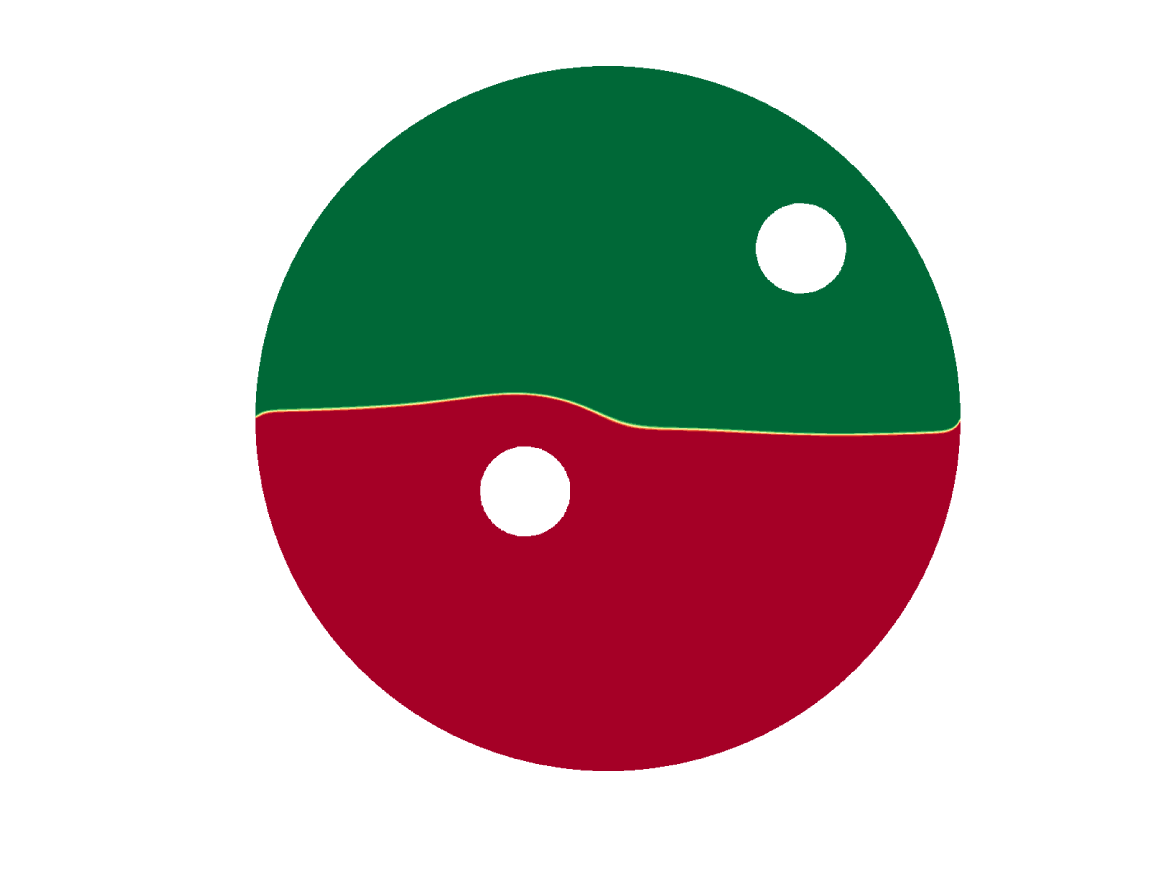}};
    \node at (8.2,-2.25){\includegraphics[trim=120 40 90 30,clip,width=0.12\textwidth]{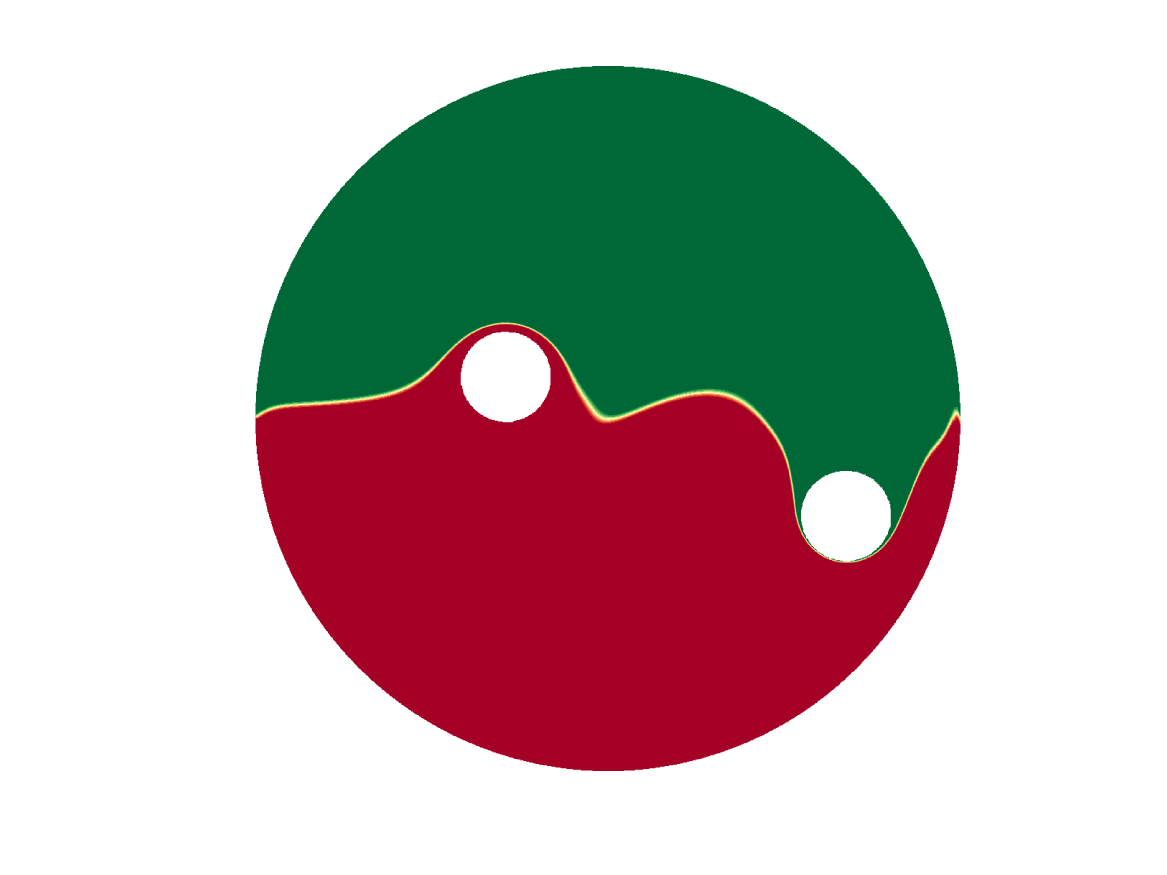}};
    \node at (13.7,-2.25){\includegraphics[trim=120 40 90 30,clip,width=0.12\textwidth]{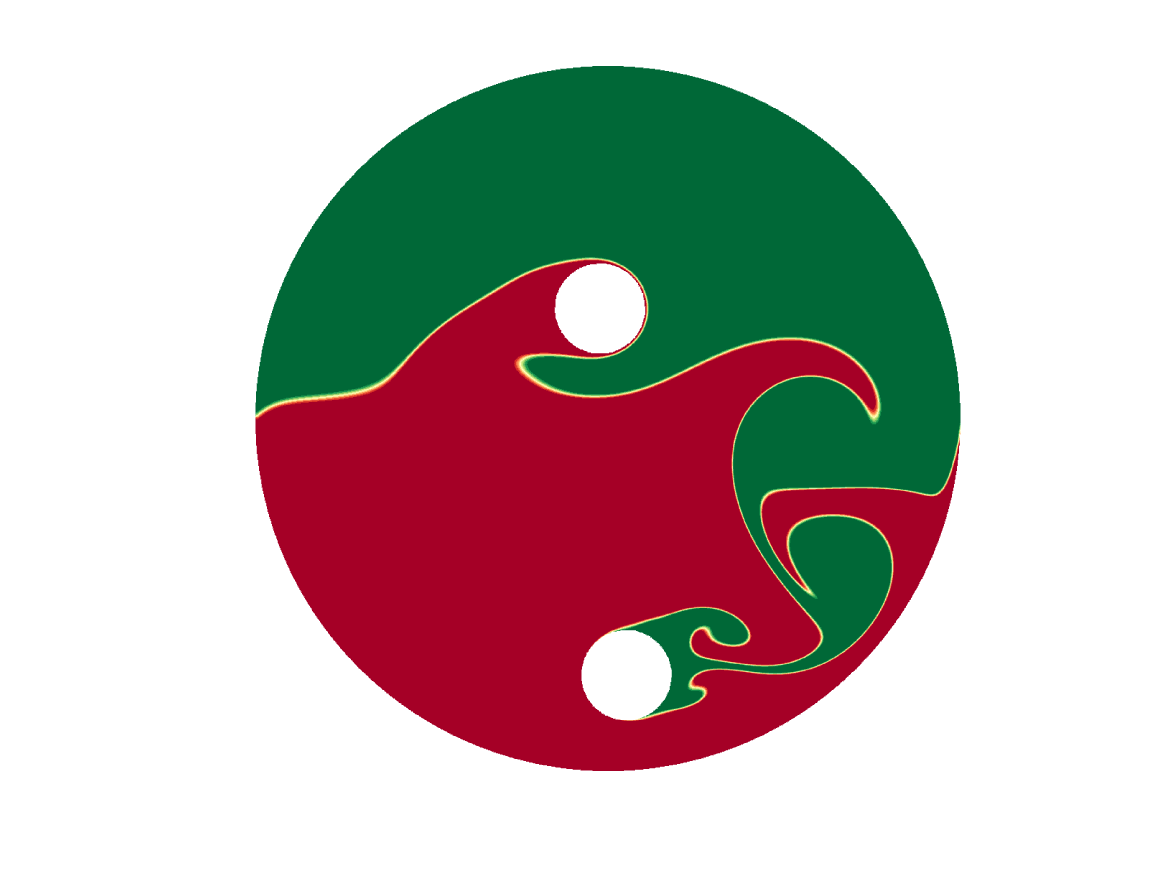}};
    \node at (0,-2.5) {(a) $\ t=2$};
    \node at (5.5,-2.5) {(b) $\ t=4$};
    \node at (11,-2.5) {(c) $\ t=6$};
    \def\y{6};
    \node at (0,-\y){\includegraphics[trim=120 40 90 30,clip,width=0.275\textwidth]{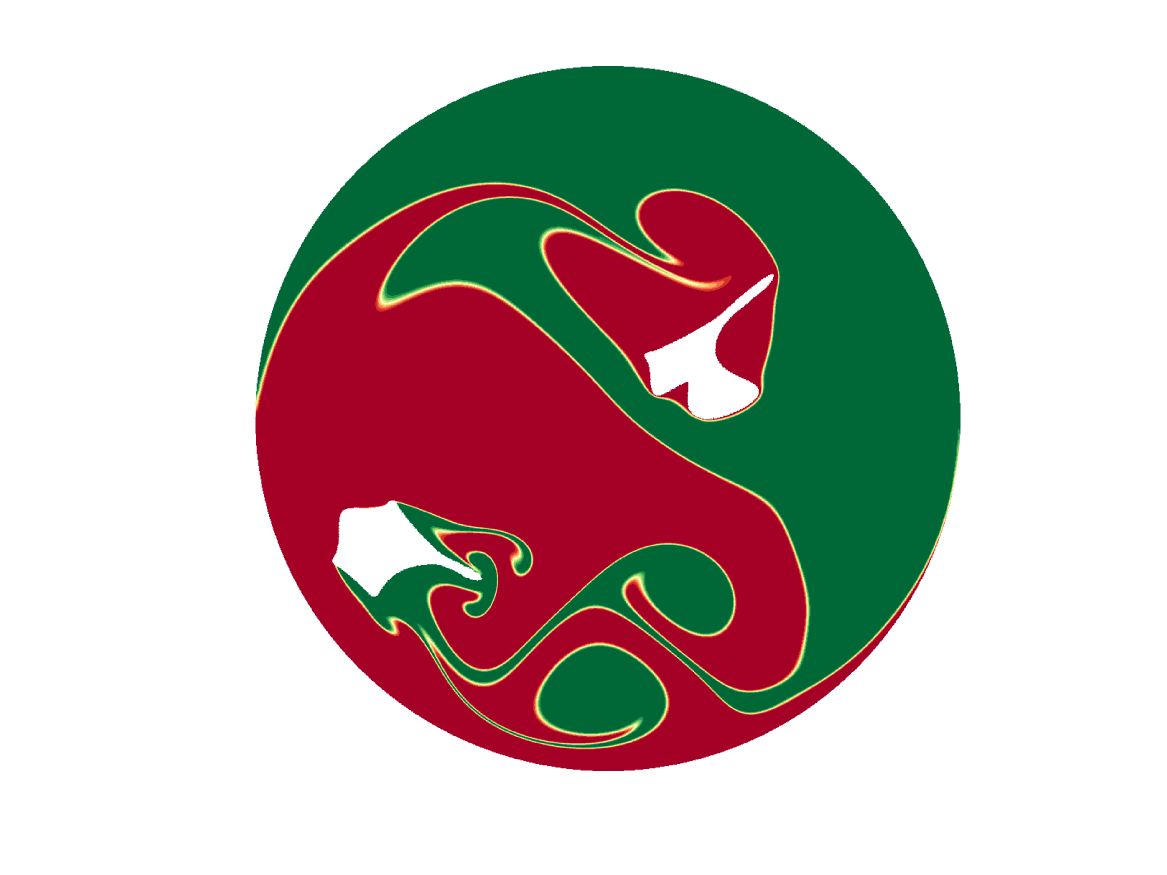}};
    \node at (5.5,-\y){\includegraphics[trim=120 40 90 30,clip,width=0.275\textwidth]{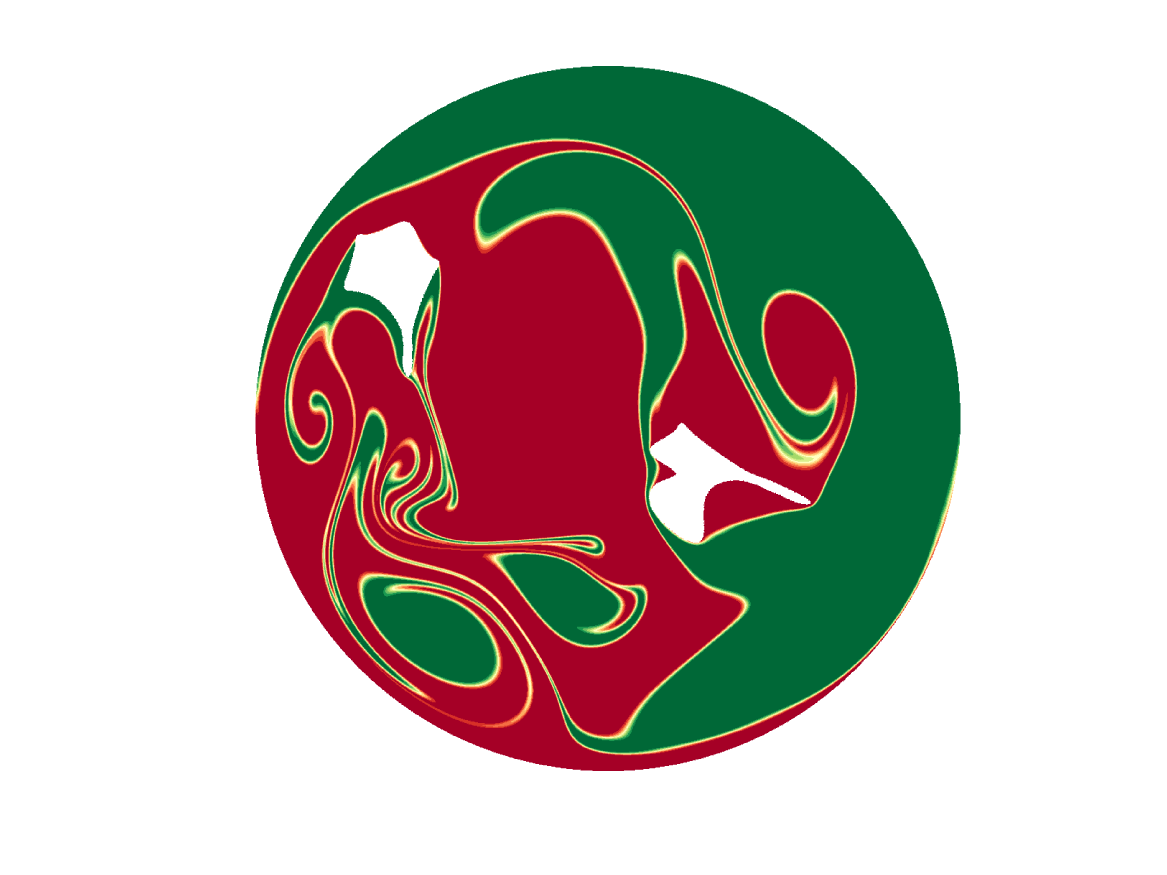}};
    \node at (11,-\y){\includegraphics[trim=120 40 90 30,clip,width=0.275\textwidth]{figures/Shape/012000.png}};
    \node at (2.7,-\y-2.25){\includegraphics[trim=120 40 90 30,clip,width=0.12\textwidth]{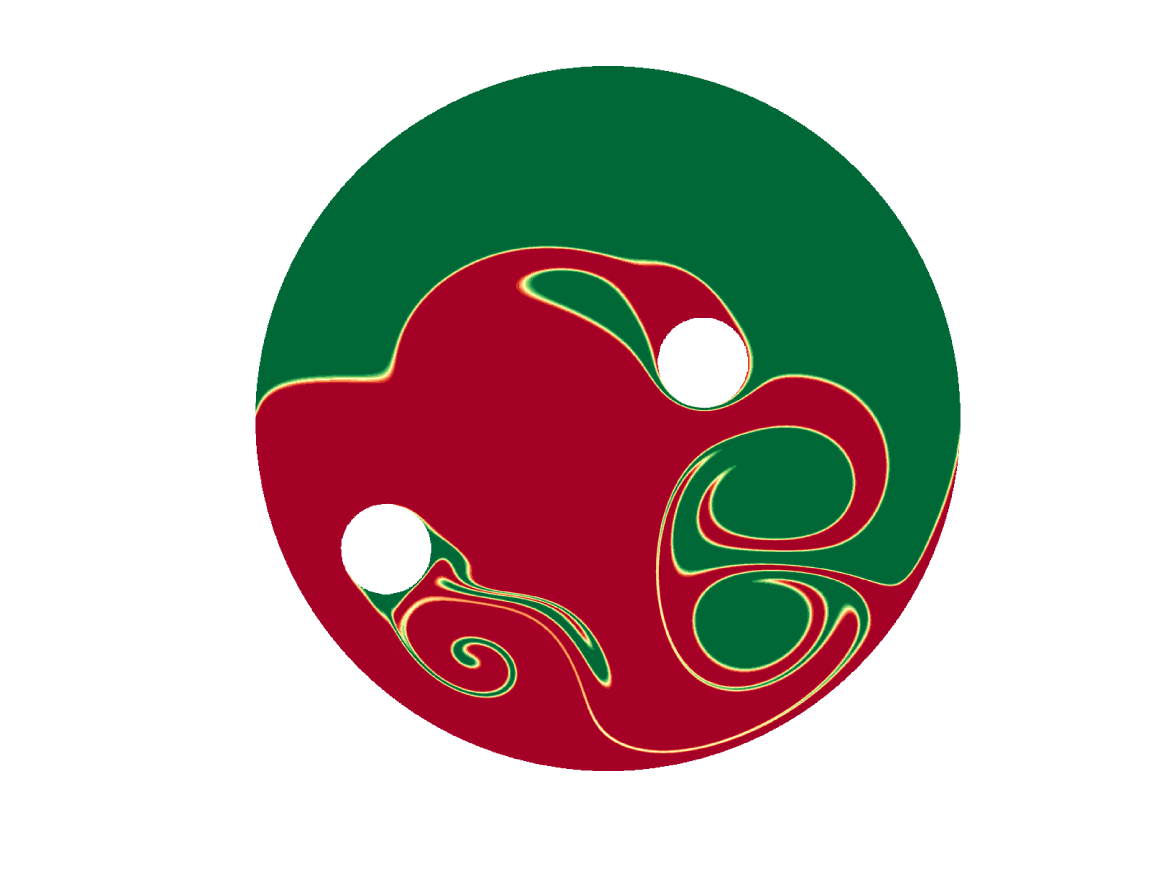}};
    \node at (8.2,-\y-2.25){\includegraphics[trim=120 40 90 30,clip,width=0.12\textwidth]{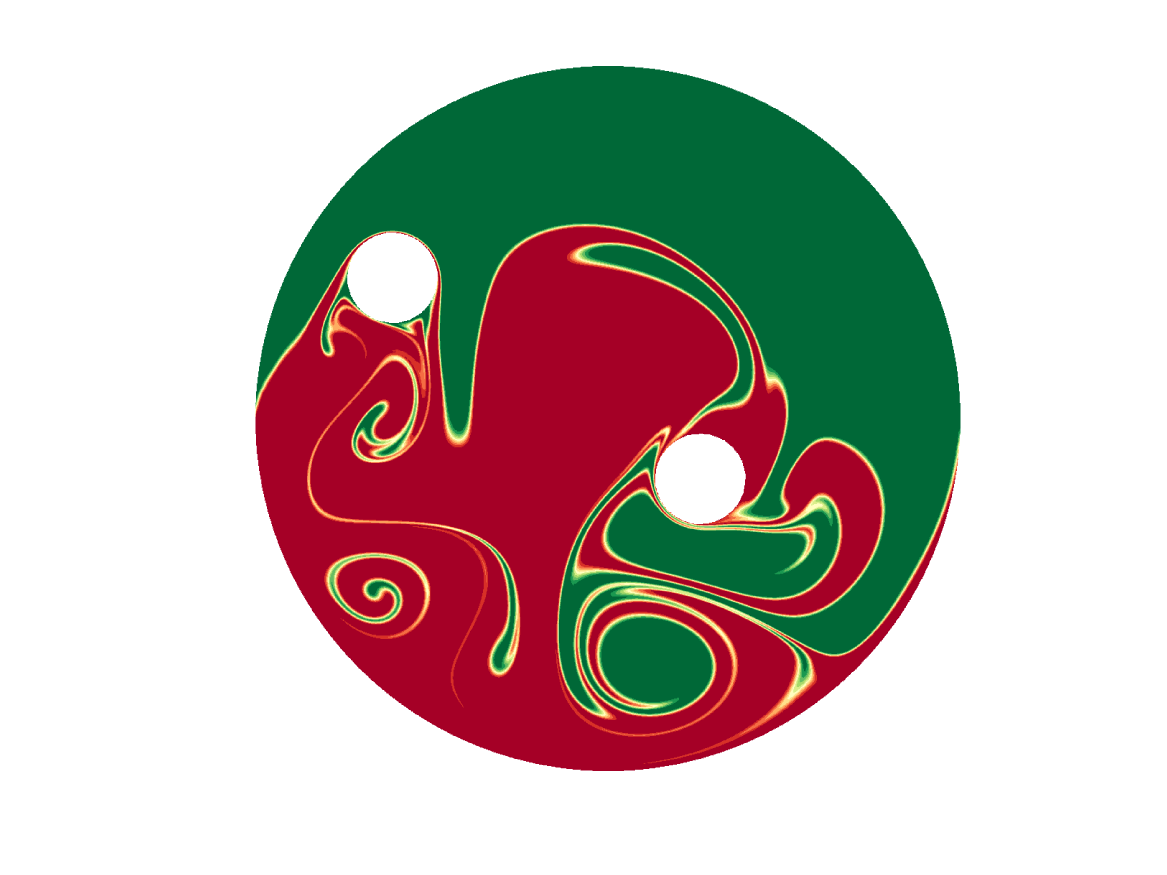}};
    \node at (13.7,-\y-2.25){\includegraphics[trim=120 40 90 30,clip,width=0.12\textwidth]{figures/Control/012000.png}};
    \node at (0,-\y-2.5) {(d) $\ t=8$};
    \node at (5.5,-\y-2.5) {(e) $\ t=10$};
    \node at (11,-\y-2.5) {(f) $\ t=12$};    
  \end{tikzpicture}
  \caption{Selected snapshots of the passive scalar $\theta$ after $11$ iterations of the direct-adjoint looping, optimizing the cross-sectional stirrer shapes to enhance the mix-norm at $T=12.$ The corresponding snapshots for the base case with cylindrical stirrers are juxtaposed to the lower right of each snapshot (in reduced size) for comparison.}
  \label{fig:Shape}
\end{figure}

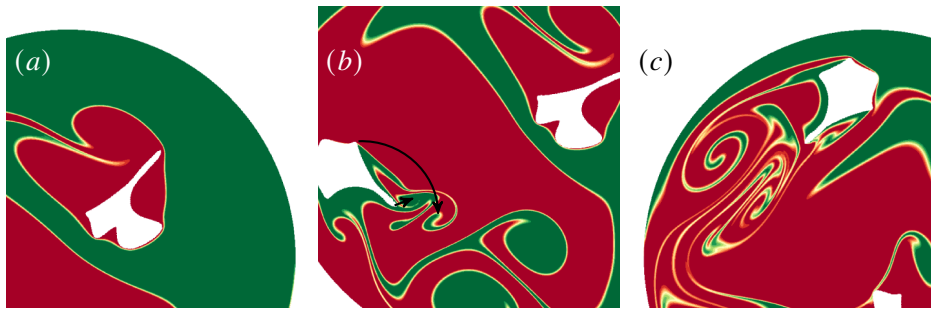
\begin{figure}[ht!]
  \centering
  \begin{tabular}{ccc}
    \begin{tikzpicture}
      \clip (0,0) rectangle (4cm,4cm);
      \node[anchor=center] at (0.6,0.5)
      {\includegraphics[width=10cm]{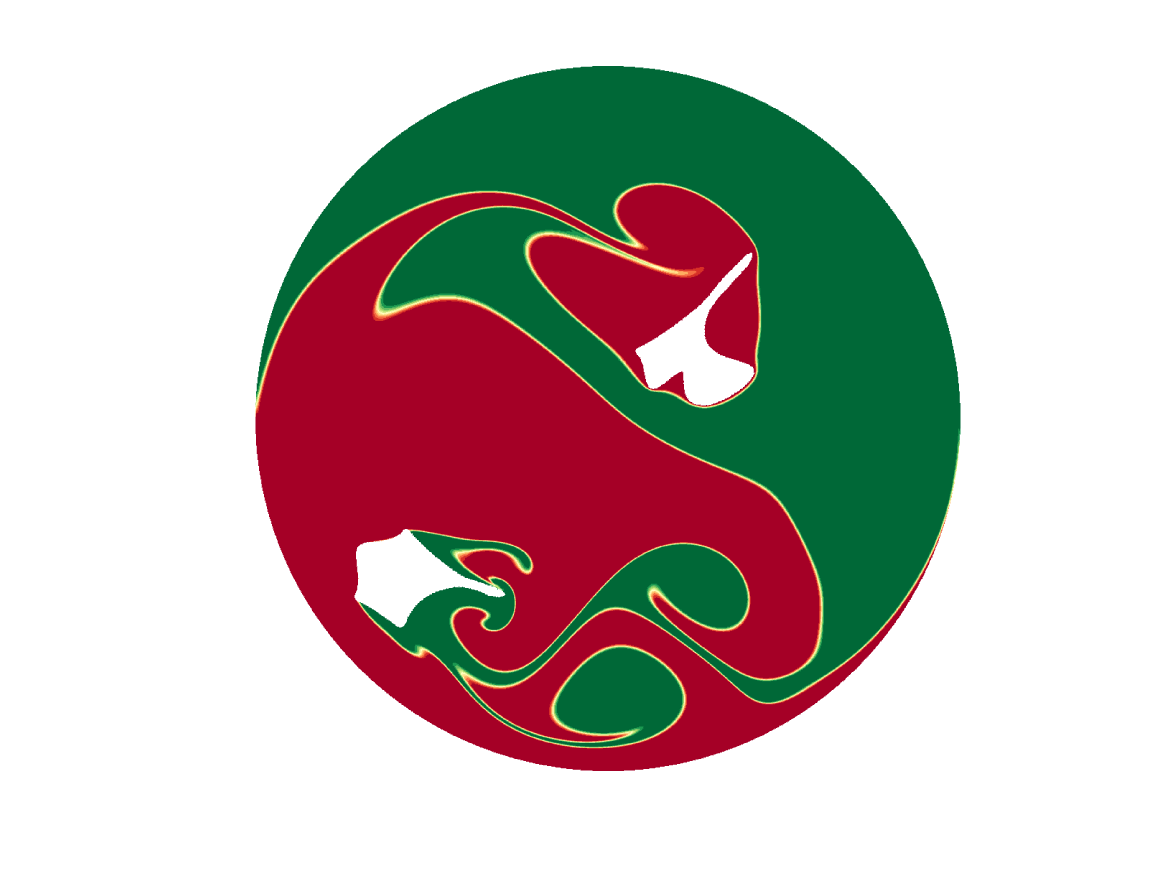}};
      \node[white] (e) at (0.35,3.25) {$(a)$};
    \end{tikzpicture} & 
    \begin{tikzpicture}
      \clip (0,0) rectangle (4cm,4cm);
      \node[anchor=center] at (2.25,2.2) {\includegraphics[width=10cm]{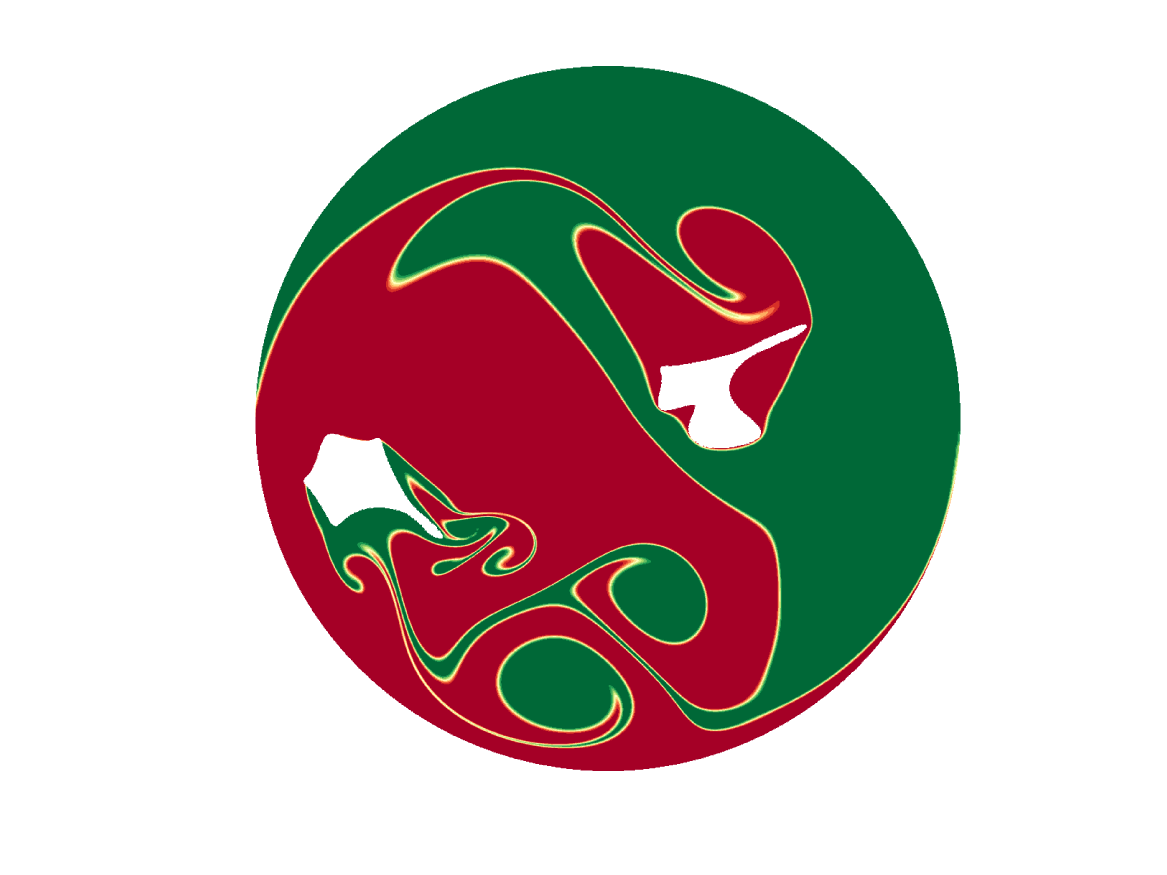}};
      \node (a) at (0.4,2.2) {};
      \node (b) at (1.6,1.1) {};
      \node (c) at (0.88,1.3) {};
      \node (d) at (1.4,1.5) {};
      \draw[->,black,thick] (a) to [bend left=45] (b) {};
      \draw[->,black,thick]  (c) to (d) {};
      \node[white] (f) at (0.35,3.25) {$(b)$};
    \end{tikzpicture}  &
    \begin{tikzpicture}
      \clip (0,0) rectangle (4cm,4cm);
      \node[anchor=center] at (3.0,0.5) {\includegraphics[width=10cm]{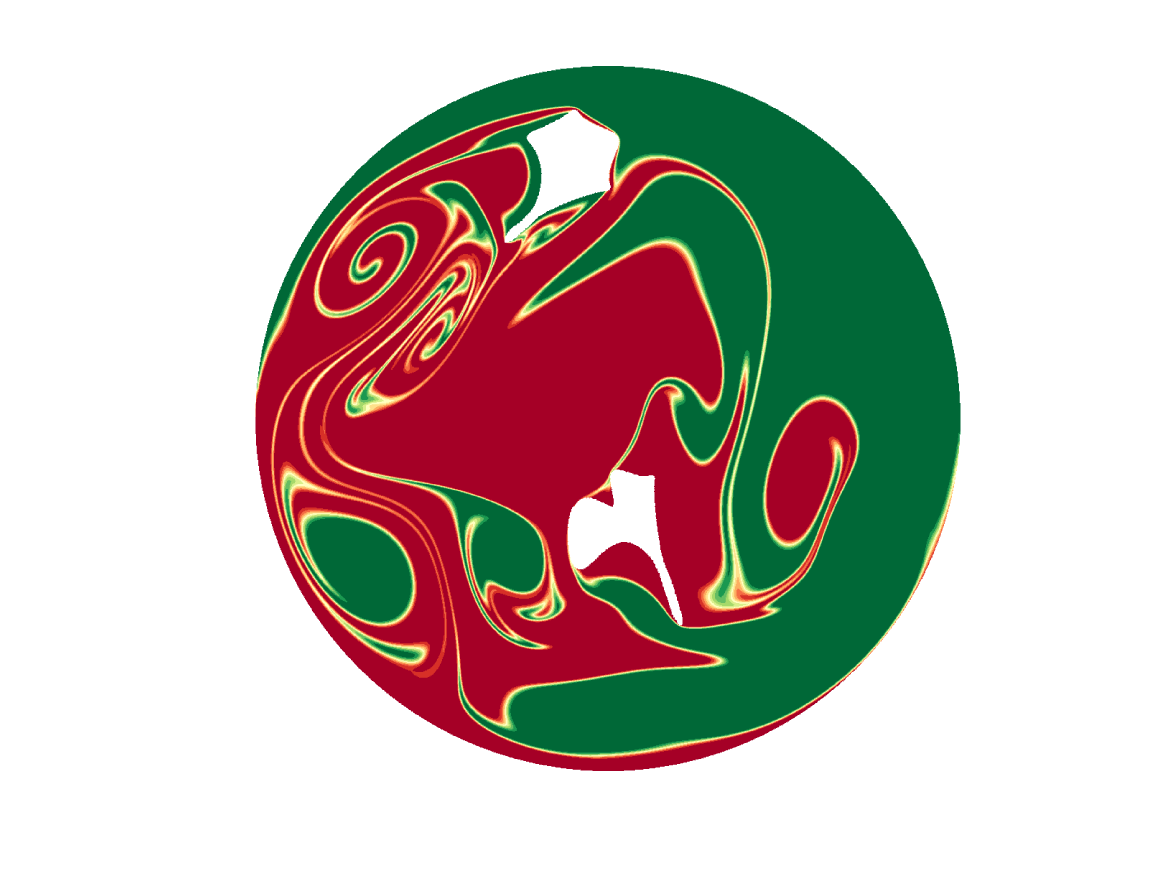}};
      \node (g) at (0.35,3.25) {$(c)$};
    \end{tikzpicture} 
    \end{tabular}
    \caption{Examples of increased passive scalar transportation with shape-optimized stirrer. (a) Modification of the inner (low-velocity) stirrer, showing an elongation perpendicular to its direction of motion. (b) Interaction of shed vortices generated by the perpendicular edge of the outer stirrer and the trailing edge protrusion, respectively, with arrows indicating the path of the vortices. (c) Generation of thin filaments and vortex interaction in the wake of the enhanced shape of the outer stirrer. See {\tt{ShapeOpt.mp4}} from the supplemental material for an animation of the mixing process.}
    \label{fig:ShapeOpt}
\end{figure}

We begin by presenting the first optimization case: optimizing the cross-sectional shape of the two moving stirrers. Given that the velocity will remain unchanged in this case, no considerations about energy constraints need to be made. On the other hand, we remark that fixing both path and velocity renders the rise of any collaborative strategies nigh impossible. As we will confirm later, these collaborative strategies are highly effective in mixing enhancement, and the limitation to only shape optimization will be reflected in relatively modest improvements in the mix-norm over the optimization horizon (see table~\ref{tab:Cases}). When optimizing the stirrers' shape, the focus is rather on mixing enhancement in the neighborhood of each stirrer. This enhancement takes the form of sharp-edged appendages that (i) transport more material between unmixed sections of the fluid, (ii) decrease the aerodynamic cross-sectional profile of the stirrers, (iii) promote the generation of vortices shedding from sharp corners and edges, and (iv) modify the trailing edge of the stirrers to increase the complexity of the trailing wake. The changes in shape during the optimization process are displayed in figure~\ref{fig:ShapeEvo}, where we observe the realization of key features mentioned above. While the outer stirrer concentrates on the generation of modified wake vortices via an extended trailing edge and sharp edges perpendicular to the travel direction, the inner (slower) stirrer adopts a bluff-body strategy by markedly extending in the direction perpendicular to its travel direction. The resulting shape will foster an increased shedding activity and an injection of substantial vorticity into the flow. The effects induced by these modified stirrers play an important role during various periods of the stirring process, and can be most clearly observed in the animations (see the supplemental material); we urge the reader to peruse the videos. These videos will also contain the dynamics beyond the optimization horizon ($T=12$) and show the subsequent passive mixing by rest inertia and diffusion. While we cannot fully do justice to this intrinsically dynamic process, we nonetheless will illustrate the stirring of a binary fluid using optimized cross-sectional stirrer shapes with the help of the static images in figures~\ref{fig:Shape} and~\ref{fig:ShapeOpt}.

Vortex generation and vortex interaction are the key fluid-dynamical tools for optimal mixing. The rather low velocity of the inner stirrer (closer to the center of the vessel) precludes the formation of significant wake vortices, let alone instigate their interaction. In light of this impediment, the inner stirrer enhances mixing by morphing into a scoop-like shape that aids in transporting homogeneous material into high-shear zones simply by a plunging action. While suboptimal to vortex-interactions (as will be shown later), this strategy results in modest, but noticeable improvements of the mix-norm over our specified optimization window. 

As observed above, over the course of $11$ direct-adjoint iterations, the shape of the inner stirrer develops an extended appendage that introduces a paddle-like feature as well as an associated cavity (see figure~\ref{fig:ShapeOpt}(a)). The extended reach of this appendage also yields, via the locked rotation of the stirrer about its centroid along the circular path, a marginally larger velocity for shedding vortices, when compared to the initial circular cross-section. The effect of this topology change is marked and can be observed when comparing the snapshots in figures~\ref{fig:Shape}(a) and (b) to their base-case equivalents. Already at this early stage in the optimization window, the interface between the binary fluids has deformed more significantly, and long, thin fluid structures have formed that are absent in the corresponding base-case solution. The formation of long and thin filaments is crucial in preparation of a mostly diffusive process, once the stirring stops. At $t=8$ (see figure~\ref{fig:Shape}(d), or figure~\ref{fig:ShapeOpt}(a) for a close-up), the advantage of an enlarged normal cross-section becomes evident: the wake structure of the inner stirrer contains significantly finer structures when compared to the wake of the initial shape. 

The outer stirrer (closer to the vessel wall) does possess the necessary velocity to take advantage of a wake-vortex-interaction strategy in order to enhance mixing. As a consequence, and in contrast to the bluff inner stirrer, it develops a more streamlined cross-sectional shape over the course of the optimization process. Nevertheless, two edges are visible at about $\pm 90^\circ$ to the traveling direction which are responsible for a continuous and vigorous vortex shedding (see figures~\ref{fig:Shape}(c-f)) and the injection of small-scale structures into the stirred fluid.  
    
We furthermore observe that the outer stirrer favors a slender protuberance near its trailing edge, roughly aligned with the direction of motion. This solid feature acts as a splitter plate for the wake, and promotes vortices of increased complexity in the flow past the stirrer. The slightly asymmetric positioning of this elongated appendage introduces oblique vortex pairs of differing strengths (see. e.g., figures~\ref{fig:Shape}(c,d) and figure~\ref{fig:ShapeOpt}(b)) which subsequently interact to form thin vortex filaments in the wake of the outer stirrer (see figures~\ref{fig:Shape}(e,f), and figures~\ref{fig:ShapeOpt}(b,c) for a close-up). The combination of multi-edged shedding of vortices and their interaction via the split wake has been identified by the direct-adjoint scheme as an effective means to decrease the mix-norm over the optimization window, to introduce small-scale structures and to promote homogeneity of the passive scalar. In this effort, the wake profile induced by the shape-modified stirrer is considerably more complex than the wake profile for the cylindrical stirrers (as can be seen from figures~\ref{fig:Comp} and~\ref{fig:ShapeOpt}c).
    
In summary, the two stirrers have been optimized in their cross-sectional shape according to the velocity regime in which they operate. The inner stirrer, in a lower-speed environment, has adopted a drag-based strategy whereby an appendage perpendicular to its direction of travel reaches into a region of larger velocity (increased distance from the center) and induces vortex shedding from its edge. An associated scooping motion, coupled to a secondary motion brought about by a cavity on the stirrer, affects the mixing towards lower mix-norm values. In contrast, the outer stirrer -- with its position in a higher-velocity range -- selects a strategy of injecting vortices into the binary fluid via an appropriately manipulated wake profile. Edges on either side of the stirrer at strategic locations generate vortices which then interact with an elongated trailing edge (and each other) to modify the wake into a more convoluted vortex street with intensified smaller scales. Despite these encouraging efforts to promote mixing, we see little evidence of collaborative strategies between structures induced by the two stirrers. For this characteristic to arise, we will need to allow for individual velocity protocols for each stirrer; this will be the case for the next optimization. 

We conclude this section with the remark that optimizing the shape for a fixed velocity along the circular paths does not alter the energy budget (even though an additional effort may be required due to an increased drag on the stirrers). As such, optimization of the stirrers' cross-sectional form may be an attractive alternative to augment mixing efficiency when an increase in energy is not feasible or advisable.  

\subsection{Velocity optimization for enhanced stirrer shapes}

After the cross-sectional shapes of the stirrers have been modified to enhance mixing, the natural next step in our quest for an improved mixing strategy is to optimize the path velocities of the modified stirrers along their circular trajectories. As the optimization results of this section are even more dynamic in nature than in the previous section, we once more prompt the reader to consult the relevant animations in the supplemental material for a full appreciation of the enhanced mixing process.

\begin{figure}[ht!]
  \centering
    \begin{tabular}{ccc}
    \begin{tikzpicture}
    \node at (0,0) {\includegraphics[width=0.3\textwidth]{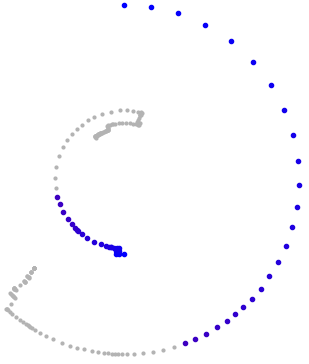}};
    \node at (-2.8,-2) {(a)};
    \node at (0,0) {$t \in [0,\ 3]$};
    \end{tikzpicture} & \phantom{123} & 
    \begin{tikzpicture}
    \node at (0,0)
    {\includegraphics[width=0.3\textwidth]{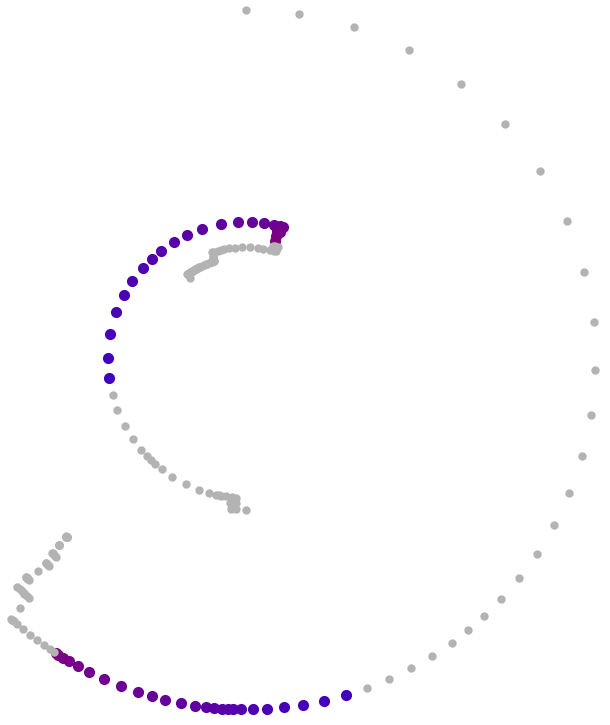}};
    \node at (-2.8,-2) {(b)};
    \node at (0,0) {$t \in [3,\ 6]$};   
    \end{tikzpicture} \\
    \begin{tikzpicture}
    \node at (0,0)
    {\includegraphics[width=0.3\textwidth]{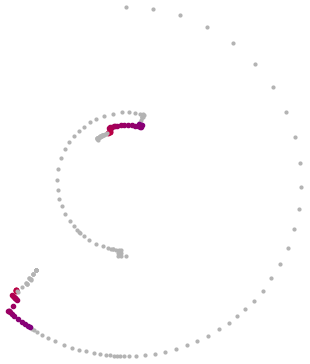}};
    \node at (-2.8,-2) {(c)}; 
    \node at (0,0) {$t \in [6,\ 9]$};
    \end{tikzpicture} &  & 
    \begin{tikzpicture}
    \node at (0,0)
    {\includegraphics[width=0.3\textwidth]{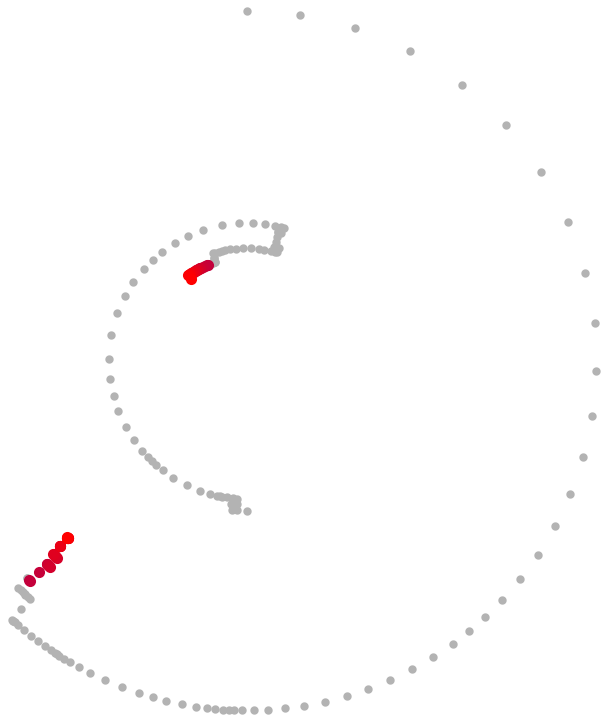}};
    \node at (-2.8,-2) {(d)};
    \node at (0,0) {$t \in [9,\ 12]$};
    \end{tikzpicture}
    \end{tabular}
    \caption{\label{fig:VelTrace} Velocity trace of the two stirrers, visualized by position samples at equispaced intervals in time. Each time the stirrers change direction, the radius is reduced by a small percentage to avoid visual overlay. The true stirrers, of course, remain on their circular paths. The entire optimization horizon is broken down into to equal temporal segments.} 
\end{figure}
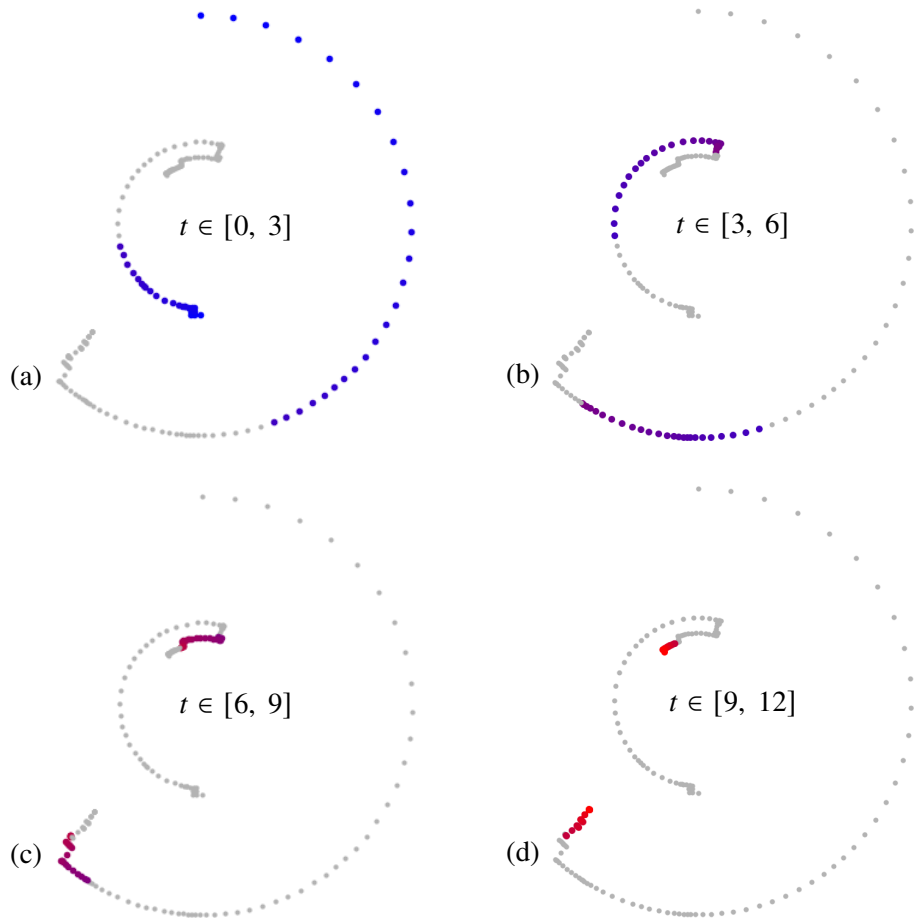

Optimizing the velocity along the circular path introduces the possibility of creating numerous start and stop vortices. While in the previous case, only one start vortex (at $t=0$) and one stop vortex (at $t=T$) resulted, we are now in a position to inject multiple vortical structures into the interior by manipulating the acceleration of the stirrers within the optimization window. In this manner, we produce 'hydrodynamic' stirring elements, beyond the two obvious solid stirrers. As a consequence of this degree of freedom, we have to impose additional bounds on the energy, velocity and acceleration of the stirrers to avoid divergences in the optimization scheme and, instead, arrive at a converged and realizable mixing strategy (see~\cite{Eggl2020} for details). But even with these bounds in place, we achieve a dramatic decrease in the mix-norm, indicating a superior mixing result. 

Once stirrers are allowed to alter their position (and velocity) in time along the path, cooperative strategies form where stirrers generate start-stop vortices which get either sheared by the wake or split by the appendages of the modified stirrers. Even after the stirrers cease to move, they often position themselves in advantageous locations and act as obstacles for the moving vortices. The consequence of either process is a swift generation of small-scale fluid structures, combined with a rapid breakdown of the heterogeneous mixture into a homogeneous blend. 

Figure~\ref{fig:VelTrace} displays the velocities along the circular paths for both stirrers. The visualization is based on a equispaced sampling in time, such that widely spaced symbols indicate fast motion, while densely spaced symbols point to slower motion. To conceptualize reverse motion, we artificially decrease the radius of the path by a small amount to avoid overlays. Figure~\ref{fig:VelTrace} breaks down the entire optimization horizon into four segments. The first segment $t\in [0,\ 3]$ shows a co-rotating outer and inner cylinder, with a strong initial acceleration of the outer cylinder followed by a noticeable slow down and an initial shaking motion of the inner cylinder. This segment is followed by an oscillatory motion (indicated by the varying density of the temporally equispaced points) of the outer cylinder, and a similar motion of the inner cylinder. The two later segments (see figure~\ref{fig:VelTrace}(c,d)) are characterized by a convulsive and reverse motion which (as we will see below) introduces start-stop vortices and interacts with the already generated vortices in the vessel. This motion is maintained until the end of the optimization window, when the two stirrers finally come to rest. 

\begin{figure}[ht!]
  \centering
  \begin{tabular}{ccc}
  \begin{subfigure}[b]{0.3\textwidth}
    \centering
    \includegraphics[trim=120 40 90 30 ,clip,width=\textwidth]{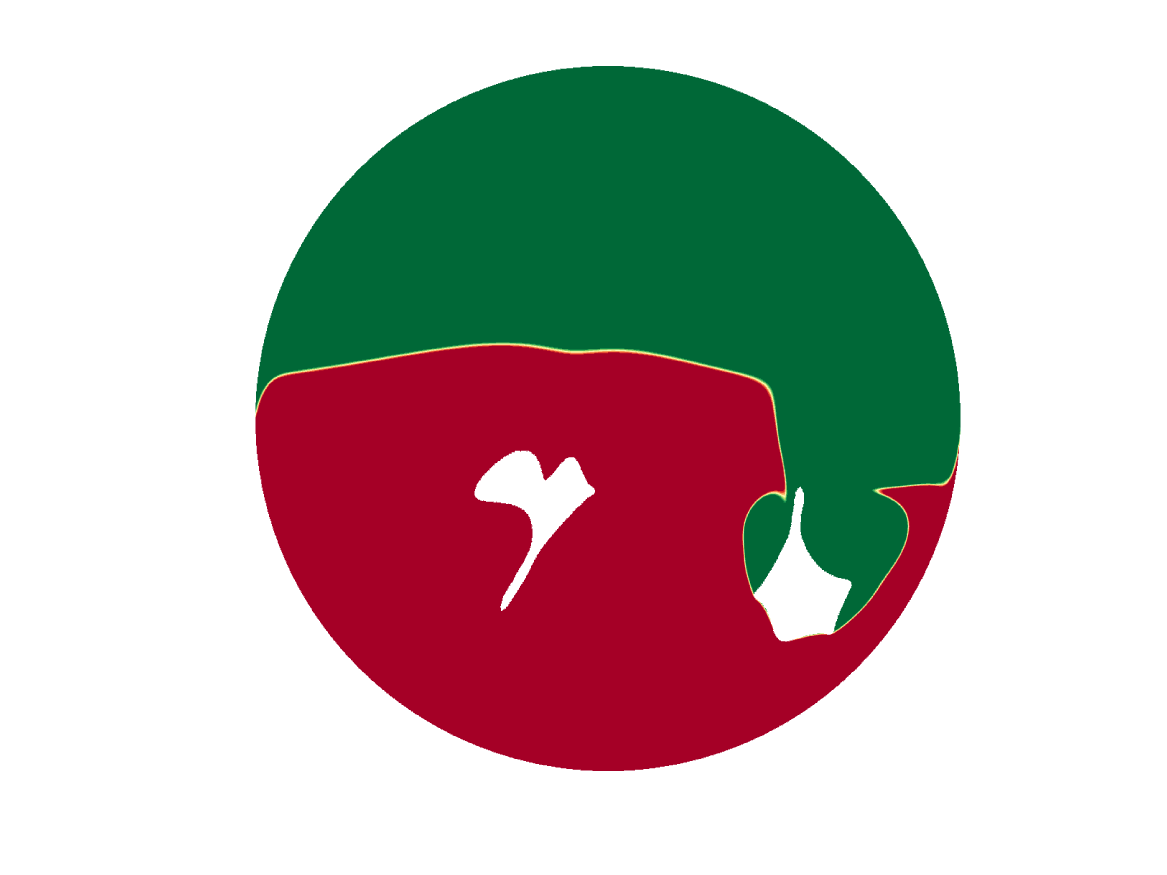}
    \caption{$t=2$}
    \label{fig:VelOpt_2}
  \end{subfigure} & 
  \begin{subfigure}[b]{0.3\textwidth}
    \centering
    \includegraphics[trim=120 40 90 30 ,clip,width=\textwidth]{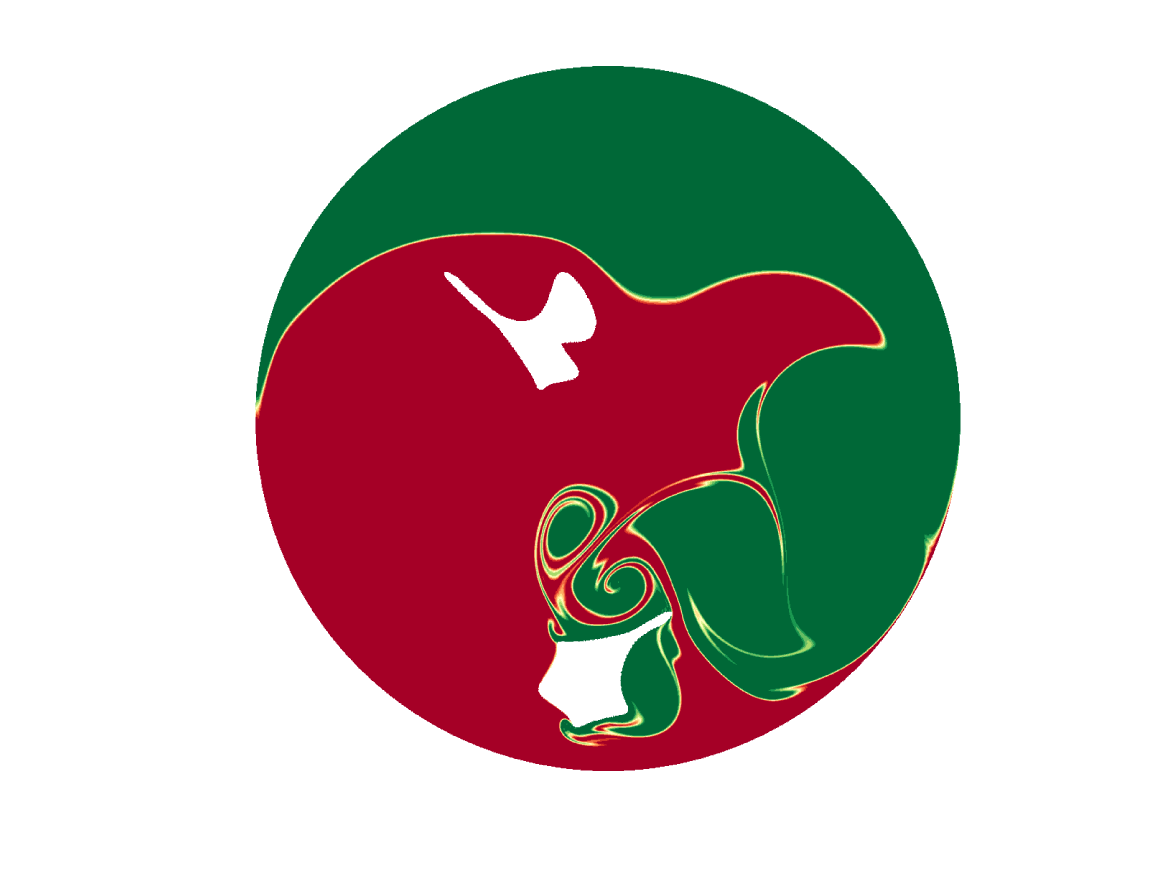}
    \caption{$t=4$}
    \label{fig:VelOpt_4}
  \end{subfigure} & 
  \begin{subfigure}[b]{0.3\textwidth}
    \centering
    \includegraphics[trim=120 40 90 30 ,clip,width=\textwidth]{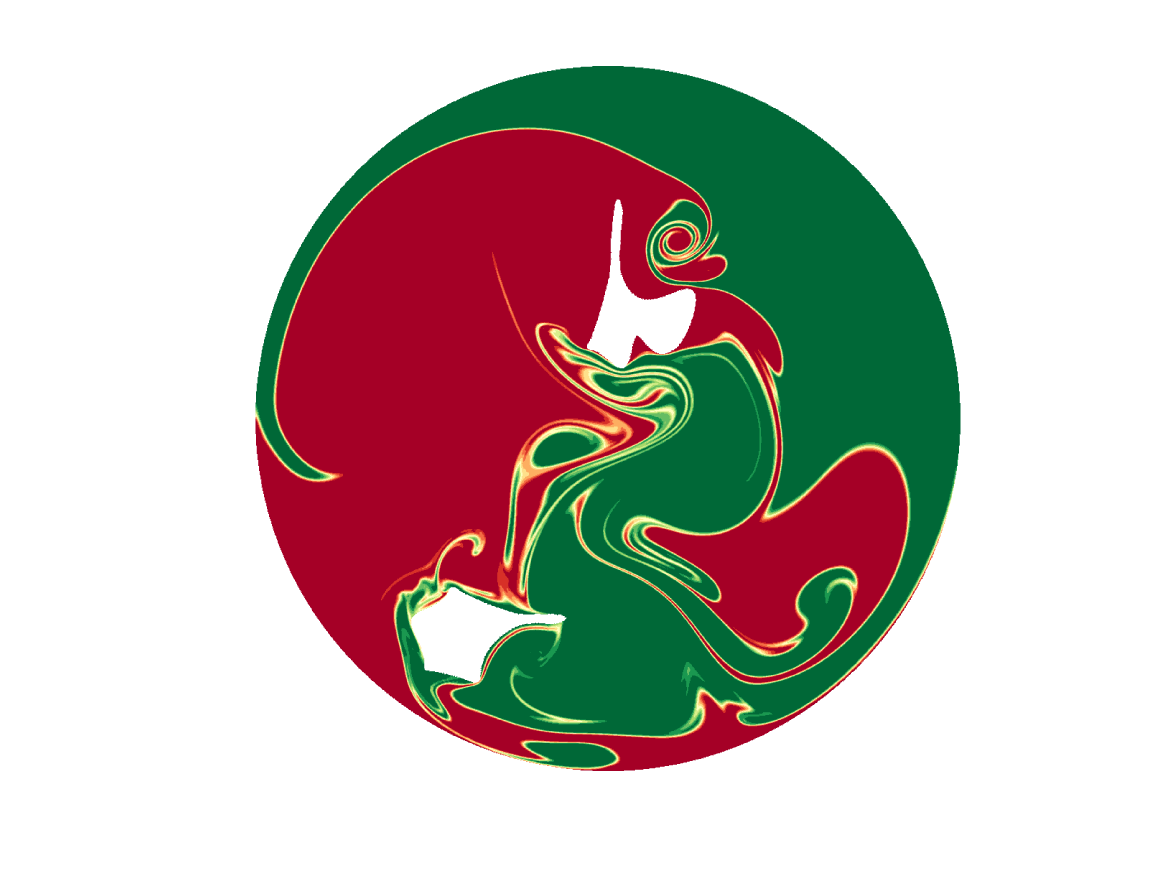}
    \caption{$t=6$}
    \label{fig:VelOpt_6}
  \end{subfigure} \\
  \begin{subfigure}[b]{0.3\textwidth}
    \centering
    \includegraphics[trim=120 40 90 30 ,clip,width=\textwidth]{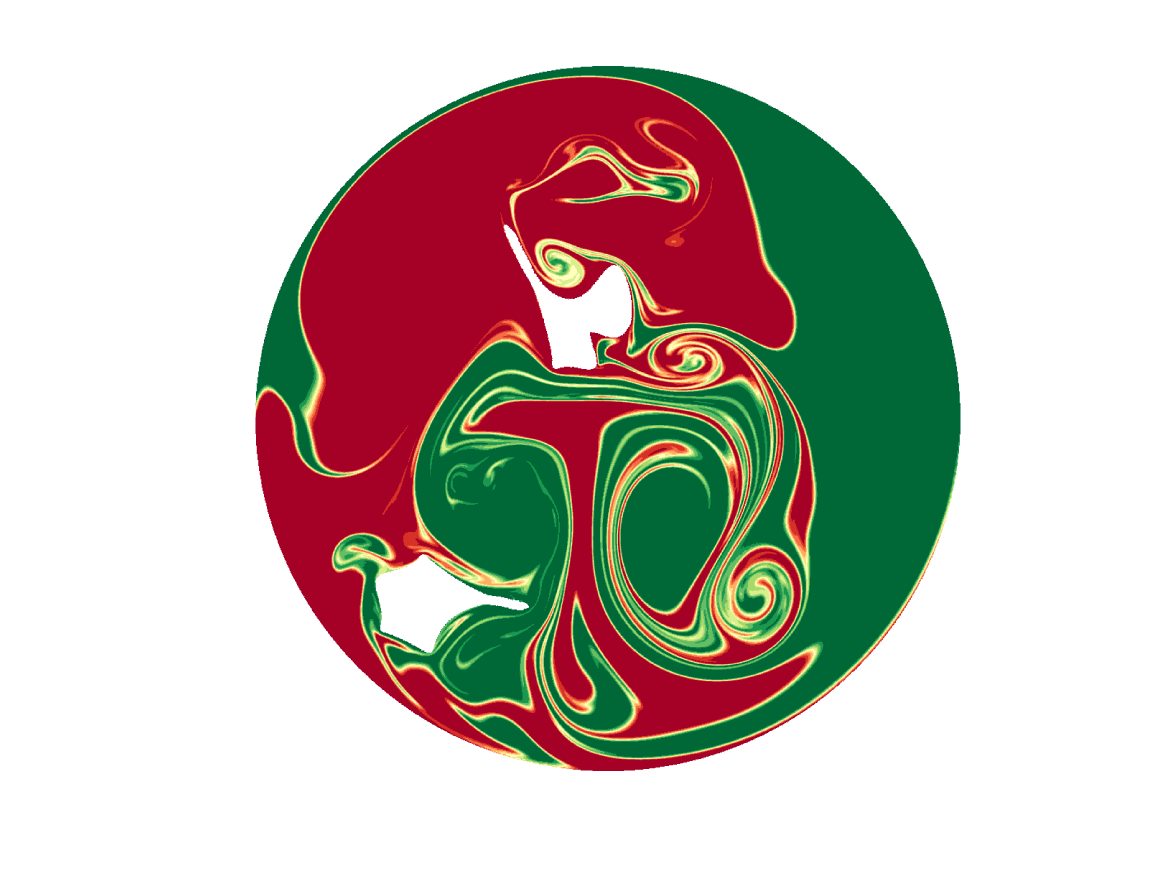}
    \caption{$t=8$}
    \label{fig:VelOpt_8}
  \end{subfigure} & 
  \begin{subfigure}[b]{0.3\textwidth}
    \centering
    \includegraphics[trim=120 40 90 30 ,clip,width=\textwidth]{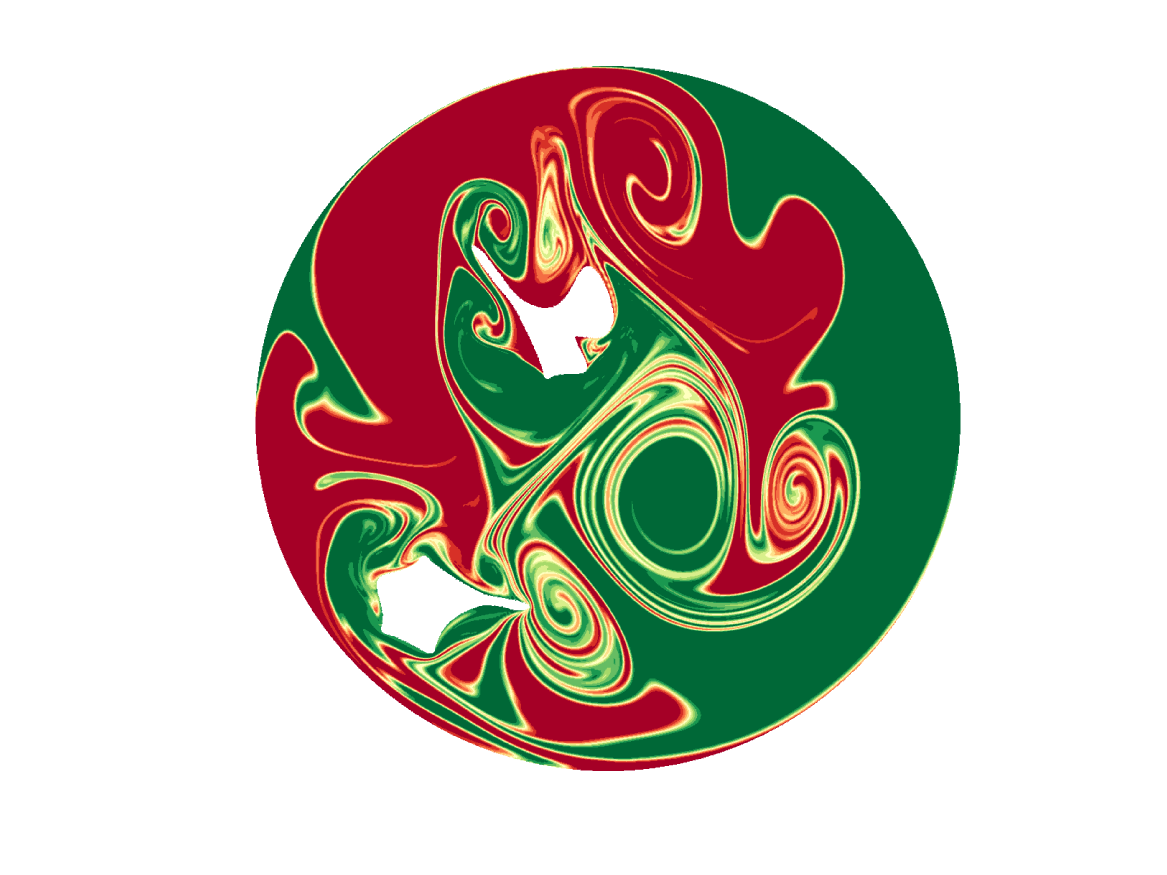}
    \caption{$t=10$}
    \label{fig:VelOpt_10}
  \end{subfigure} & 
  \begin{subfigure}[b]{0.3\textwidth}
    \centering
    \includegraphics[trim=120 40 90 30 ,clip,width=\textwidth]{figures/Velocity/012000.png}
    \caption{$t=12$}
    \label{fig:VelOpt_12b}
  \end{subfigure}
  \end{tabular}
  \caption{Selected snapshots of the final velocity optimization for the modified stirrer shapes. (a) Initial plunging through the interface by the outer stirrer, while the inner stirrer generates start-stop vortices. (b) Oscillatory motion of the outer stirrer and acceleration of the inner stirrer. (c,d) Reversal of both stirrers, allowing interaction of the stirrers with the generated vortices. (e,f) Small-amplitude rapid oscillation of the outer stirrer, while the inner stirrer gradually comes to rest. See {\tt{VelOpt.mp4}} from the supplemental material for an animation of the mixing process.}
  \label{fig:Velocity}
\end{figure}

With the modified shapes, the velocity optimization of the two stirrers yields a substantial decrease in the mix-norm and thus a better mixing result. Figure~\ref{fig:Velocity} depicts snapshots of the mixing process. The outer stirrer accelerates towards the interface and plunges through it. In the process, it creates wake vortices which further distort the interface. Towards the first quarter of the optimization window, the outer stirrer decelerates and engages in an oscillatory forward motion near the bottom of the mixing vessel. This unsteady motion induces enhanced vortex shedding which will ultimately produce thin filaments in the domain. During the same time window, the inner stirrer has generated start-stop vortices by initiating its trajectory with a rapid oscillatory and low-amplitude motion. The thin cross-stream appendage is used to full advantage in injecting vortical elements which will subsequently be further distorted by the outer stirrer (see figure~\ref{fig:Velocity}(b)). 

\begin{figure}[ht!]
  \centering
    \begin{tabular}{ccc}
    \begin{tikzpicture}
      \clip (0,0) rectangle (4cm,4cm);
      \node[anchor=center] at (1.5,3.0)
      {\includegraphics[width=10cm]{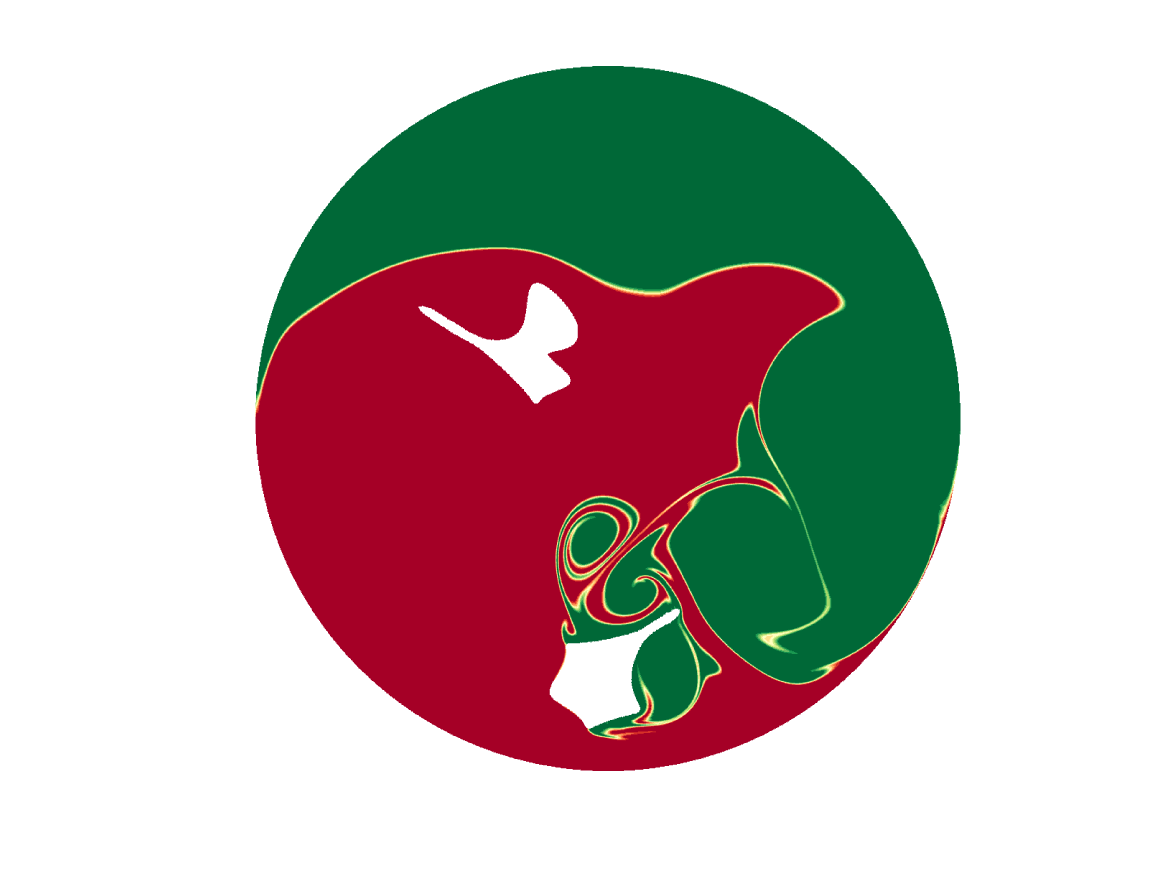}};
      \node[white,anchor=center] (e) at (3.5,3.5) {$(a)$};
      \node[anchor=center] (a) at (1.8,1.5) {};
      \node[anchor=center] (b) at (1.2,3.5) {};
      \node[anchor=center] (c) at (3,3.5) {};
      \draw[->,black,thick] (a) to [bend left=60] (b) {};
      \draw[->,black,thick,dashed] (b) to (c) {};
    \end{tikzpicture} & 
    \begin{tikzpicture}
      \clip (0,0) rectangle (4cm,4cm);
      \node[anchor=center] at (2.5,3.0) {\includegraphics[width=10cm]{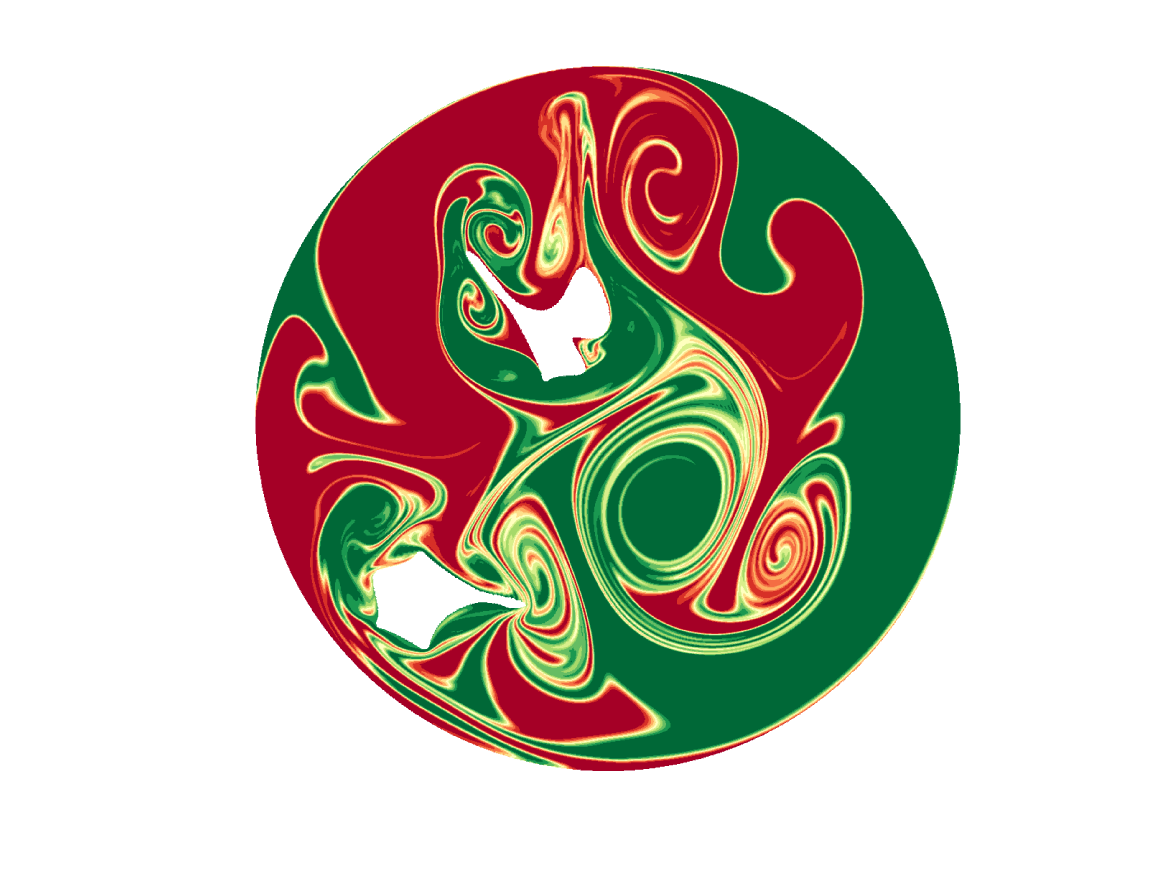}};
     \node[white] (f) at (0.5,3.5) {$(b)$};
     \node[anchor=center] (a) at (3,2.0) {};
     \node[anchor=center] (b) at (1.9,1.5) {};
     \draw[->,black,thick] (a) to (b) {};
  \end{tikzpicture}  &
  \begin{tikzpicture}
    \clip (0,0) rectangle (4cm,4cm);
    \node[anchor=center] at (3.0,0.5) {\includegraphics[width=10cm]{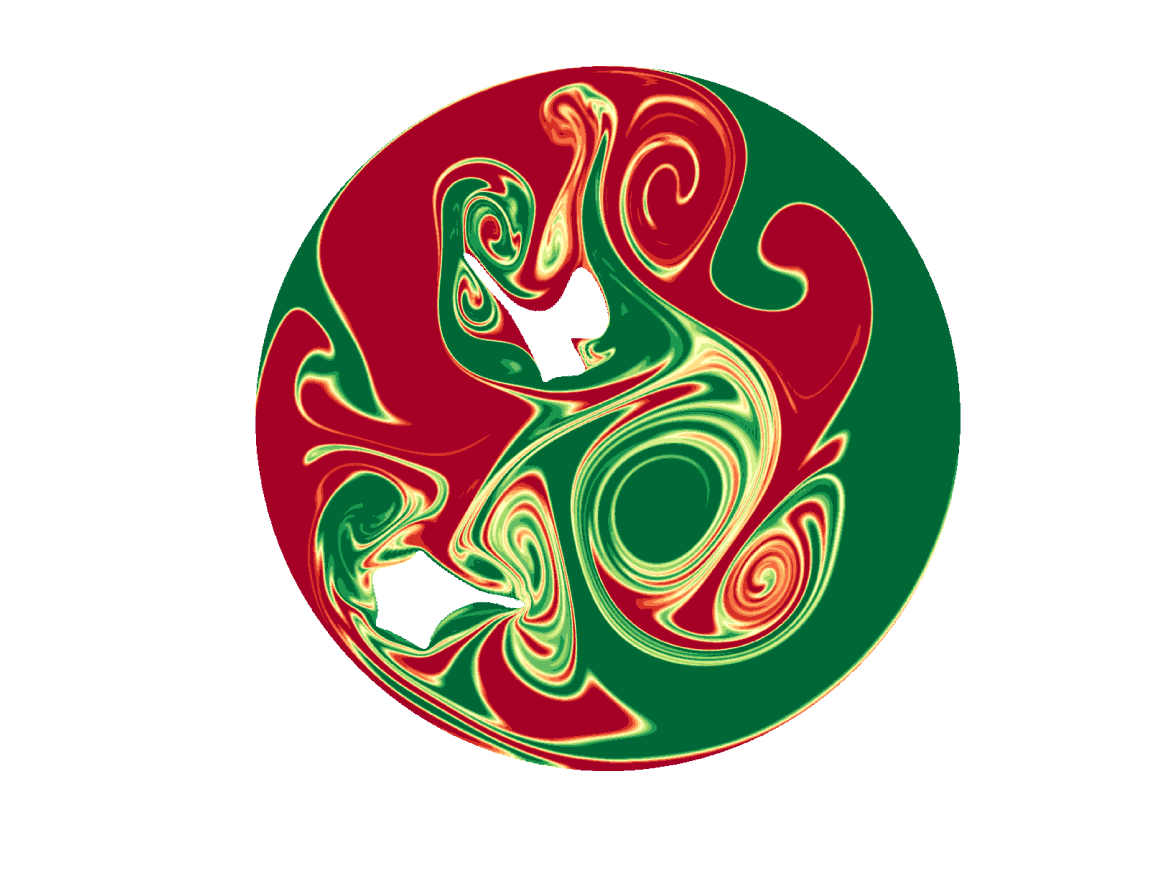}};
    \node (g) at (0.5,3.5) {$(c)$};
  \end{tikzpicture} 
  \end{tabular}
  \caption{Closeups of key features of the velocity-optimized stirring protocal. (a) Shedding of stop-vortex by the outer stirrer, aimed for impact on the inner stirrer. (b) Breakdown of large-scale vortical elements via the appendage of the outer stirrer. (c) Generation of shed vortices by oscillatory motion of the outer cylinder, near the end of the optimization horizon.}  
  \label{fig:VelOpt}
\end{figure}
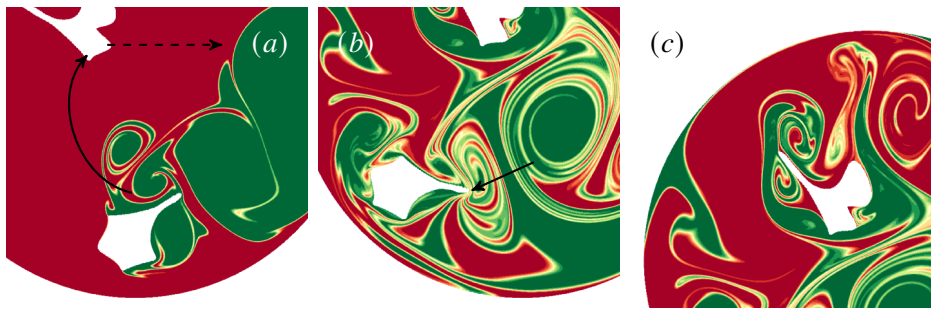
    
The goal of the optimization is the generation of a maximum of vortical element -- and their breakdown into smaller vortices -- within the optimization window. Beyond this time window, the introduced vortices will continue to interact with each other, the vessel wall and the stationary stirrers and cause further breakdown in scale due to rest inertia and ultimately diffusion. In this effort, the direct-adjoint optimization scheme takes advantage of the stirrers' edges, appendages and cavities to devise a velocity profile along the circular paths that maximizes vortex-vortex and vortex-body interactions to promote the breakdown of scales and the enhancement of mixedness. The optimal strategy is highly collaborative and, as a result, achieves significantly better mixing results than a simple shape optimization (without a concurrent velocity optimization). This tendency is the consequence of a far larger design space that is exploited by the direct-adjoint framework to produce more effective and efficient mixing protocols. 

\subsection{Combined shape-velocity optimization protocol}

The previous configuration consisted of a {\it{sequential}} application of two optimizations, for the cross-sectional shape of the stirrers and the velocity profile along circular paths for the modified shapes. In this effort, the velocity profile has been tailored to the modified shape, but no feedback from current results to the stirrers' shapes, via the shape gradient, has been allowed. In this final section, we 'unfreeze' the stirrer shapes and hence pursue a {\it{simultaneous}} optimization of the shapes and path velocities of the two stirrers. In this sense, we seek a multi-component optimum from our direct-adjoint framework.  

Both gradients -- the first based on the path velocity for each stirrer, and the second based on the cross-sectional shape of each stirrer -- aim at improving the mix-norm over the chosen optimization horizon. As these gradients enter the optimality condition, a choice has to be made as to how much of the improvement will come from a modified shape and how much will stem from an altered velocity protocol. Previous observations have confirmed a noticeably stronger overall response to velocity changes; topology changes, on the other hand, imposed a more subtle, but still perceptible effect on the mixedness of the binary fluid. A straightforward application of the direct-adjoint scheme then would favor an optimization that is chiefly focused on velocity changes, while discounting modifications of the stirrers' shapes. To counterbalance this intrinsic tendency and investigate the complex interaction of shape and velocity variations, we choose to underrelax the velocity-based gradient and give equal prominence to the shape-based gradient. 

\begin{figure}[ht!]
  \centering
  \begin{tabular}{cc}
    \begin{tikzpicture}
      \clip (0,0) rectangle (6cm,6cm);
      \node[anchor=center] at (3,3)
      {\includegraphics[width=8cm]{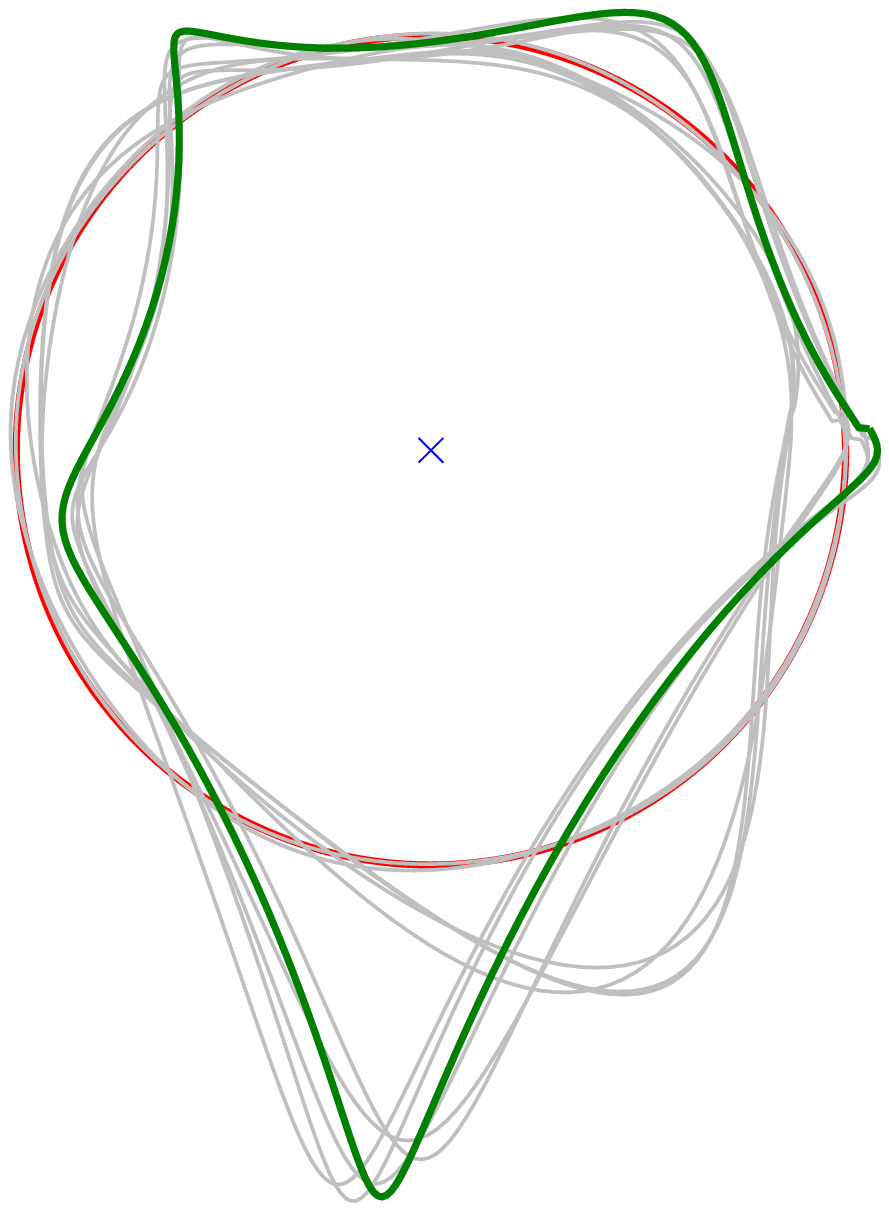}};
      \node (ar1a) at (4.0,2.6) {};
      \node (ar1b) at (2.8,0.6) {};
      \node (ar3a) at (2.4,5.0) {};
      \node (ar3b) at (2.0,5.6) {};
      \node (ar5a) at (3.8,5.0) {};
      \node (ar5b) at (4.2,5.6) {};
      \draw[->,black,thick]  (ar1a) to [bend left=60] (ar1b) {};
      \draw[->,black,thick]  (ar3a) to (ar3b) {};
      \draw[->,black,thick]  (ar5a) to (ar5b) {};
      \node at (2,1) {(a)};
    \end{tikzpicture} & 
    \begin{tikzpicture}
      \clip (0,0) rectangle (6cm,6cm);
      \node[anchor=center] at (3,3) {\includegraphics[width=8cm]{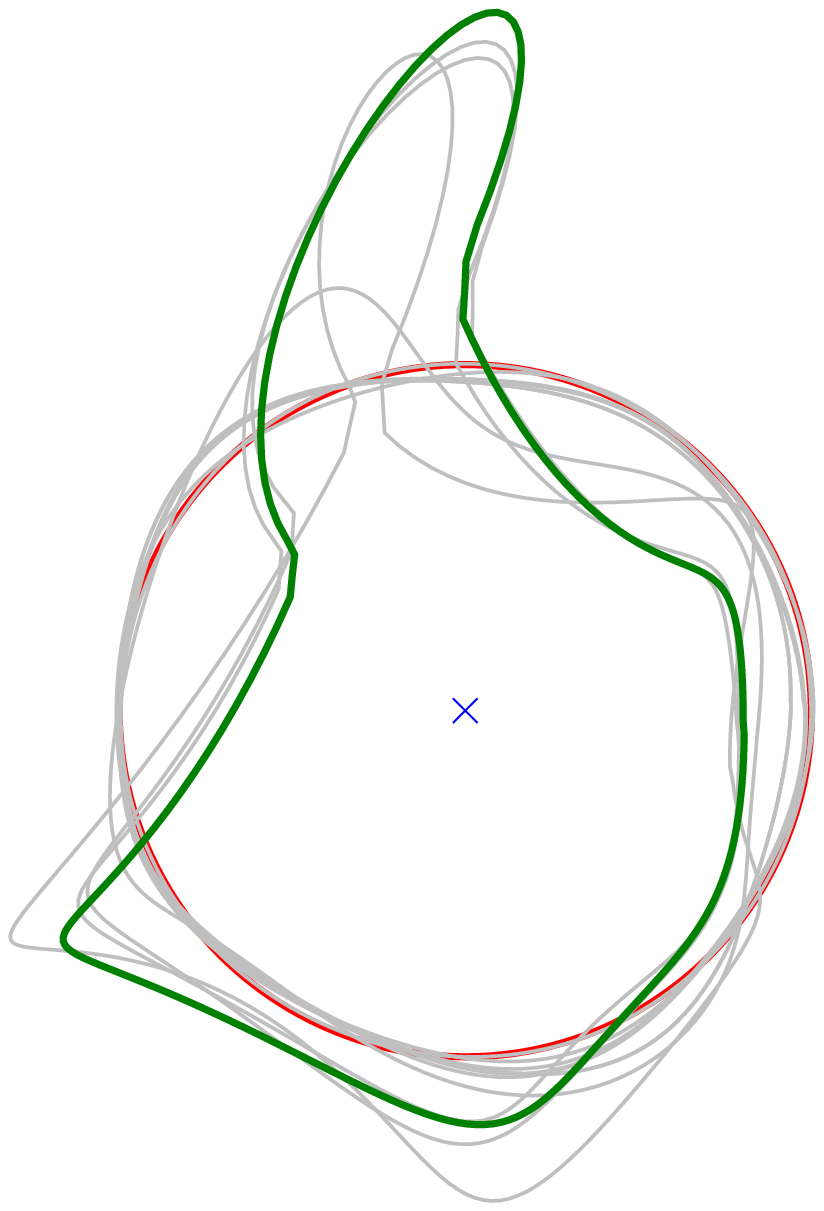}};
      \node (ar1a) at (2.4,2.0) {};
      \node (ar1b) at (1.6,1.6) {};
       \node (ar2a) at (3.2,1.4) {};
      \node (ar2b) at (3.4,0.8) {};
       \node (ar3a) at (4.1,4.0) {};
      \node (ar3b) at (3.5,3.6) {};
      \node (ar4a) at (2.2,3.5) {};
      \node (ar4b) at (3.5,5.7) {};
        \draw[->,black,thick]  (ar1a) to (ar1b) {};
    \draw[->,black,thick]  (ar2a) to (ar2b) {};
    \draw[->,black,thick]  (ar3a) to (ar3b) {};
    \draw[->,black,thick]  (ar4a) to [bend left=40] (ar4b) {};
    \node at (2,1) {(b)}; 
    \end{tikzpicture} 
    \end{tabular}
    \caption{Optimal cross-sectional shapes for both stirrers, using a combined shape-velocity optimization. The red and green curves signify the initial and final configuration, respectively, with gray shapes denoting the intermediate steps of the optimization procedure. Black arrows pinpoint features of interest, as they develop during the optimization process. (a) The outer stirrer develops an irregular pentagon shape, with one edge more pronounced. (b) The inner stirrer shows similar shape with one conspicuous extended blade.}
    \label{fig:ComboEvo}
\end{figure}

Figure~\ref{fig:ComboEvo} shows the evolution of the cross-sectional shapes of both stirrers over the course of the optimization. Starting from a circular cross-section, the optimization of the outer stirrer (figure~\ref{fig:ComboEvo}(a)) quickly develops a more complex shape with five corners, one of them more pronounced than the remaining four. The gray lines in the figure indicate a robust convergence towards a final shape after eight iterations. The inner stirrer (figure~\ref{fig:ComboEvo}(b)) undergoes a similar transformation from an initially circular cross-section to a more complicated shape, with more edges and a blade-like and curved protrusion. The gray lines, visualizing the intermediate optimization steps, show that in this case a regularization step had to be invoked to maintain a minimal thickness of the cross-sectional shape. Nonetheless, the convergence appears to be robust in this case as well. 

\begin{figure}[ht!]
  \centering
    \begin{tabular}{ccc}
    \begin{tikzpicture}
    \node at (0,0) {\includegraphics[width=0.3\textwidth]{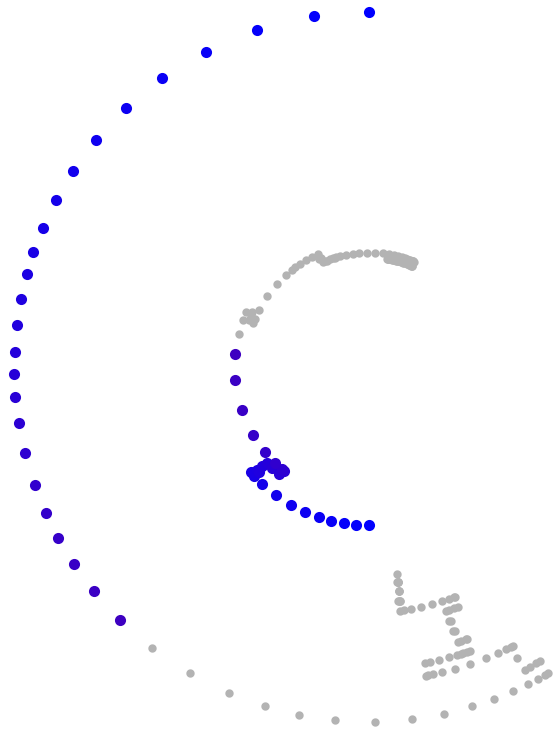}};
    \node at (-2.5,-2) {(a)};
    \node at (1,-0.15) {$t \in [0,\ 3]$};
    \end{tikzpicture} & \phantom{123} & 
    \begin{tikzpicture}
    \node at (0,0)
    {\includegraphics[width=0.3\textwidth]{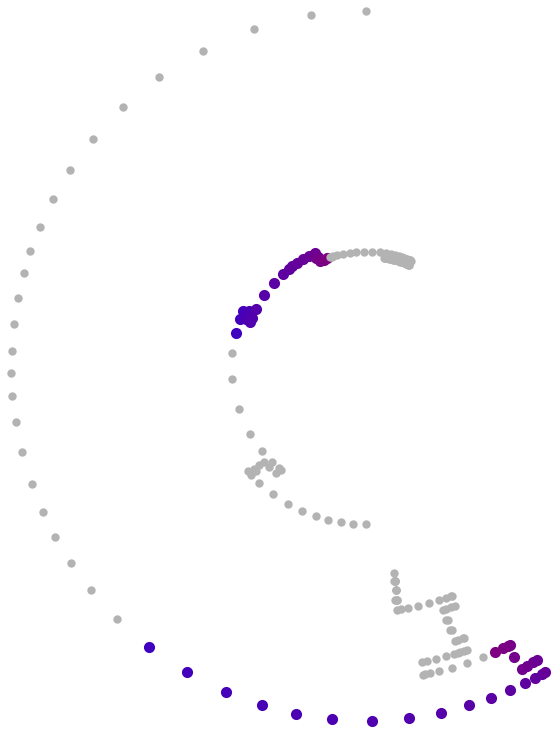}};
    \node at (-2.5,-2) {(b)};
    \node at (1,-0.15) {$t \in [3,\ 6]$};   
    \end{tikzpicture} \\
    \begin{tikzpicture}
    \node at (0,0)
    {\includegraphics[width=0.3\textwidth]{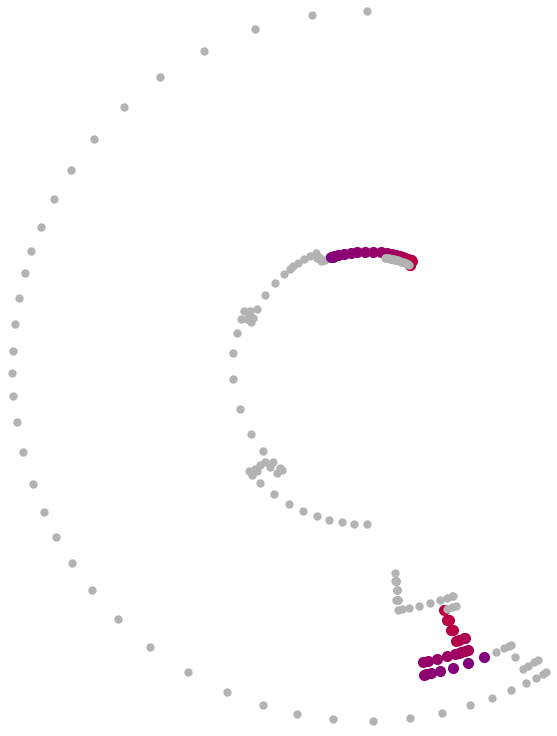}};
    \node at (-2.5,-2) {(c)}; 
    \node at (1,-0.15) {$t \in [6,\ 9]$};
    \end{tikzpicture} &  & 
    \begin{tikzpicture}
    \node at (0,0)
    {\includegraphics[width=0.3\textwidth]{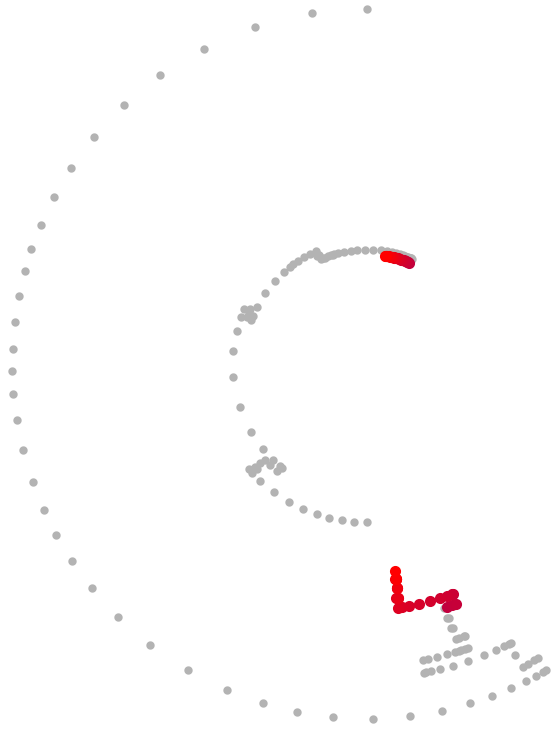}};
    \node at (-2.5,-2) {(d)};
    \node at (1,-0.15) {$t \in [9,\ 12]$};
    \end{tikzpicture}
    \end{tabular}
    \caption{\label{fig:ComboTrace} Velocity trace of the two stirrers for the combined shape-velocity optimization, visualized by position samples at equispaced intervals in time. Each time the stirrers change direction, the radius is reduced by a small percentage to avoid visual overlay. The true stirrers, of course, remain on their circular paths. The entire optimization horizon is broken down into to equal temporal segments.} 
\end{figure}

The corresponding velocities for the two stirrers -- which have been found simultaneously with the evolving shapes -- are displayed in figure~\ref{fig:ComboTrace} for the final (eighth) iteration. Again, the optimization horizon has been broken down into four temporal segments, for easier visualization. Figure~\ref{fig:ComboTrace}(a) shows that, in contrast to the previous case, the two stirrers run counter to each other, with the outer stirrer moving counter-clockwise and the inner cylinder clockwise. This choice strongly suggests the pursuit of a collaborative mixing strategies where both stirrers cooperate to introduce dynamic structures into the vessel during the early stages of the optimization horizon. 
Over the first quarter of the full window, the outer stirrer progresses towards the interface, but significantly slows down as it plunges through the interface (as evidenced by the higher density of symbols on the left edge of figure~\ref{fig:ComboTrace}(a)). This strategy causes shed vortices to overtake the stirrer and produces an enhanced distortion of the interface as a consequence. The inner stirrer follows a similar protocol, stopping and starting on its way towards the interface. The counter-moving travel of the stirrers, together with the  extended shapes, produces a close encounter of the stirrers and a strong and localized shear layer. 

The remaining temporal segments provide evidence of further vortex shedding, arising from the start-and-stop movement of the outer as well as inner stirrer. The liberal use of multiple and rapid changes in direction attest to the optimality of this strategy. The direct-adjoint scheme simply exploits this strategy and launches a multitude of shed vortices that accomplish the breakdown of heterogeneities during and -- more importantly -- beyond the optimization horizon (when the physical stirrers no longer move). 

\begin{figure}[ht!]
  \centering
  \begin{tabular}{ccc}
    \begin{subfigure}[b]{0.3\textwidth}
    \centering
    \includegraphics[trim=120 40 90 30 ,clip,width=\textwidth]{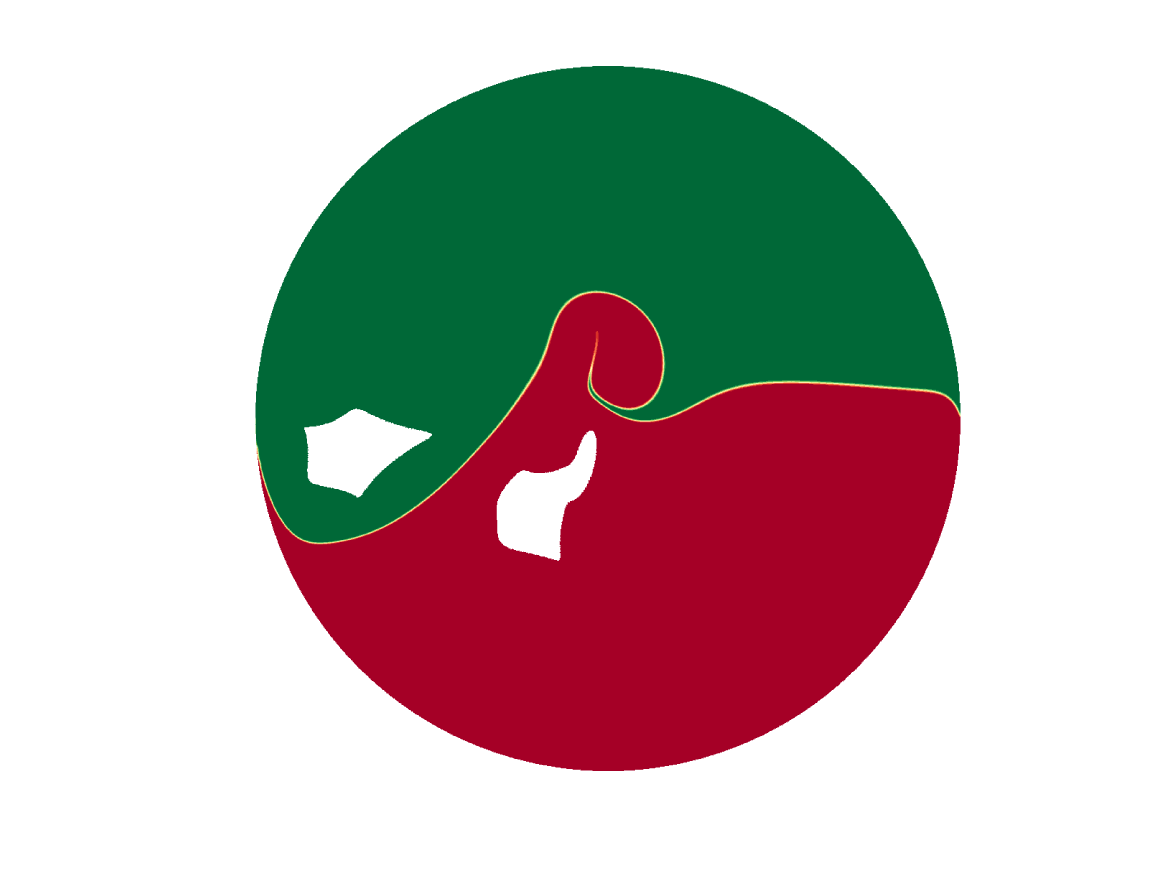}
    \caption{$t=2$}
    \label{fig:CombOpt_2}
    \end{subfigure} & 
    \begin{subfigure}[b]{0.3\textwidth}
    \centering
    \includegraphics[trim=120 40 90 30 ,clip,width=\textwidth]{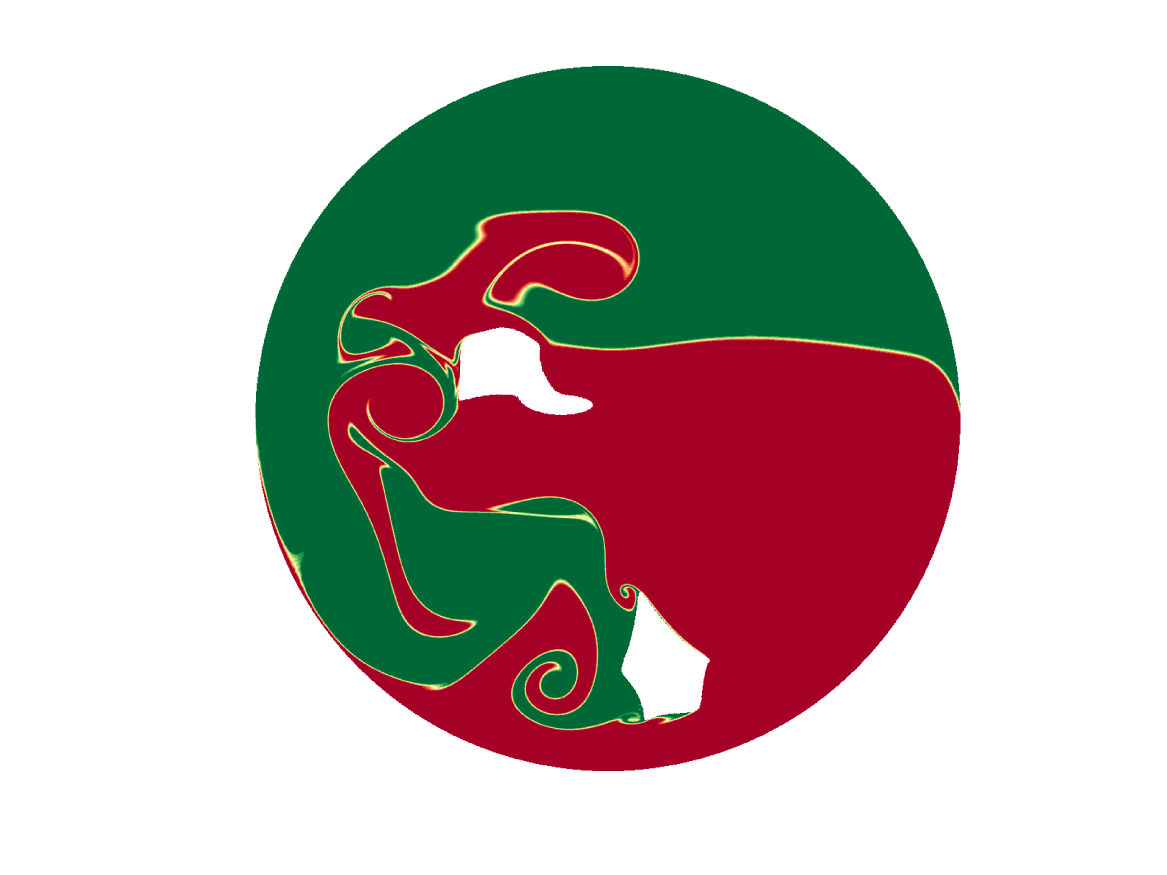}
    \caption{$t=4$}
    \label{fig:CombOpt_4}
    \end{subfigure} & 
    \begin{subfigure}[b]{0.3\textwidth}
    \centering
    \includegraphics[trim=120 40 90 30 ,clip,width=\textwidth]{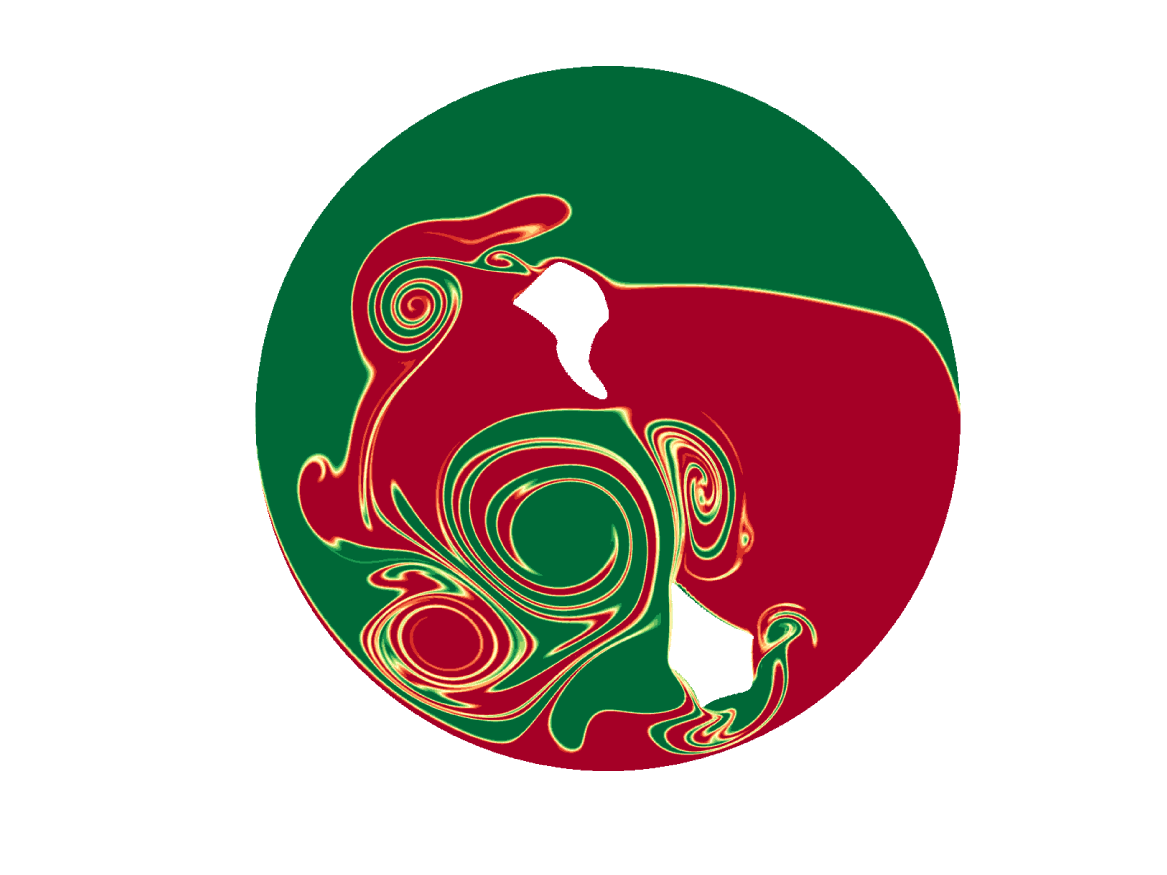}
    \caption{$t=6$}
    \label{fig:CombOpt_6}
    \end{subfigure} \\
    \begin{subfigure}[b]{0.3\textwidth}
    \centering
    \includegraphics[trim=120 40 90 30 ,clip,width=\textwidth]{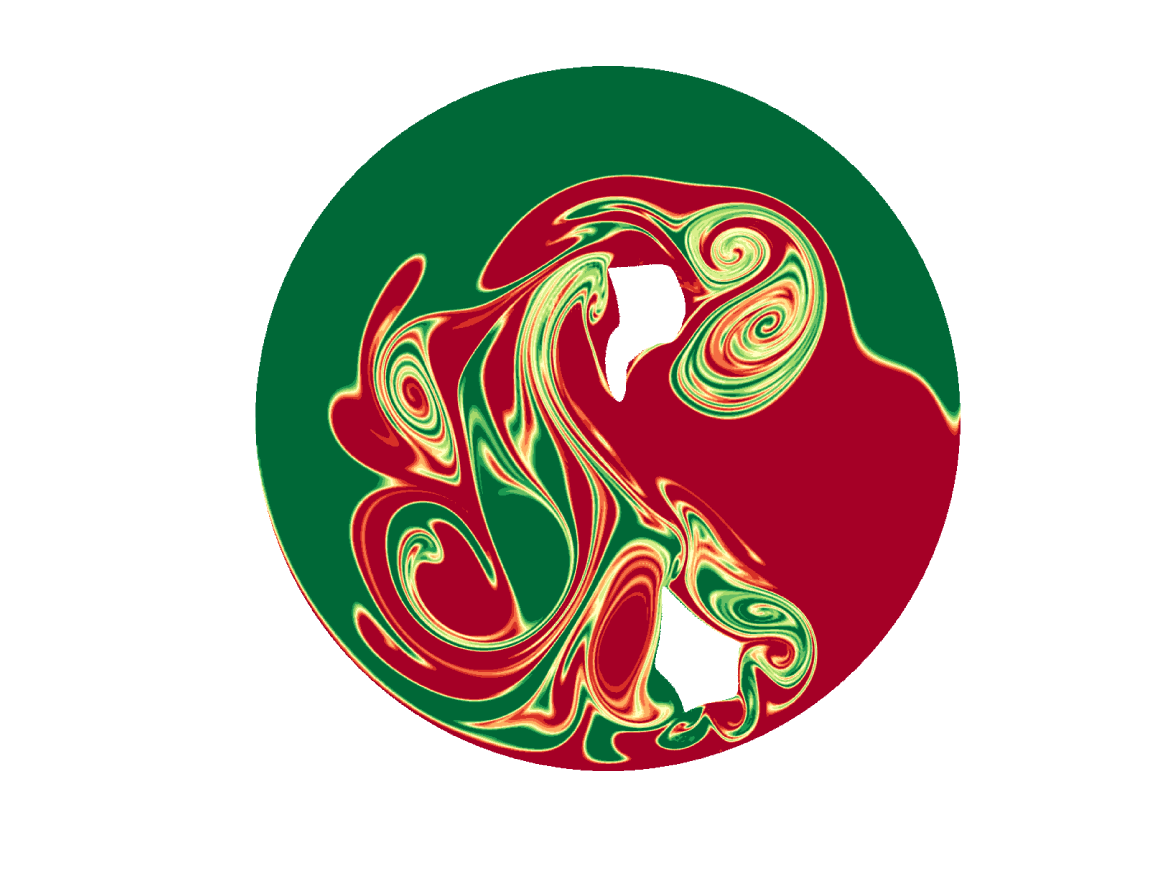}
    \caption{$t=8$}
    \label{fig:CombOpt_8}
    \end{subfigure} & 
    \begin{subfigure}[b]{0.3\textwidth}
    \centering
    \includegraphics[trim=120 40 90 30 ,clip,width=\textwidth]{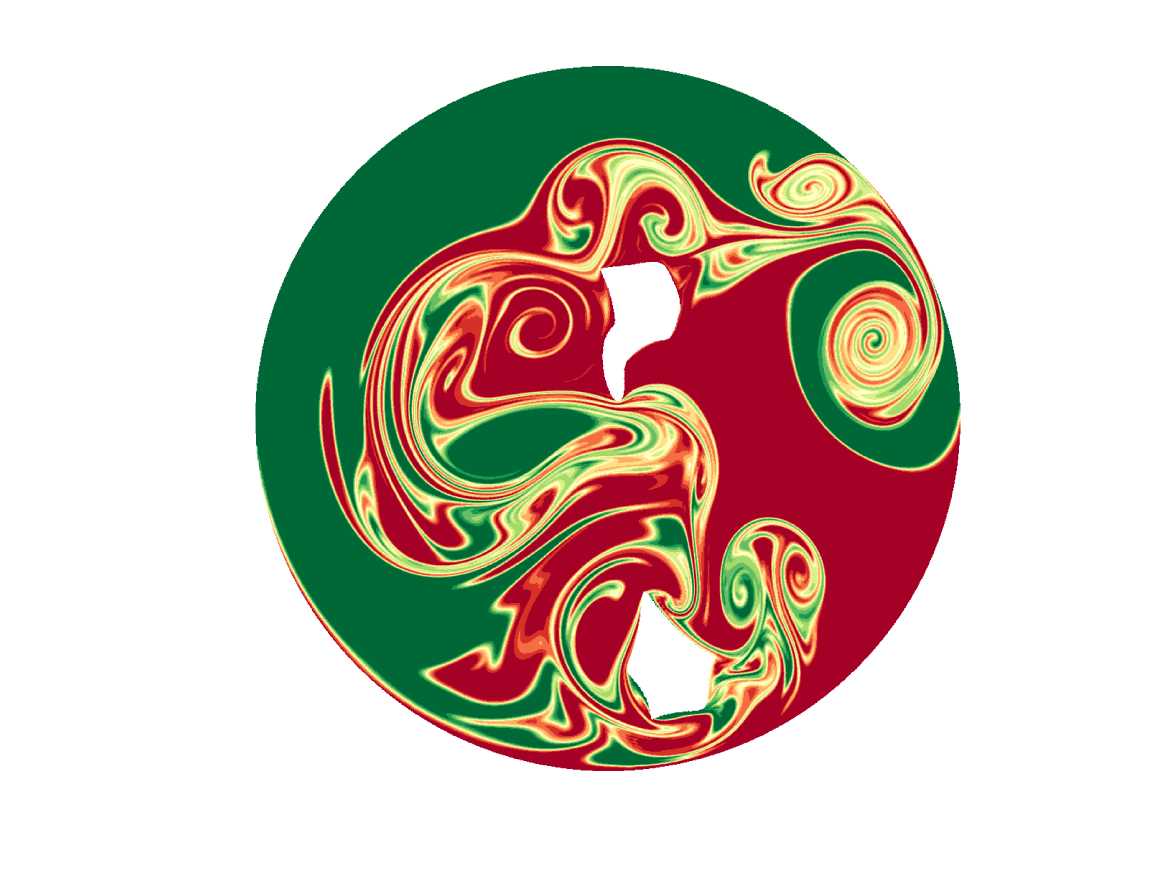}
    \caption{$t=10$}
    \label{fig:CombOpt_10}
    \end{subfigure} & 
    \begin{subfigure}[b]{0.3\textwidth}
    \centering
    \includegraphics[trim=120 40 90 30 ,clip,width=\textwidth]{figures/Combo/012000.png}
    \caption{$t=12$}
    \label{fig:CombOpt_12b}
    \end{subfigure}
  \end{tabular}
  \caption{Selected snapshots of the final combined shape-velocity optimization. (a) Initial approach of the interface by both stirrers, with deceleration-acceleration near the interface. (b) Acceleration of outer stirrer, deceleration of inner cylinder. (c) Oscillatory motion of the outer stirrer and near stagnation of inner stirrer. (d-f) Multiple reversals of outer stirrer and slow motion of inner stirrer, as the stirrers come to their respective resting position. See {\tt{ComboOpt.mp4}} from the supplemental material for an animation of the mixing process.}
  \label{fig:Combo}
\end{figure}

The success of this strategy can be assessed by inspecting the dynamic process over the course of the optimization window and beyond. Again, the animated version of this process (in the supplemental material) furnishes a far better perspective on this enhanced mixing. The still pictures in figure~\ref{fig:Combo} show that better mix-norm results have been obtained; even a qualitative assessment of the generated structures suggests a superior final result, when compared to previous cases. During the early movements of the stirrers, the close encounter of the two stirrers near $t=2$ is noteworthy. It creates a strong shear layer that is responsible for much of the subsequent interface distortion. The intensity of this shear layer is further aggravated by the elongated features of the stirrers which ensures an even closer brush of the two elements. Already at $t=4$ (see figure~\ref{fig:Combo}(b)) the initially flat interface is markedly contorted. The mixing process is subsequently supported by a complex stop-start sequence of both stirrers which further brings about the injection of spinning vortices into the stirred-up mixture. The interplay of all vortical components, large and small, and the interaction with the solid stirrers and the vessel wall causes a prevalence of thin filaments (see figure~\ref{fig:Combo}(e,f)) which ultimately succumb to diffusion. 

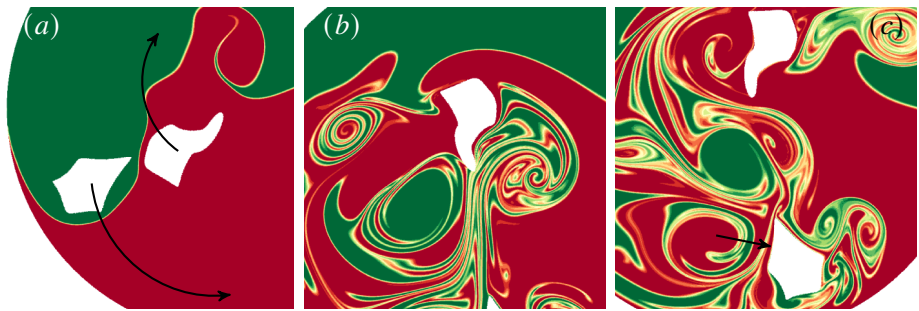
\begin{figure}[ht!]
\centering
  \begin{tabular}{ccc}
    \begin{tikzpicture}
    \clip (0,0) rectangle (4cm,4cm);
    \node[anchor=center] at (3.0,2.5) {\includegraphics[width=10cm]{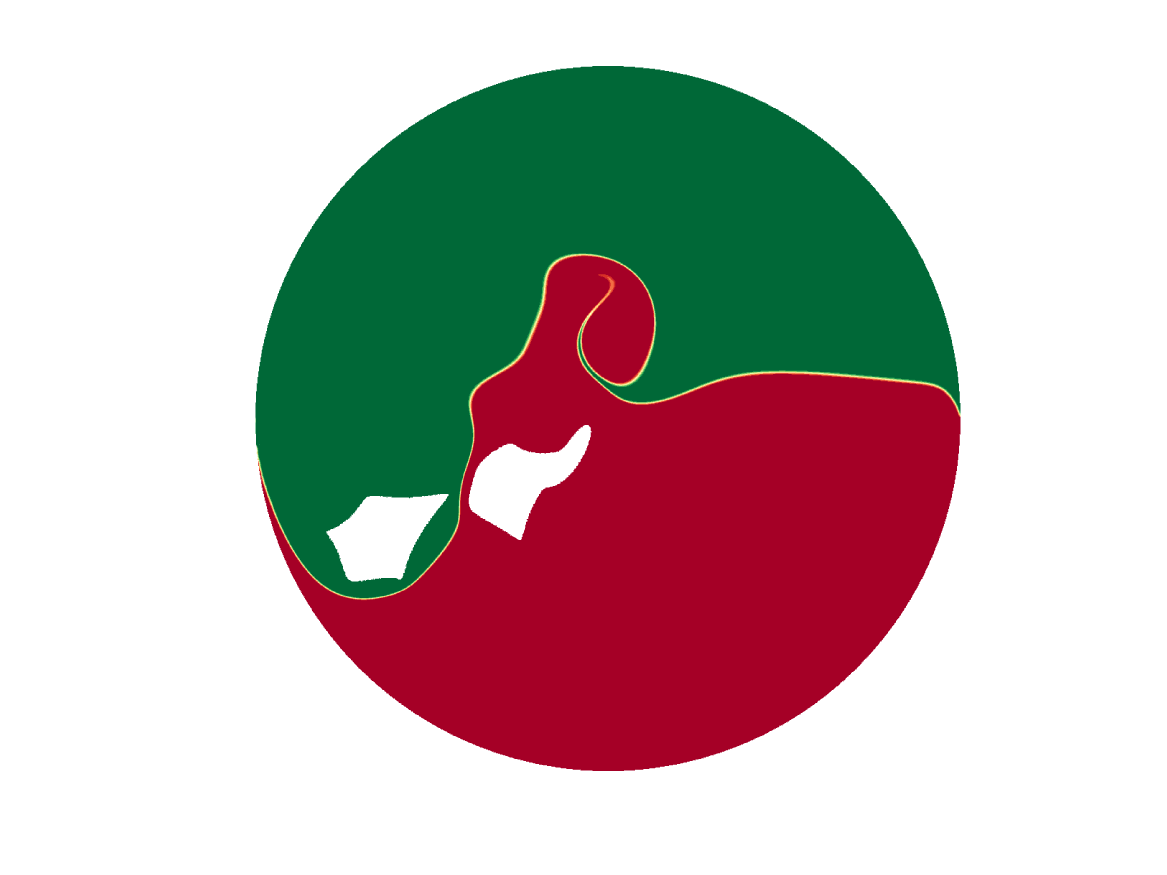}};
    \node[white] (g) at (0.65,3.75) {$(a)$};
    \node (a) at (1.3,1.8) {};
    \node (b) at (3.3,0.2) {};
    \node (c) at (2.6,2.0) {};
    \node (d) at (2.3,3.8) {};
    \draw[->,black,thick] (a) to [bend right=45] (b) {};
    \draw[->,black,thick] (c) to [bend left=45] (d) {};
    \end{tikzpicture}  &
    \begin{tikzpicture}
    \clip (0,0) rectangle (4cm,4cm);
    \node[anchor=center] at (2.0,1.5) {\includegraphics[width=10cm]{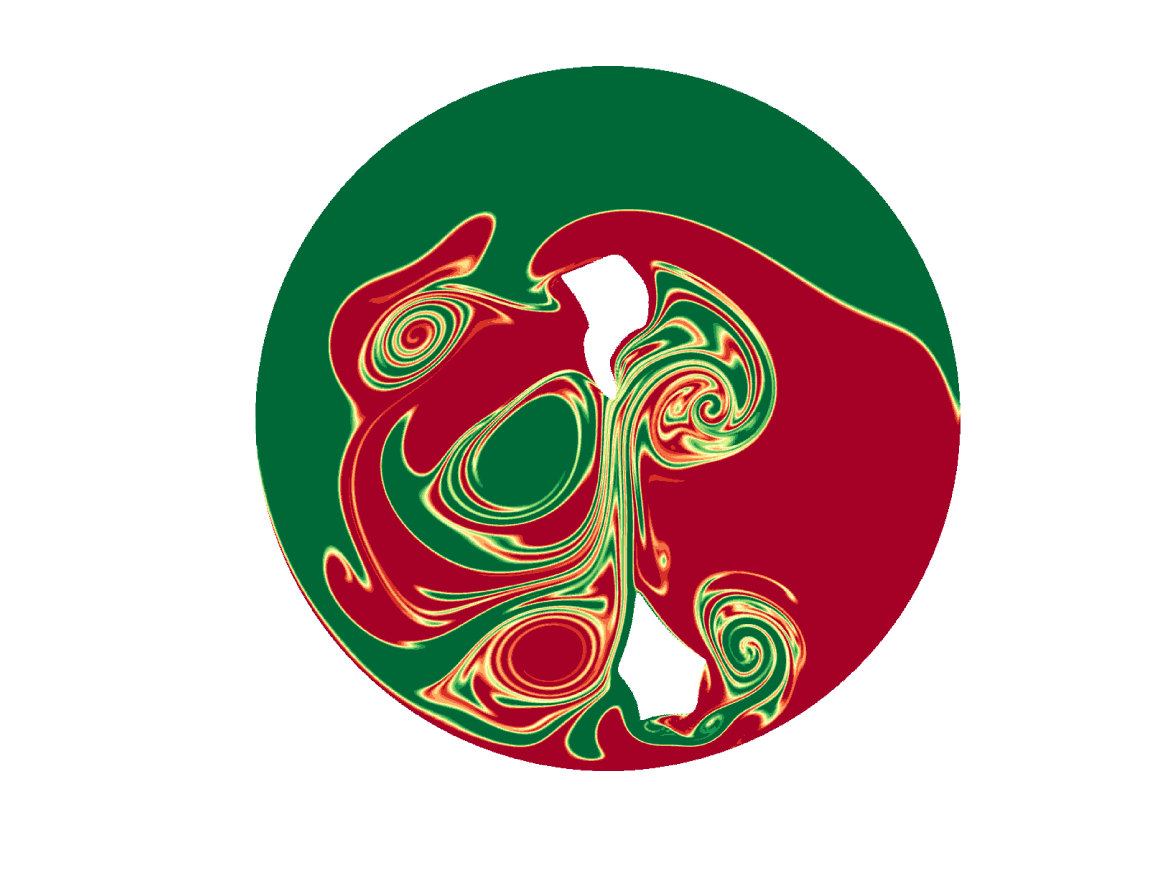}};
    \node[white] (g) at (0.5,3.75) {$(b)$};
    \end{tikzpicture} &
    \begin{tikzpicture} 
    \clip (0,0) rectangle (4cm,4cm);
    \node[anchor=center] at (1.5,2.5)
    {\includegraphics[width=10cm]{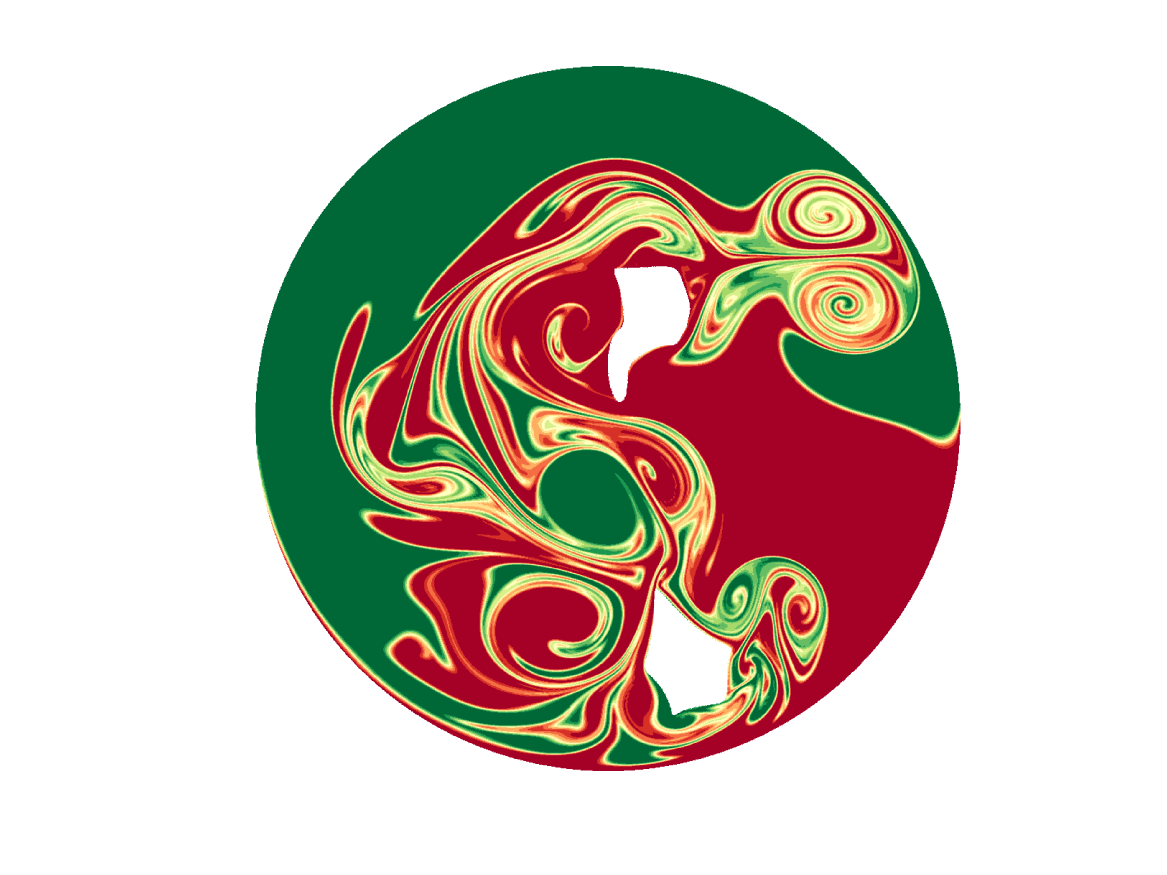}};
    \node (e) at (3.6,3.75) {$(c)$};
    \node[anchor=center] (a) at (1.2,1.0) {};
    \node[anchor=center] (b) at (2.2,0.8) {};
    \draw[->,black,thick] (a) to (b) {};
    \end{tikzpicture}
  \end{tabular}
  \caption{
  Closeups of key features of the shape-velocity-optimized stirring protocol. (a) Narrow passage of two stirrers, creating a strong shear layer and a subsequent roll-up of the interface. (b) Guiding and deflection of vortices by the inner stirrer. (c) Breakdown of vortices as they interact with the sharp protrusions of the outer stirrer.} 
  \label{fig:ComboOpt}
\end{figure}
   
Figure~\ref{fig:ComboOpt} highlights three key features from the optimal dynamic mixing process. The first detail (figure~\ref{fig:ComboOpt}(a)) showcases the narrow brush of the two stirrers early during the optimization horizon, where the two shapes nearly touch each other and pinch the interface between them. The resulting shear layer continues the mixing process in this section of the vessel long after the stirrers have moved towards other parts of the domain. The second feature (figure~\ref{fig:ComboOpt}(b)) demonstrates the impingement of vortical structures on the bespoke shapes of the stirrers, in this case the inner stirrer. The blade-like protrusion breaks down the scale of the incoming vortex, deflects its travel direction and induces in its cavity a counter-rotating smaller vortex which eventually detached from the cavity. A similar scenario can be observed in figure~\ref{fig:ComboOpt}(c) where a pair of sizeable stop-start vortices directly impinge on the bluff edge of the lower stirrer. Near the time of impact, the stirrer shifts abruptly to the left, intensifying the collision and reinforcing the subsequent breakup into smaller scales. This general scenario of vortex-generation and vortex-splitting can be observed throughout the latter part of the optimization window, and constitutes a particularly effective  mechanism chosen by the optimization scheme to enhance mixing. 
 
In summary, the composite and simultaneous optimization of the velocity protocols and stirrer shapes yielded a promising strategy for mixing enhancement. It is based on a cooperation of the two stirrers and an intricate interaction of multiple vortices at the local as well as global level. It should be kept in mind, however, that a concentration on the velocity-component of the optimization may produce the bulk of the gains in mixedness. Nonetheless, the transformed shapes of the stirrers can contribute additional, albeit more modest benefits.

\section{Summary and conclusions}
\label{sec:Conc}

A direct-adjoint optimization framework has been applied to enhance the mixing of a binary, incompressible, Newtonian fluid. Bound in a circular vessel, the fluid is mixed by two stirrers moving on two concentric circles. The cross-sectional shape of the stirrers, parameterized via Fourier series, and the velocities along the circular paths are optimized -- in series or combined -- to enhance the homogeneity of the passive scalar field, while observing a finite time horizon, a finite energy budget, and bounds on the velocities and accelerations~\citep{eggl2020mixing}. Despite an exceedingly large design space, the gradient-based procedure arrived at effective stirrer shapes and stirring protocols that yielded substantially improved mixing measures, expressed in the mix-norm of the passive scalar. 

Four cases have been considered: (i) a base case with cylindrical stirrers on prescribed circular paths, against which all subsequent optimization have been compared, (ii) an optimization of the cross-sectional shape, while carrying forward the previous path velocities, (iii) a pure velocity-optimization for the modified stirrer shapes, and (iv) a combined and weighted shape-and-velocity optimization of the mixing process. Starting from the reference case, each subsequent optimization results in a lower mix-norm over the optimization horizon of $T=12,$ and thus a more homogeneous mixture. The accomplished increase in mixing efficiency is summarized in table~\ref{tab:Cases} which lists the mix-norm values for all simulations, together with the number of direct-adjoint iterations. 

We observe that a pure shape optimization, disregarding the velocities of the stirrers, will result in a noticeable but rather moderate reduction in mix-norm. Only a decrease of $4\%$ could be achieved by manipulating the cross-sectional shapes of the stirrers. While this improvement may translate into a more sizeable saving of energy and efforts within an industrial setting, a far larger gain can be accomplished by targeting the velocities of the two stirrers. The targeting of this degree of freedom introduces the option of a collaborative strategy between the two stirrers -- an option that was largely unavailable during the pure shape-optimization. Vortex generation and interaction, plunging, unsteady vortex shedding, vortex collisions with themselves, the wall and the stirrers are at the optimizer's disposal to cast into a comprehensive mixing strategy. This optimization may lead (and has led) to counter-intuitive and complex stirring protocols that aid in the breakdown of the initial structures into small-scale elements that eventually dissipate. The efficacy of this more dynamic stirring strategy is reflected in the substantial drop in mix-norm (see table~\ref{tab:Cases}): by optimizing the velocities for the modified shapes (case (iii)), we exhibit a further $34\%$ drop in mix-norm from the previous case. By pursuing a combined optimization of the shapes and velocities of the stirrers, we can extract an additional increase in mixing efficiency, compared to the sequential optimization approach (case (iii) following case (ii)). We experience another $31\%$ decrease in mix-norm from case (iii) by using a simultaneous shape-velocity optimization. This results in the lowest value (and thus the best mixing result) for all cases considered in this study; it constitutes a $57\%$ improvement over the unoptimized base case (case (i)). 

The lowest mix-norm strategy (case (iv)) contains a variety of processes that complement each other, interact and cooperate to maximally increase the presence of thin filaments (in the passive scalar) and the dominance of shearing structures. The desired state of the fluid system at the end of the optimization window is characterized by a flow configuration that is maximally prone to rest inertia and species diffusion. The evolution of the seeded structures beyond the optimization horizon (see the supplemental material) shows a continued breakdown of scales and the convergence towards a maximally homogeneous mixture. The direct-adjoint optimization scheme proved capable and robust in finding effective and efficient stirring strategies within a small number of iterations. 

Despite this success, several challenges and opportunities remain. First, and foremost, is the inability of any gradient-based optimization approach applied to a nonlinear partial differential equation to produce global optimality. This shortcoming is due to the fact that we are dealing with a non-convex optimization problem, for which any derivative-based scheme can only guarantee a local optimum. While this issue may limit our results, is it important to realize that we achieved a substantial improvement (a $57\%$ reduction in mix-norm) which, under any industrial setting, would translate into vast savings in energy and resources.

A further challenge lies in the extension of the optimization horizon. We have chosen the parameters of our study (notably the Reynolds number, P\'eclet number and time horizon) to allow a rich dynamic range of motion, but, at the same time, avoid the influence of turbulence. As we rely on access to gradient information, the presence of turbulent fluid motion will obscure the sensitivity information needed for the shape and/or velocity optimization. 
While techniques to overcome this predicament (such as least-squares shadowing, see~\citep{Blonigan2012,blonigan2017adjoint}) are currently investigated, their cost-effectiveness applied to systems of the present complexity is still an open problem.

Besides these challenges, the current study demonstrated a great amount of flexibility in its approach, which prompts a rich set of possibilities for further optimization studies. 

These studies would include, for example, optimizing the wall shape and motion -- relinquishing the circular shape and its static nature. A moving/oscillatory and corrugated wall could join the stirrers in the generation and manipulation of vortical structures. A similar optimization could target the path of the stirrers, from concentric circles to more complicated closed curves. Introducing static baffles which break up the vortical motion may constitute yet another interesting research direction. 

An extension of the fluid model to non-Newtonian constitutive behavior is an opportunity to describe a wide range of agents used in industrial mixing~\citep[see,e.g.,][]{chhabra1999non,peng2014application}, as is the treatment of three-dimensional geometries, stirrers and fluid motion. 

The presented direct-adjoint optimization approach has shown its mettle on a variety of inertia-dominated mixing problems for a binary fluid. It is capable of addressing many of the above-mentioned generalizations and extensions, and future efforts in this direction are expected to further contribute to enhanced mixing strategies via optimized stirring protocols.

\section{Backmatter}
\paragraph{Acknowledgements}
We gratefully acknowledge discussions and exchanges with Profs. Colm Caulfield, Jean-Luc Thiffeault, Kai Schneider and Dr. Florence Marcotte.  

\paragraph{Funding Statement}
M.E. gratefully acknowledges funding through the Joachim Herz Stiftung.

\paragraph{Declaration of Interests}
The authors declare no conflict of interest.

\paragraph{Author Contributions}
Both authors created the research plan, designed the numerical experiments and wrote the manuscript. M.E. performed the code modifications and performed the numerical simulations.

\paragraph{Data Availability Statement}
N.A.

\paragraph{Ethical Standards}
The research meets all ethical guidelines, including
adherence to the legal requirements of the study country.

\paragraph{Supplementary Material}
Supplementary material (animations) are available
at ... .

\bibliography{PRFpaper}

\end{document}